\documentclass[epjST]{svjour}

\usepackage{graphicx}
\usepackage{amsfonts,amssymb,amsmath}
\usepackage{url}
\usepackage{bm}

\begin{document}

\title{The canonical equilibrium of constrained molecular models}

\author{
Pablo Echenique\inst{1,2,3}\fnmsep\thanks{echenique.p@gmail.com} \and
Claudio N. Cavasotto\inst{4,5} \and
Pablo Garc\'{\i}a-Risue\~no\inst{1,2,3}
}

\institute{
Instituto de Qu\'{\i}mica F\'{\i}sica ``Rocasolano'', CSIC, Serrano 119, 
E-28006 Madrid, Spain \and
Instituto de Biocomputaci\'on y F{\'{\i}}sica de Sistemas Complejos (BIFI), 
Universidad de Zaragoza, Mariano Esquillor s/n, Edificio I+D, E-50018 
Zaragoza, Spain \and
Departamento de F{\'{\i}}sica Te\'orica, Universidad de Zaragoza, Pedro 
Cerbuna 12, E-50009 Zaragoza, Spain \and
School of Biomedical Informatics, University of Texas Health Science 
Center at Houston, 7000 Fannin, Ste. 690, Houston, TX 77030, USA \and
Department of NanoMedicine and Biomedical Engineering,
University of Texas Health Science 
Center at Houston, 7000 Fannin, Ste. 690, Houston, TX 77030, USA
}

\abstract{
In order to increase the efficiency of the computer simulation of biological
molecules, it is very common to impose holonomic constraints on the fastest
degrees of freedom; normally bond lengths, but also possibly bond angles.
Since the maximum time step required for the stability of the dynamics is
proportional to the shortest period associated with the motions of the system,
constraining the fastest vibrations allows to increase it and, assuming that
the added numerical cost is not too high, also increase the overall efficiency
of the simulation. However, as any other element that affects the physical
model, the imposition of constraints must be assessed from the point of view
of accuracy: both the dynamics and the equilibrium statistical mechanics are
model-dependent, and they will be changed if constraints are used. In this
review, we investigate the accuracy of constrained models at the level of the
equilibrium statistical mechanics distributions produced by the different
dynamics. We carefully derive the canonical equilibrium distributions of both
the constrained and unconstrained dynamics, comparing the two of them by means
of a ``stiff'' approximation to the latter. We do so both in the case of
flexible and hard constraints, i.e., when the value of the constrained
coordinates depends on the conformation and when it is a constant number. We
obtain the different correcting terms associated with the kinetic energy
mass-metric tensor determinants, but also with the details of the potential
energy in the vicinity of the constrained subspace (encoded in its first and
second derivatives). This allows us to directly compare, at the conformational
level, how the imposition of constraints changes the thermal equilibrium of
molecular systems with respect to the unconstrained case. We also provide an
extensive review of the relevant literature, and we show that all models
previously reported can be considered special cases of the most general
treatments presented in this work. Finally, we numerically analyze a simple
methanol molecule in order to illustrate the theoretical concepts in a
practical case.
}

\maketitle

\section{Introduction}
\label{sec:introduction}

Since its first practical applications in the 1950's (see \cite{And2009JCTC}
and references therein for a detailed historical account), molecular dynamics
(MD) has become a powerful and well-established tool for the study of a wide
range of systems, including but not limited to condensed-matter materials
\cite{Rap2004Book}, fluids \cite{All2005Book,Hun2011EPJSTunp}, polymers and
biological molecules \cite{Fre2002Book}.

Among the latter, one of the most important families of systems is that of
proteins. Together with nucleic acids, proteins are one of the main
responsible of the enormous complexity and versatility that living beings
exhibit. Although they are mostly constructed as linear chains of only twenty
different monomers (the twenty \emph{proteinogenic} amino acids), and these
monomers, in turn, are made of only five types of atoms (H, C, N, O and S),
the finely tuned combination of these elements is capable of producing
remarkable `nano-machines' that perform almost every task that is complex in
biological organisms \cite{Ech2007CoP}. Thus, it is not surprising that a
great effort has been invested in the last years in devising new theoretical
and computational methods to describe and simulate these important
biomolecules. In particular, many groups are pushing forward very ambitious
scientific and technological agendas to be able to perform longer and more
accurate MD simulations of increasingly larger proteins by different
approaches \cite{Ens2007JMB,Fre2009BPJ,Sha2009Proc}. In this work, we have in
mind the long-term goal of producing more efficient methods for simulating
proteins, however, the formalism we introduce and the calculations we perform
are directly applicable to any molecular system, as long as it size makes it
manageable in present day computers.

The \emph{efficiency} of a MD simulation is customarily defined as a suitable
relation (e.g., the quotient) between its accuracy and its computational cost
\cite{Ech2008JCC}. The overall accuracy can be more finely divided into that
related to the physical model used to describe the system and the accuracy of
the method used for implementing the dynamics. The computational cost, on the
other hand, can be split into that associated with increasing the size of the
system and the cost related to the time-propagation of the dynamics
\cite{Kle2009COSB}. The fact that all these issues are intricately coupled is
easily seen if we consider examples such as the comparison between ab initio
MD based on the ground-state Born-Oppenheimer approximation
\cite{And2009JCTC,Mar2000TR,Alo2008PRL,Nik2006PRL,Ani2009CPC} and MD
simulations using as energy functions the ones known as \emph{classical force
fields}
\cite{Bro2009JCC,Jor1988JACS,Jor1996JACS,Pon2003APC,AMBER10,Pea1995CoPC}. On
the one hand, it is clear that the accuracy of the two physical models is not
the same; on the other hand, the size of the systems that can be practically
tackled is also different, since quantum chemical methods present a cost that
scales typically faster with the number of atoms than that of force fields
\cite{Jen1998Book,Ech2007MP}. Another example of this interplay between
accuracy and cost is the use of coarse-grained descriptions, which, by
selecting as interaction centers larger entities than individual atoms, and
thus changing the physical model, allow to reach larger systems and longer
time-scales than atomistic simulations using either force fields or ab initio
methods \cite{Liw2010JCTC,Cza2009JCTC,Emp2008BPJ,Han2008JCTC,dlT2011EPJSTunp}.

Yet another technique which can also be used to try to reduce the
computational cost of MD simulations by modifying the physical model, and
which is the object of this work (and of this special issue of The European
Physical Journal), is the imposition of \emph{constraints} on some of the
degrees of freedom of the system. The concept of constraints itself is very
general, and it can be used in many ways in the context of molecular modeling
and simulation: For example, in Car-Parrinello molecular dynamics
\cite{Car1985PRL}, which is a type of ab initio MD devised to mimic
ground-state Born-Oppenheimer MD, the time-dependent Khon-Sham orbitals need
to be orthonormal; a requirement that can be fulfilled imposing constraints on
their scalar product \cite{Hut2005CPC}. Also, we can use constraints to fix
some slow, representative degrees of freedom of our molecular system, also
called \emph{reaction coordinates}, in order to produce free energy profiles
along them that would have taken an unfeasibly long time to be calculated
should we have used an unconstrained simulation (see the Blue Moon Ensemble
technique in ref.~\cite{Car1989CPL} and also a number of related works in this
special issue \cite{Har2011EPJSTunp,Sch2011EPJSTunp}). In this review,
however, we are only concerned with the use of constraints in order to fix the
fastest, hardest degrees of freedom of molecular systems (like, e.g., bond
lengths or bond angles), with the objective of both reducing the effective
size of the conformational space, and allowing for larger time-steps in MD
simulations.

The first objective is worth pursuing if we acknowledge that one of the
characteristics of large, flexible molecules like proteins (and many others)
is that they have an astronomically large conformational space; an issue that
may hinder both our understanding of the problem and the possibility of
designing algorithms that could efficiently sample this space
\cite{Dil1999PS,Dill1997NSB,Lev1969Proc}. These systems additionally exhibit
movements with typical times that span a very wide time-scale, ranging from
very fast bond vibrations in the range of tenths of femtoseconds, to the more
interesting conformational changes related to biological function (like
allosteric transitions or protein folding) which can take times of the order
of the millisecond, or even the second
\cite{Dag2003NRMCB,Kar2002NSB,Sch1997ARBPBMS}. Since the time-step in MD
simulations has to be chosen at least one order of magnitude smaller than the
smallest typical time associated to the motion of the system if we want the
results to be accurate
\cite{Sch1997ARBPBMS,Bor1995TR2,Lei1995Book,Cha1985SIAMJNA,Fee1999JCC}, this
means that we need to calculate $10^{13}$-$10^{15}$ MD steps if we want to
capture the mentioned biological phenomena in our simulation.

The imposition of constraints that fix the fastest degrees of freedom holds
the promise of alleviating both these two problems, however, as any
modification of the physical model that describes our system, it may also
carry with it a loss of accuracy that renders the simulation useless. In this
work, we introduce the \emph{rigid} and \emph{stiff} constrained models in
order to be able to quantify this loss of accuracy at the level of equilibrium
statistical mechanics. In doing that, we benefit from the occasion to
thoroughly review the relevant literature, and to try to unify the different,
sometimes conflicting, vocabulary and mathematical definitions used to get a
handle on the physical concepts. This endeavour is facilitated by the fact
that the formalism used in this review is, as far as we are aware, the most
general one in the literature, therefore allowing to obtain all previously
discussed models as particular cases in which the (often implicit)
approximations can be clearly identified.

In sec.~\ref{sec:theory}, we introduce the theoretical framework, beginning by
the notation in sec.~\ref{subsec:notation}, and the Hamiltonian dynamics and
statistical mechanics of unconstrained systems in sec.~\ref{subsec:dynamics}.
Then, we describe and physically justify in sec.~\ref{subsec:constraints} what
types of constraints can be applied, both in terms of the set of coordinates
chosen to be constrained (sec.~\ref{subsubsec:coordinates_constrained}), and
in terms of whether or not their equilibrium values are considered to depend
on the conformation of the molecule
(sec.~\ref{subsubsec:flexible_vs_nonflexible}); we term the constraints
\emph{flexible} if this dependence exists, and \emph{hard} if the constrained
coordinates are assumed to take constant values. In
sec.~\ref{subsec:SM_models}, we introduce the four possible combinations of
constrained statistical mechanics models used in this work and in the
literature: the \emph{stiff} and the \emph{rigid} model, with either
\emph{flexible} or \emph{hard} constraints. In sec.~\ref{subsec:comparisons},
we comment on the mechanism for comparing the different models and introduce
the so-called \emph{Fixman potential} to do so; also, we collect the different
approximations made in the literature with respect to the most general cases
discussed in this work, and we establish the relevant relationships among
them. In sec.~\ref{sec:numerical}, we provide a numerical example in the
simple methanol molecule in order to illustrate the concepts, and we benefit
from the occasion to also review the previous numerical analyses in the
literature. Finally, in sec.~\ref{sec:conclusions}, we outline the main
conclusions of the work, and we mention possible lines of future research.

\section{Theoretical framework and relation to previous works}
\label{sec:theory}

In this section, our objective, which is in fact one of the main objectives of
this review, is to lay down the needed mathematical formalism to analyze the
equilibrium statistical mechanics of constrained molecular models, and to do
so in the most general terms possible.

Of course, we are not the first ones discussing constraints in molecular
modeling. As with any other scientific topic, there is a significant degree of
arbitrariness in any attempt to fix its historical beginning, but we can talk
about a discussion that has been going on at least for three or four decades
at the moment of the writing of this manuscript.

Therefore, \emph{what is the interest of a new account?}

We are confident to be able to convince the reader in the next subsections
that, despite the many great works that have dealt with the problem so far, no
article defines the problem with the same generality that we display here.
This general formalism, which contains all previous descriptions as particular
cases, allows us to clearly identify the different approximations that have
been assumed in the literature, to relate the many models with one another,
and to propose a consistent wording and notation which could facilitate the
clarity of future developments.

As we move along, we will indicate at each step which other choices have been
made in previous works and how they relate to the general setting.

\subsection{Notation}
\label{subsec:notation}

The system of interest (which we could call the \emph{molecule}) is a set of
$n$ mass points termed \emph{atoms}. The Euclidean coordinates of atom
$\alpha$ in a frame of reference fixed in the laboratory are denoted by
$\vec{r}_\alpha$ (Vectors of 3 components in the Euclidean 3-dimensional space
are denoted by bold symbols.), and its mass by $m_\alpha$, with
$\alpha=1,\ldots,n$. However, when no explicit mention to the atom index needs
to be made, we will use $r^T := (r^\mu)_{\mu=1}^{N}$ to denote the (row)
$N$-tuple of all the $N:=3n$ Euclidean coordinates of the system. By $r$, we
shall mean the corresponding column $N$-tuple, or just the ordered set
$(r^\mu)_{\mu=1}^{N}$ when matrix operations are not involved. These
coordinates parameterize the \emph{whole space}, denoted by $\mathcal{W}$, and
the masses $N$-tuple, $m^T := (m_\mu)_{\mu=1}^{N}$, in such a case, is formed
by consecutive groups of three identical masses, corresponding to each of the
atoms\footnote{\label{foot:indices_not-tensors} Some of the quantities
appearing in the formalism, such as the velocities $\dot{r}^\mu$, the
displacements $\delta r^\mu$, or the mass-metric tensor $G_{\mu\nu}$ (see the
following sections), change like tensors of some type under a general change
of coordinates; in such cases, the position of the indices is dictated by the
tensor type (1-time contravariant tensors in the case of $\dot{r}^\mu$ or
$\delta r^\mu$, 2-times covariant in the case of $G_{\mu\nu}$). Some other
quantities, such as $\ddot{r}^\mu$, $F_\mu$, $m_\mu$ or $r^\mu$, do not change
like any type of tensorial object under a general change of coordinates; in
such cases, the position of the indices is chosen according to notational
convenience, or to the way in which they transform under some particular
family of changes (e.g., linear ones).}.

Apart from the Euclidean coordinates, we shall also use a given set of
\emph{curvilinear coordinates} (also called sometimes \emph{general} or
\emph{generalized}), denoted by $q^T := (q^\mu)_{\mu=1}^N$, to describe the
system. The transformation between the two sets and its inverse are
respectively denoted by
\begin{subequations}
\label{eq:change_r_to_q}
\begin{align}
& r^\mu = R^\mu(q) \ , \quad \mu=1,\ldots,N \ ,
  \label{eq:change_r_to_q_a} \\
& q^\mu = Q^\mu(r) \ , \quad \mu=1,\ldots,N \ ,
  \label{eq:change_r_to_q_b}
\end{align}
\end{subequations}
and we will assume that, for the points of interest, this is a \emph{proper}
change of coordinates, i.e., that the \emph{Jacobian matrix}
\begin{equation}
\label{eq:Jacobian_r_to_q}
J^\mu_\nu := \frac{\partial R^\mu(q)}{\partial q^\nu}
\end{equation}
has non-zero determinant, being its inverse
\begin{equation}
\label{eq:Jacobian_q_to_r}
(J^{-1})^\mu_\nu = \frac{\partial Q^\mu(r)}{\partial r^\nu} \ .
\end{equation}

Although the curvilinear coordinates $q$ are a priori general, it is very
common to take profit from the fact that the typical potential energy
functions of molecular systems in absence of external fields do not depend on
the overall position or orientation, i.e., they are invariant under overall
translations and rotations, in order to choose a set of curvilinear
coordinates split into $q^T = (e^T,w^T)$, where the first six, $e^T :=
(e^A)_{A=1}^6$ are any \emph{external coordinates} that parameterize the
aforementioned global position and orientation of the
system\footnote{\label{foot:different_indices} The practice adopted in this
section of using different groups of indices, as well as different symbols,
for the individual coordinates in each one of the sets $r$, $q$, $e$, etc.,
allows us to grasp at a first glimpse the range of values in which each of the
indices vary. See table~\ref{tab:def_indices} for a summary of the indices
used, the symbols denoting the sets of coordinates, their meaning and the
spaces parameterized by them.}. The remaining $N-6$ coordinates $w^T :=
(w^a)_{a=7}^N$ are called \emph{internal coordinates} and determine the
positions of the atoms in some frame of reference fixed in the system. They
parameterize what we shall call the \emph{internal subspace}, denoted by
$\mathcal{I}$, and the coordinates $e$ parameterize the \emph{external
subspace}, denoted by $\mathcal{E}$; hence, we can split the whole space as
$\mathcal{W} = \mathcal{E} \times \mathcal{I}$ (denoting by $\times$ the
Cartesian product of sets).

The reader is probably familiar with the use of frames of reference fixed in
the system based on the center of mass and the principal axes of inertia, but,
although this is well adapted to the parameterization of the dynamics of rigid
bodies \cite{Gol2002Book}, it is less so to flexible entities like molecular
systems, where the relative distances of their constituents are not constant
in time. Following ref.~\cite{Ech2006JCC1}, we define a more suitable frame of
reference \emph{fixed in the system} to perform some of the calculations,
although it is worth stressing that most of the mathematical formalism
introduced in this work should not depend on this choice as long as the
external and internal subspaces are well separated.

\begin{figure}
\begin{center}
\includegraphics[scale=0.18]{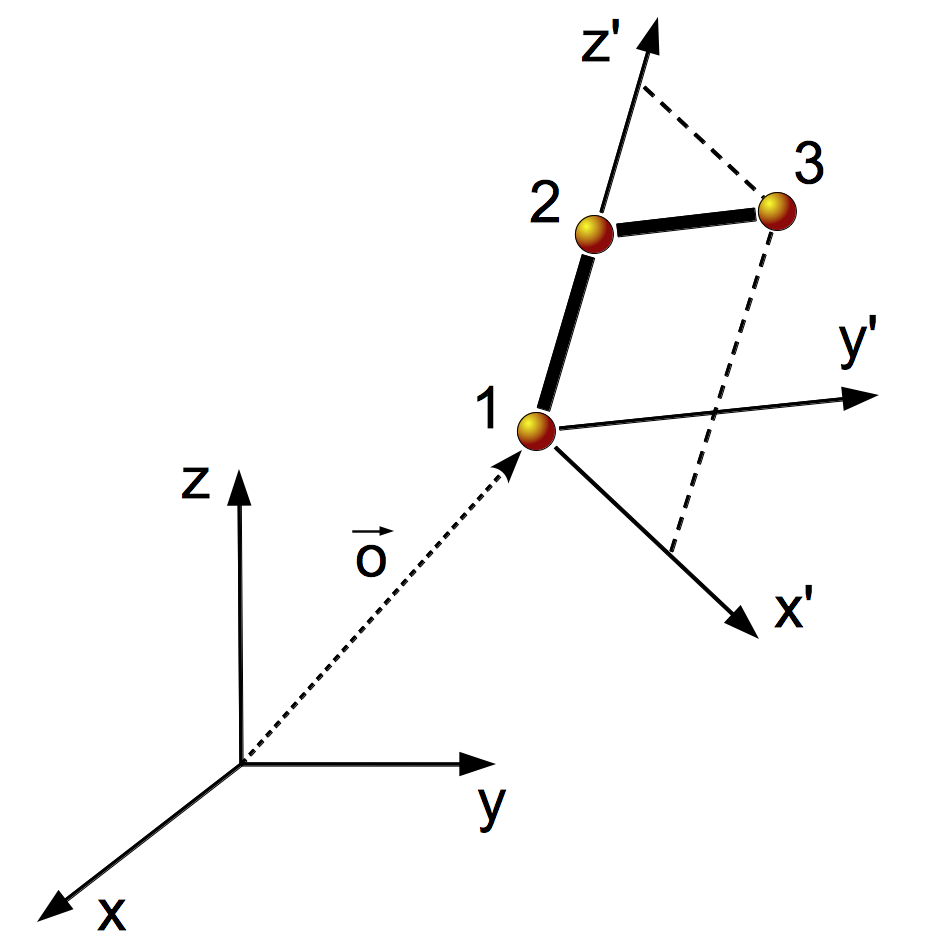}
\caption{\label{fig:axes_fixed} Definition of the frame of reference
 fixed in the system.}
\end{center}
\end{figure}

To define this frame of reference, we select three atoms (denoted by 1, 2 and
3 in Fig.~\ref{fig:axes_fixed}) in such a way that $\vec{o}$, the position in
the laboratory frame of reference of the origin of the frame of reference
fixed in the system, is the Euclidean position of atom 1 (i.e.,
$\vec{o}:=\vec{r}_{1}$). The orientation of the frame of reference
$(x^\prime,y^\prime,z^\prime)$ fixed in the system is chosen such that atom 2
lies in the positive half of the $z^\prime$-axis, and atom 3 is contained in
the $(x^\prime,z^\prime)$-plane, with projection on the positive half of the
$x^\prime$-axis (see Fig.~\ref{fig:axes_fixed}). The position of any given
atom $\alpha$ in the new frame of reference fixed in the system is denoted by
$\vec{r}_\alpha^\prime$, and we have that the previously unspecified external
coordinates are now $e^T := (e^A)_{A=1}^6 = (o_x,o_y,o_z,\phi,\theta,\psi)$,
where $(\phi,\theta,\psi)$ are three `Euler angles' that parameterize the
orientation of the `primed' axes with respect to the `unprimed' ones.

More explicitly, if $E(\phi,\theta,\psi)$ is the Euler rotation matrix (in the
ZYZ convention) that takes a free 3-vector of primed components,
$\vec{a}^\prime$, to the frame of reference fixed in the laboratory, i.e.,
$\vec{a} = E(\phi,\theta,\psi)\,\vec{a}^\prime$, its explicit expression is
the following \cite{Gol2002Book}:
\begin{equation}
\label{eq:def_E}
E(\phi,\theta,\psi) =
\underbrace{
 \left( \begin{smallmatrix}
 \scriptstyle \cos\phi & -\sin\phi& 0 \\
 \sin\phi & \cos\phi & 0 \\
 0 & 0 & 1
 \end{smallmatrix} \right)
 }_{\displaystyle \Phi(\phi)}
 \underbrace{
 \left( \begin{smallmatrix}
 -\cos\theta & 0 & \sin\theta \\
 0 & 1 & 0 \\
 -\sin\theta & 0 & -\cos\theta
 \end{smallmatrix} \right)
 }_{\displaystyle \Theta(\theta)}
 \underbrace{
 \left( \begin{smallmatrix}
 \cos\psi & -\sin\psi& 0 \\
 \sin\psi & \cos\psi & 0 \\
 0 & 0 & 1
 \end{smallmatrix} \right)
 }_{\displaystyle \Psi(\psi)}
\ .
\end{equation}

The position, $\vec{r}_\alpha^\prime$, of any given atom $\alpha$ in the axes
fixed in the system is by construction a function, $\vec{R}^\prime_\alpha(w)$,
of only the internal coordinates $w$, and the transformation from the
Euclidean coordinates $r$ to the curvilinear coordinates $q$
in~(\ref{eq:change_r_to_q_a}) may be written as follows:
\begin{equation}
\label{eq:change_r_to_q_2}
\vec{r}_\alpha = \vec{R}_\alpha(q) = \vec{o} + E(\phi,\theta,\psi)\,
 \vec{R}^\prime_\alpha(w) \ .
\end{equation}

An additional definition of subsets of the coordinates $q$ is motivated by the
imposition of \emph{constraints}. Assume that we impose $L=N-K$ independent,
holonomic and scleronomous constraints to the system:
\begin{equation}
\label{eq:constraints_1}
\sigma^{I}(q) = 0  \ , \quad I=K+1,\ldots,N \ .
\end{equation}

The usual definition \cite{Gol2002Book} is that the constraints are called
\emph{holonomic} if they can be written in the form $\sigma^I(q,t) = 0$, and
\emph{scleronomous} when the functions $\sigma^T(q) :=
\big(\sigma^I(q)\big)_{I=K+1}^N$ do not depend explicitly on time $t$. They
are \emph{independent} if their gradients $\big((\partial \sigma^I / \partial
q^\mu)(q)\big)^N_{\mu=1}$, $I=K+1,\ldots,N$, constitute $L$ linearly
independent vectors of $N$ components for every point $q$ satisfying the
constraints [i.e., for every point $q$ such that $\sigma^I(q) = 0$,
$I=K+1,\ldots,N$]; or, otherwise stated, the constraints are
\emph{independent} if the matrix with entries $(\partial \sigma^I / \partial
q^\mu)(q)$ has range $L$. In such a case, the constraints uniquely define (at
least locally) a subspace of $\mathcal{W}$ of constant dimension $N-L=K$,
called the \emph{constrained subspace} and denoted by $\mathcal{K}$.

Moreover, this independence condition allows, in the vicinity of each point
$q$ and by virtue of the Implicit Function Theorem
\cite{Dub1992Book,Wei2009Web1}, to (formally) solve
eq.~(\ref{eq:constraints_1}) for $L$ of the coordinates $q$, which we
arbitrarily place at the end of $q$, splitting the original set as $q^T =
(u^T,d^T)$, with $u^T := (u^r)_{r=1}^K$ and $d^T := (d^I)_{I=K+1}^N$. Then, in
the vicinity of each point $q$ satisfying~(\ref{eq:constraints_1}), we can
express the relations defining the constrained subspace, $\mathcal{K}$,
\emph{parametrically} by
\begin{equation}
\label{eq:constraints_2}
d^I = f^I(u) \ , \quad I=K+1,\ldots,N \ ,
\end{equation}
where the functions $f^T(u) := \big(f^I(u)\big)_{I=K+1}^N$ are the ones whose
existence the Implicit Function Theorem guarantees. The coordinates $u$ are
thus termed \emph{unconstrained} and they parameterize $\mathcal{K}$, whereas
the coordinates $d$ are called \emph{constrained} and their value is
determined at each point of $\mathcal{K}$ according
to~(\ref{eq:constraints_2}). As we will see later, in the most general case,
the functions $f$ will depend on $u$, and the constraints will be said to be
\emph{flexible}. In the particular case in which all the functions $f$ are
constant along $\mathcal{K}$, the constraints are called \emph{hard},
and all the calculations are considerably simplified.

\begin{table}
\begin{center}
\caption{Symbols and indices used for the different sets of coordinates,
  as well as the names of the spaces parameterized by them.}
\label{tab:def_indices}
\begin{tabular}{llllll}
Symbol & Indices & Range & Number & Name & Space \\
\hline \\[-8pt]
$\vec{r}$ & $\alpha,\beta,\gamma,\ldots$ & $1,\ldots,n$ & $n$ & Atoms & --- \\
$r,q$ & $\mu,\nu,\rho,\ldots$ & $1,\ldots,N$ & $N=3n$ & Whole space & $\mathcal{W}$ \\
$e$ & $A,B,C,\ldots$ & $1,\ldots,6$ & 6 & External & $\mathcal{E}$ \\
$w$ & $a,b,c,\ldots$ & $6+1,\ldots,N$ & $N-6$ & Internal & $\mathcal{I}$ \\
$s$ & $i,j,k,\ldots$ & $6+1,\ldots,K$ & $M$ & Unconstrained internal & $\Sigma$ \\
$d$ & $I,J,K,\ldots$ & $K+1,\ldots,N$ & $L=N-K$ & Constrained internal & --- \\
$u$ & $r,s,t,\ldots$ & $1,\ldots,K$ & $K$ & Unconstrained & $\mathcal{K}$ 
\end{tabular}
\end{center}
\end{table}

Of course, even if $\mathcal{K}$ is regular in all of its points, the
particular coordinates $d$ that can be solved need not to be the same along
the whole space\footnote{\label{foot:circle} One of the simplest examples of
this being the circle in $\mathbb{R}^2$, which is given by
$f(x,y):=x^2+y^2-R^2=0$, an implicit expression whose gradient is non-zero for
all $(x,y) \in \mathcal{K}$. However, if we try to solve, say, for $y$ in the
whole space $\mathcal{K}$, we will run into trouble at $y=0$; if we try to
solve for $x$, we will find it to be impossible at $x=0$. I.e., the Implicit
Function Theorem does guarantee that we can solve for \emph{some} of the
original coordinates at each regular point of $\mathcal{K}$, but sometimes the
solved coordinate has to be $x$ and sometimes it has to be $y$.}.
Nevertheless, we will assume this to be the case throughout this work, as it
is normally implicitly done in the literature
\cite{Ech2006JCC2,Hes2002JCP,Zho2000JCP}, and thus we will consider that
$\mathcal{K}$ is parameterized by the \emph{same} subset of unconstrained
coordinates $u$ in all of its points\footnote{\label{foot:regular_non-flex}
Note that this qualification is unnecessary in the hard case introduced
in sec.~\ref{subsubsec:rigid_non-flex}, where the uncoupled nature of the
functions $\sigma(q)$ makes it possible to solve for the same coordinates at
every point.}.

Although in general the functions $\sigma(q)$ in~(\ref{eq:constraints_1}) may
involve all the coordinates $q$, the already mentioned property of invariance
of the potential energy function [or the potential energy plus a term related
to the determinant of the whole-space mass-metric tensor (see
sec.~\ref{subsubsec:stiff_flex})] under changes of the external coordinates,
$e$, together with the fact (which we shall discuss later) that the potential
energy (or the aforementioned object) can be regarded as `producing' the
constraints, make physically frequent the definition of constraints affecting
only the internal coordinates, $w$:
\begin{equation}
\label{eq:constraints_1b}
\sigma^{I}(w) = 0  \ .
\end{equation}

In such a situation, which will be the one treated in this work, the
constrained coordinates, $d$, are contained into the internal ones, $w$, so
that the latter can be split as $w^T = (s^T,d^T)$, where the first
$M:=K-6=N-L-6$ ones, $s^T := (s^i)_{I=6+1}^{K}$, are called \emph{constrained
internal coordinates} and parameterize the \emph{constrained internal
subspace}, denoted by $\Sigma$. Hence, the constrained subspace can be split
as $\mathcal{K} = \mathcal{E} \times \Sigma$, and the analogue to
(\ref{eq:constraints_2}) can be written for this case as
\begin{equation}
\label{eq:constraints_2b}
d^I = f^{I}(s)  \ .
\end{equation}

Finally, if the constraints in~(\ref{eq:constraints_2b}) are used, together
with~(\ref{eq:change_r_to_q_2}), the Euclidean position of any atom may be
parameterized with the set of all unconstrained coordinates, $u$, as follows:
\begin{eqnarray}
\label{eq:transf_constrained}
\vec{r}_\alpha &=& \bm{\mathcal{R}}_\alpha(u):= 
 \vec{R}_\alpha\big(e,s,f(s)\big)
 = \vec{o} + E(\phi,\theta,\psi)
 \vec{R}^\prime_\alpha\big(s,f(s)\big)= \nonumber \\
 & =: & \vec{o} + 
  E(\phi,\theta,\psi) \bm{\mathcal{R}}^\prime_\alpha(s) \ ,
\end{eqnarray}
where the name of the transformation functions has been changed from $R$ to
$\mathcal{R}$, and from $R^\prime$ to $\mathcal{R}^\prime$, in order to
emphasize that the dependence on the coordinates is different between the two
cases.

\subsection{Hamiltonian dynamics and statistical mechanics without 
            constraints}
\label{subsec:dynamics}

Let us now briefly introduce the formalism needed to tackle the dynamics and
statistical mechanics of any molecular system such as the one described in the
previous section. We do this only to fix the notation and to write some
expressions that will be used later; the interested reader might want to check
any of the classical texts in the subject, such as
\cite{Jos1998Book,Gol2002Book,Bal1975Book}.

The central object that determines the behaviour of a classical system is the
\emph{Hamiltonian function}. If one starts by writing the \emph{Lagrangian
function} in terms of the Euclidean coordinates $r$ and velocities $\dot{r}$
(with the over-dot denoting the time derivative),
\begin{equation}
\label{eq:EuclideanL}
L_\mathrm{Euc}(r,\dot{r}) :=
 \frac{1}{2} m_\mu (\dot{r}^\mu)^2 - U(r) \ ,
\end{equation}
where $U(r)$ is the potential energy of the system and the convention
prescribing summation on repeated indices has been used, then the
\emph{Euclidean Hamiltonian} can be easily obtained through the usual Legendre
transform:
\begin{equation}
\label{eq:EuclideanH}
H_\mathrm{Euc}(r,\pi) :=
 \frac{1}{2m_\mu} (\pi_\mu)^2 + U(r) \ ,
\end{equation}
where $\pi$ are the Euclidean momenta defined as
\begin{equation}
\label{eq:Euclideanpi}
\pi_\mu := \frac{\partial L_\mathrm{Euc}}{\partial \dot{r}^\mu} =
 m_{(\mu)} \dot{r}^{(\mu)} \ , \quad \mu = 1,\ldots,N \ ,
\end{equation}
with the parentheses around the indices indicating that the convention that
prescribes summation when indices are repeated is not to be followed.


The dynamics of the system, once the Hamiltonian in eq.~(\ref{eq:EuclideanH})
is known, is given by the solutions of a set of coupled, first-order
differential equations known as \emph{Hamilton's equations}:
\begin{subequations}
\label{eq:EuclideanHamilton}
\begin{align}
\dot{r}^\mu & = \frac{\partial H_\mathrm{Euc}(r,\pi)}{\partial \pi_\mu} \ , 
 \label{eq:EuclideanHamilton_q} \\
\dot{\pi}_\mu & = - \frac{\partial H_\mathrm{Euc}(r,\pi)}{\partial r^\mu} \ .
 \label{eq:EuclideanHamilton_p}
\end{align}
\end{subequations}

If we now want to describe this dynamics in terms of the curvilinear
coordinates $q$ introduced in sec.~\ref{subsec:notation}, which are better
adapted to the covalent connectivity of molecular systems, we can start by
taking the time derivative of the change of coordinates in
eq.~(\ref{eq:change_r_to_q_a}),
\begin{equation}
\label{eq:der_change_r_to_q}
\dot{r}^\mu = \frac{\partial R^\mu(q)}{\partial q^\nu} \dot{q}^\nu \ ,
\end{equation}
and also noticing that, for any given potential energy $U(r)$ expressed as a
function of $r$, we can perform the same change of coordinates and define the
potential energy as a function of $q$ by
\begin{equation}
\label{eq:Vq}
V(q) := U\big(R(q)\big) \ ,
\end{equation}
which, as we mentioned, typically only depends on the internal ones, i.e.,
$V(q)=V(w)$.

If we take these last two expressions to eq.~(\ref{eq:EuclideanL}), we arrive
to the \emph{Lagrangian} in curvilinear coordinates, or \emph{unconstrained
Lagrangian}, as we will call it in the rest of the manuscript:
\begin{equation}
\label{eq:unconstrL}
L(q,\dot{q}) := \frac{1}{2} \dot{q}^\nu G_{\nu\rho}(q) \dot{q}^\rho 
 - V(w) \ ,
\end{equation}
where the \emph{whole-space mass-metric tensor} (MMT) $G_{\nu\rho}(q)$ is
defined as
\begin{equation}
\label{eq:G}
G_{\nu\rho}(q) := \frac{\partial R^\mu(q)}{\partial q^\nu} m_\mu
   \frac{\partial R^\mu(q)}{\partial q^\rho} \ .
\end{equation}

Following again the usual process of the Legendre transform, we can derive
from eq.~(\ref{eq:unconstrL}) the \emph{unconstrained Hamiltonian},
\begin{equation}
\label{eq:unconstrH}
H(q,p) := \frac{1}{2} p_\nu G^{\nu\rho}(q) p_\rho + V(w) \ ,
\end{equation}
where $G^{\nu\rho}(q)$ is the inverse of the whole-space MMT in
eq.~(\ref{eq:G}) and the canonical conjugate momenta are defined as
\begin{equation}
\label{eq:p}
p_\mu := \frac{\partial L}{\partial \dot{q}^\mu} =
 G_{\mu\nu}(q) \dot{q}^\nu \ .
\end{equation}

Finally, the \emph{Hamilton equations} in terms of the curvilinear
coordinates and momenta are formally identical to those in
eqs.~(\ref{eq:EuclideanHamilton}):
\begin{subequations}
\label{eq:unconstrHamilton}
\begin{align}
\dot{q}^\mu & = \frac{\partial H(q,p)}{\partial p_\mu} \ , 
 \label{eq:unconstrHamilton_q} \\
\dot{p}_\mu & = - \frac{\partial H(q,p)}{\partial q^\mu} \ ,
 \label{eq:unconstrHamilton_p}
\end{align}
\end{subequations}

Under the usual assumptions of ergodicity and equal a priori probabilities
\cite{Bal1975Book}, it can be shown that the \emph{partition function}
characterizing the statistical mechanics equilibrium of such a dynamics in the
canonical ensemble (i.e., at constant volume $V$, number of particles $n$, and
temperature $T$) is given by
\begin{equation}
\label{eq:unconstrZ}
Z = \frac{\alpha_\mathrm{QM}}{h^N} \int
	 e^{-\beta H(q,p)} \mathrm{d}q \mathrm{d}p \ ,
\end{equation}
where $h$ is Planck's constant, we denote $\beta := 1/RT$ (per mole energy
units are used throughout the article, so $RT$ is preferred over $k_B T$) and
$\alpha_\mathrm{QM}$ is a combinatorial number that accounts for quantum
indistinguishability and that must be specified in each particular case (e.g.,
for a gas of $n$ indistinguishable particles, $\alpha_\mathrm{QM}=1/n!$). With
$\int \mathrm{d}q \mathrm{d}p$, we denote integration over all positions
$q^\mu$ and their respective momenta $p_\mu$, for $\mu=1,\ldots,N$, being the
range of integration usually from $-\infty$ to $\infty$ for the momenta, and
the appropriate one for each position. For example, if a typical scheme for
defining the internal coordinates, such as the SASMIC one \cite{Ech2006JCC1},
is used, bond lengths must be integrated from $0$ to $\infty$, bond angles
from $0$ to $\pi$, and dihedral angles from $-\pi$ to $\pi$.

The corresponding \emph{equilibrium probability density function} (PDF) is
thus given by
\begin{equation}
\label{eq:unconstrPjoint}
P(q,p) = \frac{e^{-\beta H(q,p)}}
 {\int e^{-\beta H(q^\prime,p^\prime)}
  \mathrm{d}q^\prime \mathrm{d}p^\prime} \ ,
\end{equation}
being $P(q,p) \Delta q \Delta p$ interpreted as the probability of finding the
system with positions in $(q,q+\Delta q)$, and momenta in $(p,p+\Delta p)$,
for sufficiently small $\Delta q$, $\Delta p$.

In the same sense, the equilibrium average of any observable $O(q,p)$ is
\begin{equation}
\label{eq:unconstrOavg}
\langle O \rangle = \int O(q,p) P(q,p) \mathrm{d}q \mathrm{d}p \ .
\end{equation}

It is also common in the literature to study observables, or properties, which
depend only on the positions $q$ and not on the momenta $p$; the native
conformation of a protein being a notable example of this \cite{Ech2007CoP}.
In such a case, we can appeal to the well-known formula for the
$N$-dimensional Gaussian integral \cite{Pet2010Misc},
\begin{equation}
\label{eq:Gaussian_integral}
\int_{-\infty}^\infty \cdots \int_{-\infty}^\infty
 e^{-\frac{1}{2} x^\mu M_{\mu\nu} x^\nu + v_\mu x^\mu} \mathrm{d}x =
 \sqrt{\frac{(2\pi)^N}{\det M}} e^{\frac{1}{2} v_\mu M^{\mu\nu} v_\nu} \ ,
\end{equation}
where $M$ is an $N \times N$ matrix (that must be positive definite in order
for the integral to be finite), $M^{\mu\nu}$ denotes the entries of $M^{-1}$,
and $v$ is a (possibly null) $N$-tuple, to `integrate out' the momenta in the
partition function in eq.~(\ref{eq:unconstrZ}), yielding:
\begin{equation}
\label{eq:unconstrZ_2}
Z = \chi(T) \int e^{-\beta \left[ V(w)
 - T \frac{R}{2} \ln \det G(q) \right]} \mathrm{d}q \ ,	
\end{equation}
where
\begin{equation}
\label{eq:chi}
\chi(T) := \left( \frac{2\pi}{\beta} \right)^{N/2}
  \frac{\alpha_\mathrm{QM}}{h^N} \ .
\end{equation}

If we perform the same integration in the joint PDF in
eq.~(\ref{eq:unconstrPjoint}), the factor $\chi(T)$ cancels out and we arrive
to the \emph{marginal equilibrium PDF} in the space of the positions $q$:
\begin{equation}
\label{eq:unconstrPq}
P(q) = \frac{ e^{ -\beta \left[ V(w)
 - T \frac{R}{2} \ln \det G(q) \right]}}
 {\int e^{ -\beta \left[ V(w^\prime)
 - T \frac{R}{2} \ln \det G(q^\prime) \right]} \mathrm{d}q^\prime}
 =: \frac{ e^{ -\beta F(q)}}
 {\int e^{ -\beta F(q^\prime)} \mathrm{d}q^\prime} \ ,
\end{equation}
where the normal abuse of notation in probability theory has been committed,
using the same symbol, $P$, for the two different functions in
eqs.~(\ref{eq:unconstrPjoint}) and~(\ref{eq:unconstrPq}), and where we have defined
\begin{subequations}
\label{eq:unconstrFS}
\begin{align}
F(q) & := V(w) - T S^\mathrm{k}(q) \ , \label{eq:unconstrFS_F} \\
S^\mathrm{k}(q) & := \frac{R}{2} \ln \det G(q) \ . 
  \label{eq:unconstrFS_S}
\end{align}
\end{subequations}

The notation in the equations above is intentional, in the sense that $F(q)$
can be interpreted as a \emph{free} or \emph{effective} energy, since it is
obtained via the elimination of some degrees of freedom (the momenta $p$), and
it therefore describes the energetics (at least the equilibrium one) of the
remaining degrees of freedom (the positions $q$) in a sort of mean-field of
the ones that have been eliminated. This \emph{free} or \emph{effective}
character is also emphasized by the fact that $F(q)$ depends on the
temperature $T$, even if in a simple way, and the analogy can be taken one
step further if we regard $V(q)$ as an internal energy and the correcting term
as a \emph{kinetic entropy} \cite{Go1969JCP}; something which is compatible
with its being linear in $RT$.

Again, the meaning of this last \emph{marginal} PDF in
eq.~(\ref{eq:unconstrPq}) is made explicit if we note that $P(q) \Delta q$ is
the probability of finding the system with positions in $(q,q+\Delta q)$, for
sufficiently small $\Delta q$ (\emph{irrespective} of the value of the
momenta), and that the equilibrium average of any momenta-independent
observable $O(q)$ is given by
\begin{equation}
\label{eq:unconstrOavg_2}
\langle O \rangle = \int O(q) P(q) \mathrm{d}q \ .
\end{equation}

Now, as we mentioned in sec.~\ref{subsec:notation}, the potential energy of
molecular systems in absence of external fields is typically independent of
the external coordinates $e$, and this is why we have written $V(w)$ and not
$V(q)$ in all the previous expressions. The whole-space MMT $G(q)$ in
eq.~(\ref{eq:G}) on the other hand, and in particular its determinant in
eqs.~(\ref{eq:unconstrZ_2}) and~(\ref{eq:unconstrPq}), \emph{does} depend on
the external coordinates $e$. It is thus convenient, in the case that we are
also dealing with observables $O(w)$ which are independent of the external
coordinates, to try to eliminate them from the expressions if possible. One
can indeed \emph{formally} always integrate over the external coordinates (or
any other variables spanning the probability space) in order to get to the
corresponding marginal PDF in the internal space $\mathcal{I}$, i.e.,
depending only on the internal coordinates $w$. However, until recently, it
was not clear if this process could be performed \emph{analytically}
(specially for the more involved, constrained case in
sec.~\ref{subsubsec:rigid_flex}), thus yielding manageable final expressions.
In a previous work by some of us \cite{Ech2006JCC3}, we settled the issue
proving that this is in fact possible and providing the exact analytical
expressions to be used for the marginal PDF in $\mathcal{I}$.

To integrate out the external coordinates \emph{analytically} from $\det
G(q)$, we only need to realize that, since the unconstrained case can be
trivially assimilated to a constrained situation with the number of
constraints, $L=0$, all the results regarding the factorization of the induced
MMT determinant that we will introduce later in
sec.~\ref{subsubsec:rigid_flex} apply, and we have the unconstrained analogue
of eq.~(\ref{eq:detg}):
\begin{equation}
\label{eq:detG}
\det G(q) = \sin^2 \theta \det G^\prime(w) \ ,
\end{equation}
with $G^\prime(w)$ the $N \times N$ matrix obtained from eq.~(\ref{eq:gprime})
if the unconstrained internal coordinates $s$ are substituted by all the
internal coordinates $w$, and $\theta$ being one of the Euler angles
introduced in sec.~\ref{subsec:notation}.

Now, the external angle $\theta$ can be integrated both in the numerator and
in the denominator of eq.~(\ref{eq:unconstrPq}), yielding the marginal PDF in
the internal space:
\begin{equation}
\label{eq:unconstrPw}
P(w) = \frac{ e^{ -\beta F(w)}}
 {\int e^{ -\beta F(w^\prime)} \mathrm{d}w^\prime} \ ,
\end{equation}
where we have defined, analogously to eq.~(\ref{eq:unconstrFS}),
\begin{subequations}
\label{eq:unconstrFSw}
\begin{align}
F(w) & := V(w) - T S^\mathrm{k}(w) \ , \label{eq:unconstrFSw_F} \\
S^\mathrm{k}(w) & := \frac{R}{2} \ln \det G^\prime(w) \ . 
  \label{eq:unconstrFSw_S}
\end{align}
\end{subequations}

Finally, let us remark that every step taken in this section, from the
original Hamiltonian in eq.~(\ref{eq:unconstrH}) to the marginal PDF in
eq.~(\ref{eq:unconstrPw}), is \emph{exact}. The approximations will arrive
when we introduce the constrained models in the following sections.

\subsection{Types of constraints}
\label{subsec:constraints}

If we assume that the constraints we shall impose on our physical system can
be expressed as in~(\ref{eq:constraints_2b}), where $w:=(s,d)$ are typical
internal coordinates, such as the ones defined in ref.~\cite{Ech2006JCC1},
then the different types of constraints that can be imposed may be classified
according to two independent criteria, which, together, constitute a
definition of the internal constrained subspace $\Sigma$, namely, (a) What
coordinates are constrained? I.e., what coordinates comprise the set $d$? and
(b) Do the values of the constrained coordinates in $\Sigma$ depend on the
point $s$, or not?

\subsubsection{What coordinates are constrained}
\label{subsubsec:coordinates_constrained}

Regarding (a), it is typical in the literature to constrain whole groups of
coordinates of the same \emph{type}, classifying them both according to
whether they are bond lengths, bond angles, etc., and also in terms of the
atoms involved in their definition. For example, in ref.~\cite{Fre2009BPJ},
bond-lengths involving Hydrogen atoms are constrained in a 20 $\mu$s MD of the
Villin Headpiece; in ref.~\cite{Ens2007JMB}, all bond-lengths are constrained
to study the folding of a modified version of the same protein; in
ref.~\cite{Eas2010JCTC1} all bond-lengths as well as bond angles of the form
H--O--X or H--X--H, being X any atom, are constrained to simulate an
80-residue protein; the same scheme of constraints is applied in
ref.~\cite{Hes2008JCTC1}; and we have the constraining of everything but
torsion angles in the so-called \emph{torsion-angle MD}
\cite{Che2005JCC,Hin1995PRE,Mat1994PSFG,Maz1989JBMSD,Maz1997JCC}.

If the system is close to a local minimum of the potential energy $V(w)$, this
way of proceeding is based on the intuition that we can associate similar
vibrational frequencies to internal coordinates of the same type, with
little influence of the local environment surrounding the atoms that define
them. In this way, if we start by constraining the highest-frequency
coordinates, we remove their vibrations from the dynamics, and, as we
mentioned in sec.~\ref{sec:introduction}, we can use a larger time-step to
integrate the equations of motion. This is only intuitive and not rigorous
because (i) the idea is also used away from minima, where `vibrational
frequencies' are not properly defined, (ii) even close to a minimum, the
vibrational modes of molecules cannot be directly associated to individual
internal coordinates, but to linear combinations of them, and (iii) even if a
vibrational mode is almost purely associated to a given internal coordinate,
its vibrational frequency \emph{does} depend on the environment of the atoms
defining it; the question is how much.

It is therefore probably a more accurate way of proceeding to define the
constrained coordinates in an \emph{adaptive} manner, depending on the
conformation of the system; but such schemes either do not exist or they are
not very popular in the literature as far as we are aware. Therefore, although
we remark that any choice of constrained coordinates is in principle possible
as long as the resulting model is physically accurate, we will be thinking
here in the simpler case of constraining whole groups of internal coordinates
of the same type (defining \emph{type} in the common and chemically intuitive
way specified above).

Regarding this point, it is also worth remarking that, although we argued in
sec.~\ref{subsec:notation} that, by virtue of the Implicit Function Theorem,
it is locally equivalent to express the constraints implicitly, as in
eq.~(\ref{eq:constraints_1b}), or parametrically, as in
eq.~(\ref{eq:constraints_2b}), the developments that can be performed, the
final expressions, and the practical algorithms derived from them are
different in the two cases. We will base all the discussion on the formalism
in which the coordinates to be constrained, $d$, can be identified and the
constraints written as in eq.~(\ref{eq:constraints_2b}), using the
unconstrained coordinates, $u$, to parameterize the constrained subspace. In
such a case, it is most natural to assume that these coordinates, $q:=(u,d)$,
are the typical ones used in molecular simulation \cite{Ech2006JCC1}, however,
the possibility also exists to use the implicit constraints in
eq.~(\ref{eq:constraints_1b}) and work in Euclidean coordinates. In fact this
is the version that is needed for the Lagrange multipliers formulation in the
whole space and for the implementation of the vast majority of practical
algorithms, both in the flexible and hard cases discussed later
\cite{Cic1986CPR,Hes2002JCP}. Related to this, one can also choose a set of
modified curvilinear coordinates $\tilde{q}$ in which the constrained ones are
given exactly by the $\sigma$ functions in eq.~(\ref{eq:constraints_1b}), and
therefore the constraints are expressed just as $\tilde{d}=\dot{\tilde{d}}=0$.
We will make frequent references to these other choices as we advance in the
discussion.

\subsubsection{Flexible vs. hard constraints}
\label{subsubsec:flexible_vs_nonflexible}

Once the constrained coordinates $d$ have been \emph{selected}, we can ask
question (b): Do their values in $\Sigma$ depend on the point or not? I.e., do
they depend on the unconstrained internal coordinates $s$? We shall call the
constraints \emph{flexible} if the answer is `yes' and \emph{hard} if
it is `no'.

It is worth mentioning at this point that the wording used to refer to
constrained systems in the literature is multiple and often misleading. One of
the aims of this work is to clarify the mathematical definitions behind the
words and to provide a consistent vocabulary to refer to the different types
of models. What we have called \emph{flexible} constraints, for example, are
also called \emph{flexible} in
refs.~\cite{Lei1995Book,Chr2007CPC,Chr2005JCP,Hes2002JCP,Zho2000JCP,Sai2004JCP},
\emph{elastic} in~\cite{Cot2004BITNM}, \emph{adiabatic} in~\cite{Sto2003MS},
and \emph{soft}
in~\cite{Rei1996PRE,Lei1995Book,Zho2000JCP,Rei1998NA,Sto2003MS}; whereas the
\emph{hard} case is also called \emph{hard} in
refs.~\cite{Rei1996PRE,Lei1995Book,Chr2007CPC,Zho2000JCP,Sto2003MS}, just
\emph{constrained} in~\cite{Chr2005JCP}, \emph{holonomic}
in~\cite{Cot2004BITNM,Sto2003MS}, \emph{rigid}
in~\cite{Hes2002JCP,Zho2000JCP}, and \emph{fully constrained}
in~\cite{Zho2000JCP}. In the light of the more formal discussion in what
follows, the reader will appreciate that some of these terms are clearly
misleading (\emph{elastic}, \emph{holonomic} or \emph{fully constrained}),
and, in any case, so many names for such simple concepts is detrimental for
the understanding in the field. Even if the reader does not like the choice
advocated in this work, she should recognize that a unification of the
vocabulary is desirable.

The situation is further complicated by the fact that, when studying the
statistical mechanics of constrained systems, as we will also see later in
great detail, one can think about two different models for calculating the
equilibrium PDF. The names of these models often collide with the ones used
for defining the type of constraints applied. On the one hand, one can
implement the constraints by the use of very steep potentials around the
constrained subspace; a model often called \emph{flexible}
\cite{Pea1979JCP,Go1976MM,Hel1979JCP,Pec1980JCP,Alm1990MP,Ber1983Book,Per1985MM}
(the worst option due to vocabulary clashes), sometimes called \emph{soft}
\cite{Fre2002Book,Dub2010JCP}, and sometimes \emph{stiff}
\cite{Mor2004ACP,Hin1994JFM,Fre2002Book,Ech2006JCC2,vKa1984AJP,Pet2000Thesis,Ral1979JFM}
(as we advocate in this work). On the other hand, one can assume D'Alembert's
principle \cite{Gol2002Book,Gal2007Book} and hypothesize that the forces are
just the ones needed for the system to never leave the constrained subspace
during its dynamical evolution; a model normally called \emph{rigid}
\cite{Pea1979JCP,Mor2004ACP,Hin1994JFM,Go1976MM,Ech2006JCC2,Hel1979JCP,Pec1980JCP,Alm1990MP,Ber1983Book,Cha1979JCP,Per1985MM,Pet2000Thesis,Ral1979JFM},
as we do here, but sometimes also \emph{hard}
\cite{Fre2002Book,Dub2010JCP,Chr2005JCP}, which is a bad choice in terms of
clashes. This interference between the two families of wording is entirely
unjustified, since, as we will see later, the two types of constraints and the
two types of statistical mechanics models can be independently combined; one
can have either the \emph{stiff} or the \emph{rigid} model, with either
\emph{flexible} or \emph{hard} constraints.

As we briefly mentioned in sec.~\ref{subsec:notation}, the most general case
from the mathematical point of view is indeed that of flexible constraints,
where the functions $f$ in (\ref{eq:constraints_2b}) are not constant numbers,
and it is also the case that arises more naturally from physical
considerations.

Indeed, one way of \emph{justifying} a given constrained model is by comparing
it to the original, unconstrained one. This choice of the \emph{reference} is
not the only possible one; it is also legitimate to compare the constrained
models introduced in this work directly to experimental results, or to more
fundamental theoretical descriptions such as those based in quantum mechanics.
However, we (and many others
\cite{Ber1983Book,Che2005JCC,Cha1979JCP,Chr2007CPC,Cot2004BITNM,Hin1995PRE,Lei1995Book,Pas2002JCP,Pat2004JCP,Pea1979JCP,Rei1996PRE,dOt1998JCP,dOt2000MP,Kar1981MM,Per1985MM,Rei1996PRE,vGu1982MM,Zho2000JCP})
have preferred to use the classical unconstrained model as a reference to be
able to assess the influence of the imposition of constraints in an
incremental way consisting of a series of controlled steps. Although some
works have advocated the point of view that the proper justification of
constrained models should come from a quantum mechanical treatment
\cite{Hes2002JCP,Go1969JCP,Go1976MM,Maz2001CTPS,Eas2010JCTC1,Fee1999JCC,Mor2004ACP,Ral1979JFM,Sai1963SSP,Tir1996CPL},
and the comparison to experiment \cite{Tir1996CPL} or to quantum mechanics is
indeed more relevant in terms of the \emph{absolute} accuracy of the
constrained models, the approximations connecting the former to the latter are
many, thus obscuring the influence of each individual step. In particular, it
is worth mentioning refs.~\cite{Go1976MM,Maz2001Book,Hel1979JCP,Mor2004ACP}
for a clear exposition of the many arguments that appear when the comparison
is of this second, more convoluted class. This is much so that we can find
works in the literature that, based on quantum mechanical arguments, favour
the classical stiff model (introduced in sec.~\ref{subsubsec:stiff_flex})
against the rigid one (sec.~\ref{subsubsec:rigid_flex}) \cite{Go1976MM}, or
vice versa \cite{Hes2002JCP}. Before these discrepancies can be solved and a
reference classical model can be rigorously derived from quantum mechanics
(see
refs.~\cite{Fro2001CMP,dCo1982PRA,Alv1998MMTS,Alv2000MMTS,Alv2002MP,Alv2004JPCM,Cal2005JPCM,Cal2008JPCM,Hel1979JCP,vKa1984AJP}
and also the review by \'Alvarez-Estrada and Calvo in this special issue
\cite{Alv2011EPJSTunp} for the most advanced attempts to do so), we will
consider the unconstrained dynamics as a more or less `safe harbor', and as
the reference to which assess the classical constrained models discussed in
this work. It is also worth remarking that, in doing so, we are also
sidestepping the question about the accuracy of the potential energy itself
(since we compare to the unconstrained dynamics given by \emph{the same}
potential). Although this is an important issue affecting the absolute
accuracy of the results, and it is very likely that classical force fields are
not enough to describe some phenomena, e.g., those involving charge transfer
\cite{Ani2009CPC}, we can consider the two sources of error separately and
analyze each one of them at a time.

In this spirit, we may think that the fact that allows us to treat some
coordinates as constrained and some others as unconstrained is that the former
need more energy to be separated from their minimum-energy values than the
latter, and thus this separation is more unlikely to occur. In principle,
these minimum-energy values as well as the energy cost needed to separate the
system from them can be defined both \emph{including} the momenta or
\emph{excluding} them, and this can be done either at the \emph{dynamical} or
at the \emph{statistical mechanics} level.

In this work, we are mainly concerned with the \emph{equilibrium} properties
of constrained models and we point the reader interested in dynamical
considerations to
\cite{Rei1996PRE,Lei1995Book,Fix1978JCP,Cot2004BITNM,Fro2001CMP,Bor1995TR2,Bor1995TR3,Hel1979JCP,Pea1979JCP,Per1985MM,Pet2000Thesis,vKa1985PR},
as well to
refs.~\cite{dlT2011EPJSTunp,Ske2011EPJSTunp,Baj2011EPJSTunp,Mad2011EPJSTunp}
in this same special issue. In brief, we want to remark that the approach
followed in this work is in general simpler and it is subject to less
uncertainties than dynamical analyses, but, on the other hand, its application
must be circumscribed to the computation of equilibrium averages.

In equilibrium statistical mechanics, all the information is contained in the
different partition functions [such as~(\ref{eq:unconstrZ})] and PDFs [such
as~(\ref{eq:unconstrPjoint})], and all observable quantities are calculated
through integrals, such as~(\ref{eq:unconstrOavg}). Therefore, it is most
natural to look for the regions in which the quantity under the integral sign
is non-negligible, and define our constrained subspace exactly as those
regions. Assuming that the observables $O(q,p)$, $O(q)$ are smooth enough
functions of their arguments, the constrained subspace so defined corresponds
to the values of the constrained coordinates that maximize the corresponding
PDFs for every value of the rest of variables. Once this information is used
to define the constrained subspace [i.e., to answer question (b)], the issue
about how to sample the conformations of the system to produce the appropriate
equilibrium PDF becomes an algorithmic one (we can use Monte Carlo techniques,
MD, etc.). If the algorithm used to do the sampling is, as it is often the
case, a dynamical one, then nothing in the discussed approach guarantees that
the trajectories obtained are similar in any way to those of the unconstrained
system. The only thing that we can be sure of is that the equilibrium averages
will be indeed consistent with the approximations involved in the definition
of the constrained subspace as that of maximum probability.

In this framework, we have several options to define the most probable values
of the constrained coordinates $d$, i.e., the functions $f$ in
eq.~(\ref{eq:constraints_2}) or~(\ref{eq:constraints_2b}): The most immediate
one is to look at the equilibrium PDF of the unconstrained system in
eq.~(\ref{eq:unconstrPjoint}), before having eliminated the momenta from the
description. The values that maximize this PDF are those that minimize the
unconstrained Hamiltonian for each value of the rest of the coordinates, $u$,
and all the momenta, $p$. This can be formally accounted for by stating that
the functions $f$ take the values that make $H(u,d,p)$ a local minimum for
each fixed $u$ and $p$, i.e.,
\begin{equation}
\label{eq:minimumH}
H(u,f,p) \leq H(u,d,p) \ , \ \forall d \in \mathcal{D}(f) \ ,
\end{equation}
where $\mathcal{D}(f)$ is a suitable open set in $\mathbb{R}^L$ containing the
point $f$, and therefore the functions $f$ must satisfy
\begin{equation}
\label{eq:derH}
\frac{\partial H}{\partial d^I}(u,f,p) = 0 \ , \quad
 I=K+1,\ldots,N \ ,
\end{equation}
with the corresponding Hessian,
\begin{equation}
\label{eq:HessianH}
\frac{\partial^2 H}{\partial d^I \partial d^J}(u,f,p) \ ,
\end{equation}
being a positive definite $L \times L$ matrix.

Now, since the left-hand side of eq.~(\ref{eq:derH}) is precisely minus the
right-hand side of the Hamilton equation in eq.~(\ref{eq:unconstrHamilton_p})
for the time derivative $\dot{p}_I$ of the canonical momentum associated to
$d^I$, we have that this kind of constraint is equivalent to asking that
\begin{equation}
\label{eq:derH_p0}
\dot{p}_I = 0 \quad \Rightarrow \quad p_I(t) = p_I(t_0) \ , \quad
 \forall t \ .
\end{equation}

Using the definition of the momenta in eq.~(\ref{eq:p}), this condition can be
seen to define the following non-holonomic constraint involving the 
velocities:
\begin{equation}
\label{eq:derH_non-holonomic}
G_{I\mu}(q)\dot{q}^\mu = C_I \ ,
\end{equation}
being $C_I$ a constant number dependent on the initial conditions. In fact, if
we look at the implicit definition of the functions $f$ in
eq.~(\ref{eq:minimumH}) or~(\ref{eq:derH}), we see that they must depend on
the momenta, i.e., they are $f(u,p)$ and not $f(u)$, which already clashes
with the holonomic scheme introduced in sec.~\ref{subsec:notation}.

Although some interpretation problems are associated to non-holonomic
constraints \cite{Fla2005AJP}, and they are known to not accept in general a
closed Hamiltonian formalism \cite{Koo1997ReMP,vdS1994ReMP,Mol2010Prep}, they
are perfectly valid in the algorithmic sense discussed before (i.e., as a tool
to define the integration region in which the integrated quantity is
non-negligible in equilibrium statistical mechanics). In fact, when their form
is linear in the velocities, as it is the case of
eq.~(\ref{eq:derH_non-holonomic}), the corresponding equations of motion
obtained from D'Alembert's principle are very simple
\cite{Koo1997ReMP,vdS1994ReMP,Mol2010Prep}, and, even if they cannot be
expressed as the Hamilton equations of a given Hamiltonian function, it has
been shown that their equilibrium distribution can be nevertheless computed
\cite{Tuc2001JCP}. In \cite{Hes2002JCP}, for example, this type of flexible
constraints are used to build a practical algorithm that is shown to be
time-reversible and to present good energy conservation properties, whereas it
is not symplectic; as expected from the non-Hamiltonian character of the
dynamics. However, it remains to be proved, maybe using the techniques in
\cite{Tuc2001JCP}, what is the equilibrium PDF produced by this dynamics.
Since in this work we are interested in the statistical mechanics of
constrained models, we prefer not to deal with this case related to the
minimization of $H(q,p)$ until these fundamental problems are solved.

A different option to define the constrained subspace appears if we consider
the frequent case of momenta-independent observables $O(q)$. In this
situation, as we discussed before, the momenta can be `integrated out' from
the description, and equilibrium averages are computed as in
eq.~(\ref{eq:unconstrOavg_2}), using the \emph{marginal} PDF defined in
eq.~(\ref{eq:unconstrPq}). Following the same ideas that we applied in the
momenta-dependent case, we see that the region of position space in which the
integrated object in eq.~(\ref{eq:unconstrOavg_2}) will be non-negligible is
the one in which the equilibrium PDF is maximal; or, equivalently, that in
which the quantity $F(u,d) := V(s,d)-TS^\mathrm{k}(u,d)$ defined in
eq.~(\ref{eq:unconstrFS}) is made a local minimum for each value of the
unconstrained coordinates $u$, i.e., we ask the functions $f$ that define
the constrained subspace to satisfy
\begin{equation}
\label{eq:minimumF}
F\big(u,f(u)\big) \leq F(u,d) \ , 
 \ \forall d \in \mathcal{D}\big(f(u)\big) \ ,
\end{equation}
where $\mathcal{D}\big(f(u)\big)$ is again a suitable open set in
$\mathbb{R}^L$ containing the point $f(u)$, and therefore we have that
\begin{equation}
\label{eq:derF}
\frac{\partial F}{\partial d^I}\big(u,f(u)\big) :=
\frac{\partial V}{\partial d^I}\big(s,f(u)\big) -
T\frac{\partial S^\mathrm{k}}{\partial d^I}\big(u,f(u)\big) = 0 \ , \quad
 I=K+1,\ldots,N \ ,
\end{equation}
with the corresponding Hessian,
\begin{equation}
\label{eq:HessianF}
\frac{\partial^2 F}{\partial d^I \partial d^J}\big(u,f(u)\big) \ ,
\end{equation}
being a positive definite $L \times L$ matrix.

In contrast with the previous definition, this one (which has never been used
in the literature as far as we are aware) \emph{does} produce constraints that
are holonomic, i.e., the functions $f$ that are now implicitly defined by the
equations above depend only on position-like degrees of freedom (the
unconstrained coordinates $u$), and the constrained subspace is now defined by
the holonomic relations in eq.~(\ref{eq:constraints_2}). Assuming D'Alembert's
principle, these constraints produce a dynamics which is Hamiltonian, which is
therefore amenable to symplectic integration methods, and whose equilibrium
PDF can be easily computed using the same standard ideas that we develop in
sec.~\ref{subsec:SM_models}.

In this work, however, for the sake of simplicity, we choose to define the
flexible constrained subspace in a third, different way which is nevertheless
related to both of the definitions just discussed. Simply stated, the function
of the positions whose minimum we use to implicitly define the value $f$ of
the constrained coordinates $d$ is neither the Hamiltonian $H(q,p)$ nor the
free energy $F(q)$, but the potential energy $V(w)$, i.e., we ask that, for
each fixed value $s$ of the unconstrained coordinates,
\begin{equation}
\label{eq:minimumV}
V\big(s,f(s)\big) \leq V(s,d) \ , 
 \ \forall d \in \mathcal{D}\big(f(s)\big) \ ,
\end{equation}
where $\mathcal{D}\big(f(s)\big)$ is a suitable open set in $\mathbb{R}^L$
containing the point $f(s)$, and we have that
\begin{equation}
\label{eq:derV}
\frac{\partial V}{\partial d^I}\big(s,f(s)\big) = 0 \ , \quad
 I=K+1,\ldots,N \ ,
\end{equation}
with the corresponding Hessian,
\begin{equation}
\label{eq:HessianV}
\mathcal{H}^\Sigma_{IJ}(s) :=
\frac{\partial^2 V}{\partial d^I \partial d^J}\big(s,f(s)\big) \ ,
\end{equation}
being a positive definite $L \times L$ matrix.

Some points about this last definition are worth remarking before we move on:

\begin{itemize}

\item This choice, which is also used in
refs.~\cite{Go1969JCP,Zho2000JCP,Mor2004ACP,Rei1996PRE,Cot2004BITNM,dOt1998JCP,dOt2000MP,Hel1979JCP,Lei1995Book,Pet2000Thesis,Ral1979JFM}
and in all of the previous works by our group in this topic
\cite{Ech2006JCC2,Ech2006JCC3,Ech2010Sub}, can immediately be seen to come
from either of the two previous definitions if some terms are neglected. In
the momenta-dependent case, we need to neglect the derivative of the kinetic
energy with respect to the constrained coordinates $d$ in eq.~(\ref{eq:derH})
to arrive to eq.~(\ref{eq:derV}); in the case in which the constraints are
defined by minimizing the free energy $F(q)$, the derivative of
$S^\mathrm{k}(q)$ in eq.~(\ref{eq:derF}) needs to be neglected to obtain the
condition in eq.~(\ref{eq:derV}). Using that, for any $N \times N$ matrix
$A(x)$ dependent on a parameter $x$ \cite{Pet2010Misc},
\begin{equation}
\label{eq:derDetA}
\frac{\partial \det A}{\partial x} = \det A 
 \cdot \mathrm{tr} \left(A^{-1} \frac{\partial A}{\partial x} \right) \ ,
\end{equation}
and also that, trivially,
\begin{equation}
\label{eq:derAinverse}
A^{-1}A = \mathbb{I} \quad \Rightarrow \quad
\mathrm{tr} \left(A^{-1}A\right) = N \quad \Rightarrow \quad
\mathrm{tr} \left(\frac{\partial A^{-1}}{\partial x} A \right) +
\mathrm{tr} \left(A^{-1} \frac{\partial A}{\partial x} \right) = 0 \ ,
\end{equation}
the reader can check that the mentioned condition on the kinetic energy
implies the one on $S^\mathrm{k}(q)$ (as expected), but the reverse does not 
hold in general.

\item As we will show in the following sections, if suitable internal
coordinates are used \cite{Ech2006JCC1}, the expression of $\det G(q)$ as a
function of the coordinates is actually very simple. Therefore, not only
all the basic ideas in this work are applicable both to the definition of
constraints using $F(q)$ and the one using $V(w)$, but it is also rather
straightforward, technically speaking, to add the term associated to the
minimization of $S^\mathrm{k}(q)$ to the mix. We plan to do this in future
works.

\item Since, as we mentioned, the potential energy function of molecular
systems in the absence of external fields does not depend on the external
coordinates, the constrained values $f$ defined in this last case do not
depend on the whole set of unconstrained coordinates $u$, but only on the
internal ones $s$; and the relevant constrained subspace is not $\mathcal{K}$
but $\Sigma$ (see sec.~\ref{subsec:notation}).

\begin{figure}[!t]
\begin{center}
\includegraphics[scale=0.3]{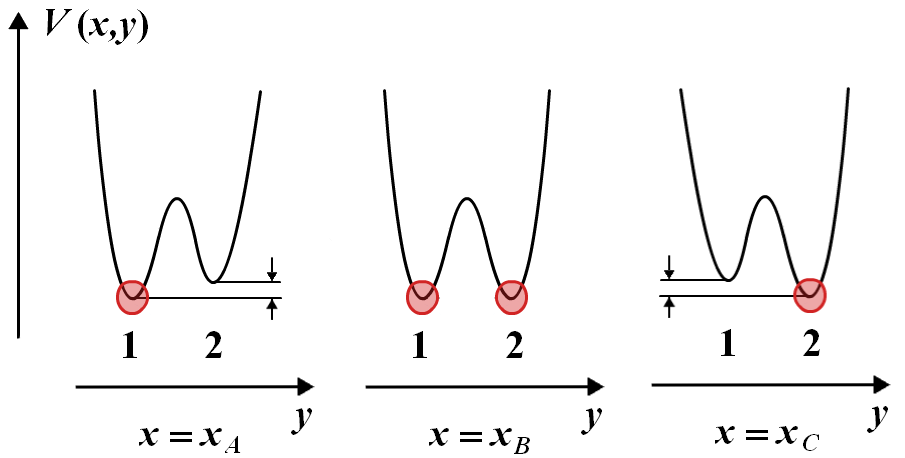}
\caption{\label{fig:2D_minima_flip} Schematic depiction of an unphysical flip
due to the use of the global minimum criterium for defining the constrained
subspace. Three sections of the potential energy surface are shown at three
different values of $x$. At $x_A$, the minimum 1 is the global one, whereas,
at $x_C$, it is 2. At $x_B$, the potential energy of 1 and 2 are the same, and
the global minimum criterium becomes ambiguous.}
\end{center}
\end{figure}

\item For this last case, but also for the previous two, there are at least
two reasons why the definition based on a \emph{local} minima is more
convenient that a possible definition based on a \emph{global} one (such as
the one we erroneously advocated in a previous work by some of us
\cite{Ech2006JCC2}): First, the problem of global minimization is a difficult
one; for general functions it is known to be NP complete
\cite{Wei2009Ebook,Kir1983Sci}, and therefore, even if we wished for some
reason to define the constrained subspace as the place where some function
attains its global minimum, there is no guarantee that we will find the
practical means to compute it. Secondly, such a definition would very likely
cause unphysical effects: Imagine for the sake of simplicity that we have a
system with two position-like degrees of freedom, denoted by $(x,y)$, and
their respective momenta. If we identify the coordinate $y$ as `stiff' and
want to constraint it to its \emph{global} minimum-energy value for each value
of the unconstrained coordinate $x$, we could face the situation schematically
depicted in fig.~\ref{fig:2D_minima_flip}. In such a case, as we smoothly
progress along the coordinate $x$, we can see that the global minimum
definition might become ambiguous at some value $x=x_B$ and then the system
can instantaneously change the value of its coordinate $y$ to one which is
close in energy to the previous one but far away in conformational space. The
flip is unphysical simply because, if the barrier separating the minima is
high enough (something we shall need in order to consider $y$ stiff), then it
will be a very unlikely event that an unconstrained trajectory actually
performs the flip, and even if it does, it will not do it instantly (and with
a discontinuity in the constrained coordinates) as in the flawed constrained
approximation based on global minimization. This situation is not academic but
actually very common when we have different molecular isomers, such as the cis
and trans forms of the peptide bond in proteins \cite{Ech2007CoP}.

\end{itemize}

Using the definition based on the minimization of $V(w)$ from now on, we can
now relate the possible answers to question (a) in
sec.~\ref{subsubsec:coordinates_constrained} to the ideas introduced in this
section: Following the reasoning that has led us to this point and which is
behind the definition of the constrained subspace, a selected set of
constrained coordinates, $d$, will be a priori a good choice [i.e., a good
answer to (a)] if, for `small' variations $\Delta d$ in $d$, it happens that
$V\big(s,f(s) + \Delta d \big) - V\big(s,f(s)\big) \gg RT$. If this is so, the
statistical weight, which is proportional to $\exp\big(-V(s,d)/RT\big)$, will
become negligibly small as soon as we separate from the constrained subspace
$\Sigma$ by any relevant amount. Albeit intuitively appealing (so much that we
ourselves argued in this line in a previous work \cite{Ech2006JCC2}), we will
leave this argument at this point remarking that the scary quotes around
`small' actually hide real definition problems, as anyone can easily see if
she considers that, to begin with, some internal coordinates have length units
and some others are angles. We are at the moment pursuing a more
mathematically involved and more rigorous definition of the properties that a
proper `stiff' constrained coordinate must have, but notice that this only
affects the \emph{quantitative} details pertaining what do we mean when we
write `small'; the basic, intuitive idea that the constrained coordinates are
difficult to change due to energetic reasons is still valid.

Of course, the answer to question (b) discussed in this section, i.e., whether
or not (or how much) the equilibrium values $f(s)$ of the constrained
coordinates depend on the point in $\Sigma$ given by $s$, is also determined
(by definition) by the nature of the potential energy used to model the
system. Using Quantum Mechanics-based potentials, it has already been proved
in small peptides that, for a given choice of constrained coordinates, their
equilibrium values do depend on the conformation $s$ significantly
\cite{Yu2001JMS,Ech2006JCC2,Sch1995BP,Sch1997JMS}. In this work, we will show
\emph{en passant}, as some previous works have already done
\cite{Hes2002JCP,Che2005JCC,Hag1979JACS,Lei1995Book,Zho2000JCP},
that the functions $f(s)$ also depend on $s$ in the case of the much simpler
classical force fields
\cite{Bro2009JCC,Jor1988JACS,Jor1996JACS,Pon2003APC,AMBER10,Pea1995CoPC}
typically used for MD of proteins and nucleic acids. This is hardly surprising
if we think that the force fields potential energy function typically has the
following form:
\begin{equation}
\label{eq:Vff}
V_\mathrm{FF}(w) := 
   \frac{1}{2}\sum_{\alpha = 1}^{N_l}
      K_{l_\alpha}(l_\alpha-l_\alpha^0)^2 + 
   \frac{1}{2}\sum_{\alpha = 1}^{N_\theta}
      K_{\theta_\alpha}(\theta_\alpha - \theta_\alpha^0)^2
   + V_\mathrm{FF}^\mathrm{t}(\phi_\alpha) +
      V_\mathrm{FF}^\mathrm{lr}(w) \ ,
\end{equation}
where $l_\alpha$ are bond lengths, $\theta_\alpha$ are bond angles, and
$\phi_\alpha$ are dihedral angles. The integer $N_l$ is the number of bond
lengths, $N_\theta$ the number of bond angles, and the quantities
$K_{l_\alpha}$, $K_{\theta_\alpha}$, $l_\alpha^0$ and $\theta_\alpha^0$ are
constant numbers. The term denoted by $V_\mathrm{FF}^\mathrm{t}(\phi_\alpha)$
is a commonly included torsional potential that depends only on the dihedral
angles $\phi_\alpha$\footnote{\label{foot:dihedralV} The terms in
$V_\mathrm{FF}^\mathrm{t}(\phi_\alpha)$ associated to `soft' dihedrals, such
as the Ramachandran angles in proteins, are typically trigonometric functions;
whereas those associated to `stiff' dihedrals, such as the peptide-bond
dihedral angle $\omega$, are typically harmonic.}. On the other hand,
$V_\mathrm{FF}^\mathrm{lr}$ normally comprises long-range interactions such as
Coulomb or van der Waals, which involve $O(n^2)$ terms related to all possible
atom pairs, each one of them depending on the difference of the atomic
positions $\vec{r}_\alpha - \vec{r}_\beta$.

It is precisely this last dependence the one that couples all internal
coordinates with one another, the one that makes force fields non-trivial, and
the one that makes the minimum-energy values of, say, bond lengths and bond
angles, \emph{not the constant numbers} $l_\alpha^0$ and $\theta_\alpha^0$ in
the force field but some $s$-dependent quantities.

Although it becomes clear that the most natural way of thinking about the
physical origin of constraints presented in this section suggests that they
must be (at least in some degree) flexible, i.e., that the functions $f(s)$ in
eq.~(\ref{eq:constraints_2b}) must indeed depend on the unconstrained internal
coordinates $s$, the most common approach (by far) in the computational
chemistry literature
\cite{Aba1989JBMSD,Abe1984CC,Alm1990MP,Edb1986JCP,Dub2010JCP,Maz1989JBMSD,Maz1991JCoP,Maz1997JCC,Nog1983JPSJ,Ber1983Book,Cha1979JCP,Che2005JCC,Cic1986CPR,Fix1974PNAS,Fre2002Book,Go1976MM,Hin1994JFM,Hin1995PRE,Kar1981MM,Li2009PRL,Pas2002JCP,Pat2004JCP,Pea1979JCP,Per1985MM,Ryc1977JCoP,Sch2003MP,vGu1977MP,vGu1980MP,vGu1982MM,vGu1989Book}
has been to consider them as constants, or \emph{hard}; existing a few works,
most of them recent, dealing with the flexible case
\cite{Hes2002JCP,Hes2004JCP,Chr2005JCP,Chr2007CPC,Lei1995Book,Rei1996PRE,Sai2004JCP,Zho2000JCP},
including some previous works by us
\cite{Ech2006JCC2,Ech2006JCC3,Ech2010Sub} (although it is worth remarking that
the basic idea and the fact that it comes more naturally from physical
considerations had been mentioned as early as 1969 by G\=o and Scheraga
\cite{Go1969JCP}). This is partly due to numerical convenience but it also
might be related to the great influence that classical force fields have had
in the area. Indeed, if we take a look at the generic expression in
eq.~(\ref{eq:Vff}) of such an energy function, we see that the quantities,
say, $l_\alpha^0$ and $\theta_\alpha^0$, appearing in the harmonic energy
terms associated to bond lengths and bond angles, are \emph{readily available}
to be elected as the candidate constant numbers to which equate this
coordinates should we want to consider them as `stiff' and constrain them;
even if they are not, as mentioned, the actual minimum-energy values of the
bond lengths and bond angles in any given conformation of the molecule. If,
instead of a force field, we consider a \emph{less explicit} energy function,
such as the one produced by the ground-state Born-Oppenheimer approximation
for the electrons in quantum chemistry \cite{Ech2007MP,Sza1996Book}, then no
candidate number appears before our eyes as the constants to be used in the
constrained model, and the flexible way of proceeding is even more natural.

A different but related misconception, probably also stemming from the
prevalence of force fields in the literature about constraints, is the idea
that the steeper the potential energy function is when the system leaves the
constrained subspace, the better approximation the hard one becomes
\cite{Hel1979JCP,Go1976MM}. This is indeed true if the potential has a form
such as the one in eq.~(\ref{eq:Vff}), because, as the `spring-constants'
$K_{l_\alpha}$ and $K_{\theta_\alpha}$ get larger and larger, the associated
harmonic terms overcome the long-range interactions, which were, as we argued,
the ones responsible for the minimum-energy values of bond lengths and bond
angles not being the constant numbers $l_\alpha^0$ and $\theta_\alpha^0$ but
conformation-dependent quantities. However, this particular form of the
potential energy function is not the only possible choice, and one could
perfectly argue that harmonic terms such as the following ones, which
contradict the previously stated argument,
\begin{equation}
\label{eq:non-flex_harmonic}
\frac{1}{2}\sum_{\alpha = 1}^{N_l}
      K_{l_\alpha}\big(l_\alpha-l_\alpha^0(s)\big)^2 + 
   \frac{1}{2}\sum_{\alpha = 1}^{N_\theta}
      K_{\theta_\alpha}\big(\theta_\alpha - \theta_\alpha^0(s)\big)^2
\end{equation}
may model the actual physics of the system more accurately in some cases.
Indeed, using this kind of functional form, we can have infinitely steep
potentials around the constrained subspace at the same time that the
minimum-energy values of the constrained coordinates do depend on $s$.

In sec.~\ref{subsec:comparisons} more information is provided about the use of
particular energy functions in the literature and its influence on the
flexible vs. hard issue, as well as on other approximations. We also point the
reader to the discussion about the choice of coordinates in
secs.~\ref{subsubsec:rigid_flex} and sec.~\ref{subsubsec:stiff_flex}, which is
very much related to the issue of the form of the potential energy function in
the vicinity of the constrained subspace. Finally, in
sec.~\ref{sec:numerical}, we analyze the difference between flexible and hard
constraints in a simple methanol molecule, and some works are mentioned in
which this same comparison has been performed in other practical cases.

\subsection{Statistical Mechanics models}
\label{subsec:SM_models}

\subsubsection{Rigid model with flexible constraints}
\label{subsubsec:rigid_flex}

The reasoning based in energetic considerations in the previous section
justifies to substitute the \emph{fact} that the system is more likely to
evolve close to a given region of its internal space, namely, the internal
constrained subspace $\Sigma$, by the \emph{approximation} that it evolves
\emph{only} in that region.

When imposing constraints on a physical system to enforce this approximation,
it is very common (specially in the theoretical physics literature
\cite{Gol2002Book,Jos1998Book} but also in computational chemistry works
\cite{Fre2002Book,All2005Book}) to assume not only that the constraints are
\emph{exactly} fulfilled at any given time, but also that \emph{D'Alembert's
principle} holds, i.e., that the force of constraint \emph{does no work} (it
is entirely orthogonal to the constrained subspace $\Sigma$). It can be shown
that, assuming these hypotheses, the dynamics of the system, described now in
terms of the unconstrained coordinates $u$ (i.e., the external ones, $e$, plus
the unconstrained internal ones, $s$) is the solution of the Hamilton equations
associated to the following \emph{rigid Hamiltonian} \cite{Gal2007Book}:
\begin{equation}
\label{eq:rigidH}
H_\mathrm{r}(u,\eta) := \frac{1}{2} \eta_r g^{rs}(u) \eta_s
 + V_\Sigma(s) \ ,
\end{equation}
where the restriction $V_\Sigma(s)$ of the potential energy $V(w)$ to the
constrained subspace $\Sigma$ is defined as
\begin{equation}
\label{eq:VSigma}
V_\Sigma(s) := V\big(s,f(s)\big) \ ,	
\end{equation}
being the functions $f(s)$ the ones that define the constraints in
eq.~(\ref{eq:constraints_2b}) and whose origin has been discussed in the
previous section. Of course, in the most general case, these functions do
depend on $s$ and we are therefore working in the \emph{flexible} setting.

Inserting these constraints into the kinetic energy in the unconstrained
Lagrangian in eq.~(\ref{eq:unconstrL}), as well as their time derivatives,
\begin{equation}
\label{eq:der_constraints_2b}
\dot{d}^I = \frac{\partial f^I(s)}{\partial s^i} \dot{s}^i \ ,
\end{equation}
we obtain the \emph{rigid kinetic energy}:
\begin{equation}
\label{eq:rigidK}
K_\mathrm{r}(u,\dot{u}) := \frac{1}{2} \dot{u}^r g_{rs}(u) \dot{u}^s \ ,
\end{equation}
where $g_{rs}(u)$ is the \emph{induced mass-metric tensor} (or 
\emph{pull-back}), given by
\begin{equation}
\label{eq:g}
g_{rs}(u) := G_{rs}^\mathcal{K}(u)
   + \frac{\partial f^I(u)}{\partial u^r} G_{Is}^\mathcal{K}(u)
   + G_{rJ}^\mathcal{K}(u) \frac{\partial f^J(u)}{\partial u^s}
   + \frac{\partial f^I(u)}{\partial u^r} G_{IJ}^\mathcal{K}(u)
     \frac{\partial f^J(u)}{\partial u^s} \ ,
\end{equation}
being $G_{\mu\nu}^\mathcal{K}(u)$ simply the restriction of the whole-space
MMT in eq.~(\ref{eq:G}) to the constrained subspace $\mathcal{K} = \mathcal{E}
\times \Sigma$, i.e.,
\begin{equation}
\label{eq:GK}
G_{\mu\nu}^\mathcal{K}(u) := G_{\mu\nu}\big(e,s,f(s)\big) \ .
\end{equation}

Now eqs.~(\ref{eq:VSigma}) and (\ref{eq:rigidK}) allows us to construct
the \emph{rigid Lagrangian},
\begin{equation}
\label{eq:rigidL}
L_\mathrm{r}(u,\dot{u}) := \frac{1}{2} \dot{u}^r g_{rs}(u) \dot{u}^s
 - V_\Sigma(s) \ ,
\end{equation}
which finally produces the rigid Hamiltonian in eq.~(\ref{eq:rigidH}) via the
usual Legendre transform, defining the \emph{canonical conjugate momenta} as
\begin{equation}
\label{eq:eta}
\eta_r := \frac{\partial L_\mathrm{r}}{\partial \dot{u}^r} =
   g_{rs}(u) \dot{u}^s \ .
\end{equation}

The constant-energy dynamics produced by the rigid Hamiltonian in
eq.~(\ref{eq:rigidH}) (via the corresponding Hamilton equations) can be
implemented in a computer using, for example, the algorithms described in
refs.~\cite{Zho2000JCP,Hes2002JCP,Chr2005JCP}. If constant-temperature
dynamics needs to be performed in this setting, one can use any of the methods
available for hard rigid simulations
\cite{Ryc1977JCoP,Cic1986CPR,Dub2010JCP,Fer1985MP}, probably after some small
adaptation to the flexible case \cite{Hes2002JCP,Hes2011Priv}. See also
refs.~\cite{Baj2011EPJSTunp,Mad2011EPJSTunp,Yu2011EPJSTunp} in this special
issue for more information about constrained simulations in ensembles other
than the microcanonical one.

With regard to the associated equilibrium statistical mechanics, and making
the same considerations we made for the unconstrained case in
sec.~\ref{subsec:dynamics}, we can write the corresponding \emph{rigid
partition function} in the canonical ensemble as
\begin{equation}
\label{eq:rigidZ}
Z_r = \frac{\alpha_\mathrm{QM}}{h^K} \int
 e^{-\beta H_\mathrm{r}(u,\eta)} \mathrm{d}u \mathrm{d}\eta \ .
\end{equation}

Hence, the \emph{equilibrium PDF} is given by
\begin{equation}
\label{eq:rigidPjoint}
P_\mathrm{r}(u,\eta) = \frac{e^{-\beta H_\mathrm{r}(u,\eta)}}
 {\int e^{-\beta H_\mathrm{r}(u^\prime,\eta^\prime)}
  \mathrm{d}u^\prime \mathrm{d}\eta^\prime} \ ,
\end{equation}
which has the analogous meaning as the one attributed to
eq.~(\ref{eq:unconstrPjoint}) in sec.~\ref{subsec:dynamics}, being the
equilibrium average of any observable $O(u,\eta)$:
\begin{equation}
\label{eq:rigidOavg}
\langle O \rangle_\mathrm{r} = \int O(u,\eta) P_\mathrm{r}(u,\eta)
  \mathrm{d}u \mathrm{d}\eta \ .
\end{equation}

Again, if we are interested in observables that are dependent only on the
positions $u$, we can use eq.~(\ref{eq:Gaussian_integral}) to integrate out
the momenta in eq.~(\ref{eq:rigidZ}), yielding
\begin{equation}
\label{eq:rigidZ_2}
Z_r = \chi_\mathrm{r}(T) \int e^{-\beta \left[ V_\Sigma(s)
 - T \frac{R}{2} \ln \det g(u) \right]} \mathrm{d}u \ ,	
\end{equation}
where
\begin{equation}
\label{eq:rigidchi}
\chi_\mathrm{r}(T) := \left( \frac{2\pi}{\beta} \right)^{K/2}
  \frac{\alpha_\mathrm{QM}}{h^K} \ .
\end{equation}

If we perform the same integration in the joint PDF in
eq.~(\ref{eq:rigidPjoint}), the factor $\chi_\mathrm{r}(T)$ cancels out and we
arrive to the \emph{marginal equilibrium PDF} in the space of the positions
$u$:
\begin{equation}
\label{eq:rigidPu}
P_\mathrm{r}(u) = \frac{ e^{ -\beta \left[ V_\Sigma(s)
 - T \frac{R}{2} \ln \det g(u) \right]}}
 {\int e^{ -\beta \left[ V_\Sigma(s^\prime)
 - T \frac{R}{2} \ln \det g(u^\prime) \right]} \mathrm{d}u^\prime} \ ,
\end{equation}
being the equilibrium average of any momenta-independent observable $O(u)$
\begin{equation}
\label{eq:rigidOavg_2}
\langle O \rangle_\mathrm{r} = \int O(u) P_\mathrm{r}(u) \mathrm{d}u \ .
\end{equation}

The same invariance of the potential energy of molecular systems with respect
to the external coordinates, $e$, that we used in sec.~\ref{subsec:dynamics}
to integrate these degrees of freedom out is here applicable as well, and this
is why we have written $V_\Sigma(s)$ and not $V_\mathcal{K}(u)$ in all the
previous expressions. The induced MMT $g(u)$ in eq.~(\ref{eq:g}) on the other
hand, and in particular its determinant in eqs.~(\ref{eq:rigidZ_2})
and~(\ref{eq:rigidPu}), does depend on the external coordinates. As in
sec.~\ref{subsec:dynamics}, one can always \emph{formally} integrate over the
external coordinates in order to get to the corresponding marginal PDF in
$\Sigma$, i.e., depending only on the unconstrained internal coordinates $s$.
However, until recent, it was not clear if this process could be performed
\emph{analytically} for general flexible constraints, thus yielding manageable
final expressions. In a previous work \cite{Ech2006JCC3}, we settled the issue
proving that this is in fact possible and providing the exact analytical
expressions to be used for the marginal PDF in $\Sigma$ (this case with
constraints is much more involved that the unconstrained one mentioned in
sec.~\ref{subsec:dynamics}, which had been proved long ago by G\=o and
Scheraga using a different development \cite{Go1976MM}).

We showed that the determinant of the reduced MMT $g(u)$ can be written as
follows:
\begin{equation}
\label{eq:detg}
\det g(u) = \sin^2 \theta \det g^\prime(s) \ ,
\end{equation}
where $\theta$ is one of the external Euler angles and the
externals-independent matrix $g^\prime(s)$ is given by
\begin{equation}
\label{eq:gprime}
g^\prime(s) = \sum_\alpha m_\alpha 
\left(\begin{array}{cc|c@{\hspace{8pt}}c@{\hspace{8pt}}c}
\mathbb{I}^{(3)} &
v(\bm{\mathcal{R}}_\alpha^\prime) &
\cdots &
\displaystyle \frac{\partial \bm{\mathcal{R}}_\alpha^\prime}
                   {\partial s^j} &
\cdots \\[10pt]
v^T(\bm{\mathcal{R}}_\alpha^\prime) &
J(\bm{\mathcal{R}}_\alpha^\prime) &
\cdots &
\displaystyle \frac{\partial \bm{\mathcal{R}}^\prime_{\alpha}}
                   {\partial s^j} \times
\bm{\mathcal{R}}^\prime_{\alpha} & \cdots \\[14pt]
\hline
\vdots & \vdots & & \vdots & \\
\displaystyle
\frac{\partial \bm{\mathcal{R}}_\alpha^{\prime\,T}}{\partial s^i} &
\displaystyle \left(\frac{\partial \bm{\mathcal{R}}^\prime_{\alpha}}
                         {\partial s^i} \times
\bm{\mathcal{R}}^\prime_{\alpha} \right)^T &
\cdots &
\displaystyle
\frac{\partial \bm{\mathcal{R}}^{\prime\,T}_{\alpha}}{\partial s^i}
\frac{\partial \bm{\mathcal{R}}^\prime_{\alpha}}{\partial s^j} &
\cdots \\[-2pt]
\vdots & \vdots & & \vdots & 
\end{array} \right) \ ,
\end{equation}
where $\mathbb{I}^{(3)}$ is the $3 \times 3$ identity matrix, and the matrices
$v$ and $J$ are defined as
\begin{equation}
\label{eq:vRalpha}
v(\bm{\mathcal{R}}_\alpha^\prime) :=
\left( \begin{array}{ccc}
0 & -\mathcal{Z}_\alpha^\prime & \mathcal{Y}_\alpha^\prime  \\
\mathcal{Z}_\alpha^\prime & 0 & -\mathcal{X}_\alpha^\prime \\
-\mathcal{Y}_\alpha^\prime & \mathcal{X}_\alpha^\prime & 0
\end{array} \right) \ ,
\end{equation}
and
\begin{equation}
\label{eq:JRalpha}
J(\bm{\mathcal{R}}_\alpha^\prime) :=
\left( \begin{array}{ccc}
\mathcal{Y}_\alpha^{\prime \, 2} + \mathcal{Z}_\alpha^{\prime \, 2} &
- \mathcal{X}_\alpha^\prime \mathcal{Y}_\alpha^\prime &
- \mathcal{X}_\alpha^\prime \mathcal{Z}_\alpha^\prime \\
- \mathcal{X}_\alpha^\prime \mathcal{Y}_\alpha^\prime &
\mathcal{X}_\alpha^{\prime \, 2} + \mathcal{Z}_\alpha^{\prime \, 2} &
- \mathcal{Y}_\alpha^\prime \mathcal{Z}_\alpha^\prime \\
- \mathcal{X}_\alpha^\prime \mathcal{Z}_\alpha^\prime &
- \mathcal{Y}_\alpha^\prime \mathcal{Z}_\alpha^\prime &
\mathcal{X}_\alpha^{\prime \, 2} + \mathcal{Y}_\alpha^{\prime \, 2} \\
\end{array} \right) \ ,
\end{equation}
respectively, being $\mathcal{X}_\alpha^\prime$, $\mathcal{Y}_\alpha^\prime$
and $\mathcal{Z}_\alpha^\prime$ the three Euclidean components of
$\bm{\mathcal{R}}_\alpha^\prime$ (see sec.~\ref{subsec:notation} for a
precise definition of these quantities).

This result allows to factorize the exponential in eqs.~(\ref{eq:rigidZ_2})
and~(\ref{eq:rigidPu}), and to perform a second integration to get rid of the
uninteresting external coordinates $e$, thus arriving to the \emph{marginal
equilibrium} PDF in the space parameterized by the unconstrained internal
coordinates $s$, i.e., $\Sigma$:
\begin{equation}
\label{eq:rigidPs}
P_\mathrm{r}(s) = \frac{e^{-\beta F_\mathrm{r}(s)}}
  {\int e^{-\beta F_\mathrm{r}(s^\prime)} \mathrm{d}s^\prime} \ ,
\end{equation}
where we have defined
\begin{subequations}
\begin{align}
F_\mathrm{r}(s) & := V_\Sigma(s)
 - T S_\mathrm{r}^\mathrm{k}(s) \label{eq:rigidF} \ , \\
S_\mathrm{r}^\mathrm{k}(s) & :=
  \frac{R}{2} \ln \det g^\prime(s) \label{eq:rigidSk} \ .
\end{align}
\end{subequations}

Again, the notation in the equations above is intentional, and we regard
$F_\mathrm{r}(s)$ as a \emph{free} or \emph{effective} energy, and
$S_\mathrm{r}^\mathrm{k}(s)$ as a sort of a \emph{kinetic entropy}.

Using $P_\mathrm{r}(s)$ above, the equilibrium average of any
observable $O(s)$ that depends only on $s$ can be calculated as
\begin{equation}
\label{eq:rigidOavg_3}
\langle O \rangle_\mathrm{r} = \int O(s) P_\mathrm{r}(s) \mathrm{d}s \ .
\end{equation}

Finally, let us remark that the derivation in this section is equivalent to
the ones in
refs.~\cite{Cic1986CPR,Car1989CPL,Rei1996PRE,dOt1998JCP,dOt2000MP,Mor2004ACP,Fix1974PNAS,Pet2000Thesis,Ral1979JFM,Sch2003MP},
being the only difference that, in those works, different generalized
coordinates $\tilde{q}$ are chosen in such a way that the constraint functions
in eq.~(\ref{eq:constraints_1b}),
\begin{equation}
\label{eq:sigma}
\sigma^I(w) := d^I - f^I(s) \ , \quad I=K+1,\ldots,N \ ,
\end{equation}
are among them. Therefore, in these new coordinates $\tilde{q}$, the
constraints and their time derivatives are not expressed as in
eqs.~(\ref{eq:constraints_2b}) and~(\ref{eq:der_constraints_2b}), but simply
as
\begin{subequations}
\label{eq:constraints_2b_qtilde}
\begin{align}
\tilde{d}^I & = 0 \ , \quad I=K+1,\ldots,N \ ,
 \label{eq:constraints_2b_qtilde_a} \\
\dot{\tilde{d}}^I & = 0 \ , \quad I=K+1,\ldots,N \ .
 \label{eq:constraints_2b_qtilde_der_b}
\end{align}
\end{subequations}

Despite the earned simplifications that this choice brings, we have preferred
to use coordinates which do not include the functions $\sigma$ among them
because typical internal coordinates used in molecular simulations in fact
\emph{do not} include these functions \cite{Ech2006JCC1}, and therefore we can
more easily relate to the chemically intuitive quantities represented by them
(bond lengths, bond angles, etc.) in the formalism. However, since many works
use the modified coordinates $\tilde{q}$, we will make a stop now and then to
mention how this choice affects the final expressions and conclusions in this
work.

The transformation to the new coordinates is actually very simple:
\begin{subequations}
\label{eq:q_to_qtilde}
\begin{align}
& \tilde{u}^r = \tilde{U}^r(q) := u^r \ , \quad r=1,\ldots,K \ ,  
  \label{eq:q_to_qtilde_u} \\
& \tilde{d}^I = \tilde{D}^I(q) := d^I - f^I(s) \ , \quad I=K+1,\ldots,N \ ,  
  \label{eq:q_to_qtilde_d}
\end{align}
\end{subequations}
and so it is its inverse:
\begin{subequations}
\label{eq:qtilde_to_q}
\begin{align}
& u^r = U^r(\tilde{q}) := \tilde{u}^r \ , \quad r=1,\ldots,K \ ,  
  \label{eq:qtilde_to_q_u} \\
& d^I = D^I(\tilde{q}) := \tilde{d}^I + f^I(\tilde{s})
  \ , \quad I=K+1,\ldots,N \ .
  \label{eq:qtilde_to_q_d}
\end{align}
\end{subequations}

We will see in what follows that the fact that this transformation affects
only the definition of the constrained coordinates, $d$, produces remarkable
properties and very simple rules for most of the transformed objects.

This can already be seen in the structure of the Jacobian matrix:
\begin{equation}
\label{eq:Jacobian_q_to_qtilde}
(J^\mu_\nu) :=
 \left( \frac{\partial Q^\mu(\tilde{q})}{\partial \tilde{q}^\nu} \right) =
\left(\begin{array}{ccc|c@{\hspace{8pt}}c@{\hspace{8pt}}c}
1 & & 0 & & & \\
  & \ddots & & & 0 & \\
0 & & 1 & & & \\
\hline
 & \vdots & & 1 & & 0 \\
\cdots &
\displaystyle \frac{\partial f^I(\tilde{s})}{\partial \tilde{s}^i}
 & \cdots & & \ddots & \\
 & \vdots & & 0 & & 1
\end{array} \right) \ ,
\end{equation}
and its inverse:
\begin{equation}
\label{eq:Jacobian_qtilde_to_q}
\big((J^{-1})^\mu_\nu\big) :=
 \left( \frac{\partial \tilde{Q}^\mu(q)}{\partial q^\nu} \right) =
\left(\begin{array}{ccc|c@{\hspace{8pt}}c@{\hspace{8pt}}c}
1 & & 0 & & & \\
  & \ddots & & & 0 & \\
0 & & 1 & & & \\
\hline
 & \vdots & & 1 & & 0 \\
\cdots &
\displaystyle - \frac{\partial f^I(s)}{\partial s^i}
 & \cdots & & \ddots & \\
 & \vdots & & 0 & & 1
\end{array} \right) \ ,
\end{equation}
both of which have unit determinant: $\det J = \det J^{-1} = 1$.

It is also worth mentioning that, if the constraints are defined as hard (see
the next section), then the $\sigma$ functions in eq.~(\ref{eq:sigma}) become
\begin{equation}
\label{eq:sigma_non-flex}
\sigma^I(w) := d^I - d^I_0 \ , \quad I=K+1,\ldots,N \ ,
\end{equation}
being $d^I_0$ constant numbers, and in such a case the simplicity of the
change of coordinates is even higher, being the Jacobian and its inverse
in the expressions above both exactly equal to the identity matrix.

In any case, either if the generalized velocities are related as in
eq.~(\ref{eq:der_constraints_2b}) or if they are zero as in
eq.~(\ref{eq:constraints_2b_qtilde_der_b}) and in
refs.~\cite{Cic1986CPR,Car1989CPL,Rei1996PRE,dOt1998JCP,dOt2000MP,Mor2004ACP,Fix1974PNAS,Pet2000Thesis,Ral1979JFM,Sch2003MP},
it is convenient to bear in mind that momenta and velocities are only
proportional when the matrix in the kinetic energy is diagonal (e.g., in
Euclidean coordinates), and the statement sometimes found in the literature
\cite{Rei1998NA,Cha1979JCP,Hel1979JCP,Ber1983Book,Fix1974PNAS,vGu1980MP,vGu1989Book}
about the cancelation of the momenta associated to the constrained coordinates
is wrong in general.

Regarding the calculations in this section, the use of the modified
coordinates $\tilde{q}$ leads to a simpler expression for the induced MMT in
eq.~(\ref{eq:g}):
\begin{equation}
\label{eq:g_qtilde}
\tilde{g}_{rs}(\tilde{u}) = \tilde{G}_{rs}^\mathcal{K}(\tilde{u}) \ ,
\end{equation}
i.e., it is simply the unconstrained-unconstrained block of the whole-space
MMT $\tilde{G}$ evaluated in the constrained subspace (see
sec.~\ref{subsec:comparisons} for the implications of this in the
calculation of Fixman's potential).

Moreover, if we use eq.~(\ref{eq:G_q_to_qtilde}) for the transformation of
the whole-space MMT $\tilde{G}$ together with the change of coordinates in
eq.~(\ref{eq:qtilde_to_q}), we can prove that
\begin{eqnarray}
\label{eq:g_qtilde2}
\tilde{g}_{rs}(\tilde{u}) & = & \tilde{G}_{rs}^\mathcal{K}(\tilde{u}) 
 := \tilde{G}_{rs}(\tilde{u},0)
  \nonumber \\
 & = & G_{rs}\big(Q(\tilde{u},0)\big) 
   + \frac{\partial f^I(\tilde{u})}{\partial \tilde{u}^r}
     G_{Is}\big(Q(\tilde{u},0)\big)
   + G_{rJ}\big(Q(\tilde{u},0)\big)
     \frac{\partial f^J(\tilde{u})}{\partial \tilde{u}^s} \nonumber \\
 & & \mbox{} + \frac{\partial f^I(\tilde{u})}{\partial \tilde{u}^r}
     G_{IJ}\big(Q(\tilde{u},0)\big)
	 \frac{\partial f^J(\tilde{u})}{\partial \tilde{u}^s} \nonumber \\
 & = & G_{rs}\big(\tilde{u},f(\tilde{u})\big) 
   + \frac{\partial f^I(\tilde{u})}{\partial \tilde{u}^r}
     G_{Is}\big(\tilde{u},f(\tilde{u})\big)
   + G_{rJ}\big(\tilde{u},f(\tilde{u})\big)
     \frac{\partial f^J(\tilde{u})}{\partial \tilde{u}^s} \nonumber \\
 & & \mbox{} + \frac{\partial f^I(\tilde{u})}{\partial \tilde{u}^r}
     G_{IJ}\big(\tilde{u},f(\tilde{u})\big)
	 \frac{\partial f^J(\tilde{u})}{\partial \tilde{u}^s}
 = g_{rs}(\tilde{u}) \ ,
\end{eqnarray}
i.e., the induced MMT is not modified upon the change of coordinates from
$q$ to $\tilde{q}$; it is the same as the original one component-wise.

This directly implies that the calculations associated to the factorization
and elimination of the external coordinates in ref.~\cite{Ech2006JCC3}, which
led us to expressions~(\ref{eq:detg}), (\ref{eq:gprime}), and related
quantities, still apply in the new modified coordinates. The only change to be
done is to place a `tilde' over the appropriate symbols.

\subsubsection{Rigid model with hard constraints}
\label{subsubsec:rigid_non-flex}

As we mentioned in sec.~\ref{subsubsec:flexible_vs_nonflexible}, the most
common constrained dynamics considered in the literature is not the rigid one
with flexible constraints described in the previous section, but the one with
\emph{hard} constraints. This option, which is numerically simpler,
and which can be accurate enough for certain applications, is described in
the following paragraphs.

The key difference between the two models is that, in the hard case,
the constrained coordinates $d$ are not equal to some functions $f(s)$ of the
unconstrained internal coordinates, as expressed in
eq.~(\ref{eq:constraints_2b}), but to some constant numbers $d_0$:
\begin{equation}
\label{eq:constraints_non-flex}
d^I = d^I_0 \ , \quad I=K+1,\ldots,N \ .
\end{equation}

This makes all partial derivatives in eqs.~(\ref{eq:der_constraints_2b})
and~(\ref{eq:g}) zero, and converts the induced MMT $g(u)$ simply in the
unconstrained-unconstrained sub-block of the restriction to $\mathcal{K}$ of
the whole-space MMT, i.e.,
\begin{equation}
\label{eq:g_non-flex}
\bar{g}_{rs}(u) = \bar{G}_{rs}^\mathcal{K}(u) := G_{rs}(e,s,d_0) \ ,
\end{equation}
where we have used an over-bar to denote hard objects. This fact has important
implications regarding the Fixman's compensating potential, introduced in
sec.~\ref{subsec:comparisons}. Also notice, that, contrarily to
eq.~(\ref{eq:g_qtilde}), which only held in the flexible case if the modified
coordinates $\tilde{q}$ were used, we now have this property in the original
molecular coordinates $q$.

Another, more subtle consequence of the use of hard constraints
affects the calculations of the partial derivatives $\partial
\bm{\mathcal{R}}_\alpha^\prime / \partial s^i$, which are a key ingredient of
the matrix $g^\prime(s)$ in eq.~(\ref{eq:gprime}). If we realize that the
functions $\bm{\mathcal{R}}_\alpha^\prime(s)$ are implicitly defined in
eq.~(\ref{eq:transf_constrained}) as
\begin{equation}
\label{eq:Rcalprime}
\bm{\mathcal{R}}_\alpha^\prime(s) := 
 \vec{R}_\alpha^\prime\big(s,f(s)\big) \ ,
\end{equation}
the sought derivatives can be computed using
\begin{equation}
\label{eq:derRcalprime}
\frac{\partial \bm{\mathcal{R}}_\alpha^\prime}{\partial s^i}(s) =
 \frac{\partial \vec{R}_\alpha^\prime}{\partial s^i}\big(s,f(s)\big)
+\frac{\partial \vec{R}_\alpha^\prime}{\partial d^I}\big(s,f(s)\big)
 \frac{\partial f^I}{\partial s^i}(s) \ .
\end{equation}

In the flexible case, all terms in the sum are in principle non-zero, and
the fact that the derivatives $\partial f^I / \partial s^i$ must be typically
found numerically requires the use of complicated algorithms to obtain this
quantities (see sec.~\ref{sec:numerical} and ref.~\cite{Ech2010Sub}).
In the hard case, on the other hand, all these derivatives are zero
and eq.~(\ref{eq:derRcalprime}) reduces to
\begin{equation}
\label{eq:derRcalprime_non-flex}
\frac{\partial \bm{\mathcal{R}}_\alpha^\prime}{\partial s^i}(s) =
 \frac{\partial \vec{R}_\alpha^\prime}{\partial s^i}(s,d_0) \ ,
\end{equation}
which contains only explicit derivatives which can typically be computed
analytically. The efficient calculation of the MMT, MMT determinants, and
related quantities in
refs.~\cite{Pat2004JCP,Abe1984CC,Nog1983JPSJ,Fix1974PNAS,Pas2002JCP,Pea1979JCP,Per1985MM},
where they consider only hard constraints, is implicitly based on this fact.

Apart from these (important) differences, the dynamics of the rigid model with
hard constraints is derived following the same steps we took in the previous
section, and the rigid Hamiltonian in eq.~(\ref{eq:rigidH}) is still
applicable to this case as long as we compute the induced MMT as indicated in
eq.~(\ref{eq:g_non-flex}). This dynamics can be implemented in a computer
using many different algorithms (see
refs.~\cite{Ryc1977JCoP,Cic1986CPR,Dub2010JCP}, and references therein; and
also \cite{Elb2011EPJSTunp} in this special issue), and, again, under the
usual assumptions of ergodicity and equal a priori probabilities, its
statistical mechanics equilibrium can be described in a formally identical way
to the flexible case, via eqs.~(\ref{eq:rigidPs}), (\ref{eq:rigidF})
and~(\ref{eq:rigidSk}).

Despite this formal identity, however, it has been shown that the flexible and
hard dynamics produce different physical results (see
refs.~\cite{Chr2007CPC,Hes2002JCP,Rei1996PRE,Rei1995PD,Zho2000JCP} and also
sec.~\ref{sec:numerical}), and indeed in
sec.~\ref{subsubsec:flexible_vs_nonflexible} we built a strong case for
considering the latter as an approximation of the former. In
sec.~\ref{sec:numerical}, we will additionally illustrate how this difference
in the dynamics produce measurable discrepancies at the level of equilibrium
statistical mechanics in a methanol molecule.

\subsubsection{Stiff model with flexible constraints}
\label{subsubsec:stiff_flex}

In the previous two sections, we discussed the dynamics as well as the
canonical equilibrium of the so-called \emph{rigid} model, in which not only
the constraints are assumed to hold exactly but also D'Alembert's principle,
which hypothesizes that the components of the forces of constraint tangent to
the constrained subspace $\Sigma$ are zero \cite{Gol2002Book,Gal2007Book}.

Nevertheless, if we accept the unconstrained dynamics as the reference against
which to assess the accuracy of any approximate model (see the discussion
supporting this choice in sec.~\ref{subsubsec:flexible_vs_nonflexible}), we
must realize that, if the system is in contact with a thermal bath at
temperature $T$, even if there exist strong forces that drive it to the
constrained subspace $\Sigma$, there also exist random fluctuations that will
inevitably take the system away from this region. Moreover, even in the case
in which we can assume that the potential energy outside $\Sigma$ tends to
infinity (a case in which the constraints hold exactly by simple energy
conservation arguments \cite{Arn1989Book}), it can be proved that the force on
$\Sigma$ resulting from this limit procedure not only consists of the
orthogonal component needed to keep the system in the constrained subspace,
but also of a non-zero tangent component that must be added to the expected
gradient of $V_\Sigma(s)$, i.e., D'Alembert's principle, even in the
infinitely steep confining potential case, is not satisfied in general
\cite{Gal2007Book,Fro2001CMP,Bor1995TR2,Bor1995TR3,Rub1957CPAM,vKa1984AJP}.

These two in general violated assumptions confirm what we already suspected:
that the rigid dynamics and statistical mechanics equilibrium are just an
approximation of the real, unconstrained ones (even if we use the more
physically sound flexible constraints). Therefore, if the rigid dynamics is
to be used in practical simulations in order to gain the advantages discussed
in sec.~\ref{sec:introduction}, it becomes necessary to assess how much error
these approximations introduce.

Of course, in order to compare the rigid statistical mechanics equilibrium PDF
$P_\mathrm{r}(s)$ in eq.~(\ref{eq:rigidPs}) to the unconstrained one in $P(w)$
in eq.~(\ref{eq:unconstrPw}) (or the free energies in their respective
exponents), we should arrive to some objects that are defined on the same
space, being the most natural candidates $P_\mathrm{r}(s)$ itself and the
marginal unconstrained PDF in the space, $\Sigma$, spanned by the
unconstrained internal coordinates $s$. This last function is obtained through
the integration of the constrained coordinates $d$ as usual:
\begin{equation}
\label{eq:unconstrPs}
P(s):=\int P(w) \mathrm{d}d \ .
\end{equation}

However, in the general case this is only formal, i.e., for a general
potential $V(w)$ and general curvilinear coordinates $q$, this integral cannot
be calculated analytically. One of the available options is to compute it
numerically, with the help of methods for calculating free energy differences
\cite{Cic2005CPC,Car1989CPL,Tor1977JCoP,Har2011EPJSTunp,Sch2011EPJSTunp}. In
this work, on the other hand, we have chosen to develop a statistical
mechanics model, termed \emph{stiff}, whose effective free energy in $\Sigma$
can be analytically obtained and compared to the rigid one in a more direct
and insightful way than the result of numerically performing the integral in
eq.~(\ref{eq:unconstrPs}), at the same time that it could be considered an
approximation to the unconstrained equilibrium containing weaker assumptions
than the rigid one. Contrarily to the rigid model, the stiff one \emph{does}
account for the statistical weights of the conformations close to (but out of)
$\Sigma$, at the same time that it allows for the existence of velocities that
are orthogonal to the constrained subspace. Also, in the derivation of the
stiff model, D'Alembert's principle is never invoked. These arguments,
together with the knowledge of the controlled steps that take us from the
unconstrained case to the stiff model, suggest that one could expect the stiff
equilibrium to be closer to the unconstrained one than the rigid equilibrium
not only conceptually but also in quantitative terms. Although in this article
we will assume this to be true and the stiff model will be used as the
reference against which to assess the accuracy of any constrained model,
numerical confirmation will be pursued in future works.

The first step in the derivation of the stiff model consists of Taylor
expanding the potential energy $V(s,d)$ at a fixed point $s$, in terms of the
constrained coordinates $d$, around the point $d=f(s)$:
\begin{eqnarray}
\label{eq:VTaylor}
V(s,d) & = & V\big(s,f(s)\big) + 
   \left[\frac{\partial V}{\partial d^I}\big(s,f(s)\big)\right]
   \big(d^I - f^I(s)\big) \\
 & + & \frac{1}{2} \left[\frac{\partial^2 V}
    {\partial d^I \partial d^J}\big(s,f(s)\big)\right]
    \big(d^I - f^I(s)\big)\big(d^J - f^J(s)\big)
  + O\big(\big(d-f(s)\big)^3\big) \ , \nonumber
\end{eqnarray}
or, realizing that to evaluate in $\big(s,f(s)\big)$ is to evaluate in
the internal constrained subspace $\Sigma$,
\begin{eqnarray}
\label{eq:VTaylor_2}
V(s,d) & = & V_\Sigma(s) + 
   \partial_I V_\Sigma (s)
   \big(d^I - f^I(s)\big) \\
 & + & \frac{1}{2} \mathcal{H}_{IJ}^\Sigma(s)
    \big(d^I - f^I(s)\big)\big(d^J - f^J(s)\big)
  + O\big(\big(d-f(s)\big)^3\big) \ , \nonumber
\end{eqnarray}
where we have used the notation for the constrained Hessian in
eq.~(\ref{eq:HessianV}), and we have introduced the notation $\partial_I
V_\Sigma (s)$ for the gradient of $V(w)$ restricted to $\Sigma$.

The same expansion can be performed on $S^\mathrm{k}(w)$, yielding
\begin{eqnarray}
\label{eq:SkTaylor}
S^\mathrm{k}(s,d) & = & S^\mathrm{k}_\Sigma(s) + 
   \partial_I S^\mathrm{k}_\Sigma(s)
   \big(d^I - f^I(s)\big) \\
 & + & \frac{1}{2} \partial^2_{IJ} S^\mathrm{k}_\Sigma(s)
    \big(d^I - f^I(s)\big)\big(d^J - f^J(s)\big) 
  + O\big(\big(d-f(s)\big)^3\big) \ , \nonumber
\end{eqnarray}
and both of them can be introduced into the integral that appear in the
definition of the marginal PDF in eq.~(\ref{eq:unconstrPs}):
\begin{equation}
\label{eq:unconstrPsfromPw}
P(s) := \int P(w) \mathrm{d}d =
 \frac{\int e^{ -\beta \left[ V(w) - T S^\mathrm{k}(w) \right]} \mathrm{d}d}
 {\int e^{ -\beta \left[ V(w^\prime)
 - T S^\mathrm{k}(w^\prime) \right]} \mathrm{d}s^\prime \mathrm{d}d^\prime}
 \ .
\end{equation}

Note that expanding the term associated to the determinant of the whole-space
MMT $\det G$ in the marginal PDF in positions space is in principle more
accurate that the procedure followed in
refs.~\cite{Mor2004ACP,dOt1998JCP,dOt2000MP,Cic1986CPR,Fix1974PNAS,Fre2002Book,Kar1981MM,Li2009PRL,Pet2000Thesis,Ral1979JFM,Sch2003JCP}
or in our own work in ref.~\cite{Ech2006JCC2}, where the dependence of $\det
G$ on the constrained coordinates $d$ is simply neglected, evaluating this
determinant directly in $d=f(s)$ and thus effectively truncating the expansion
in eq.~(\ref{eq:SkTaylor}) at zero order (often implicitly). In fact, this
point and the inclusion of flexible constraints in the mix are the only issues
that make the formalism in this work more general than the ones in
refs.~\cite{Ech2006JCC2,Mor2004ACP,dOt1998JCP,dOt2000MP,Pet2000Thesis,Ral1979JFM}.

Now, if we assume that the constrained coordinates are indeed `stiff', i.e.,
that, as mentioned in sec.~\ref{subsubsec:flexible_vs_nonflexible}, for
`small' variations $\Delta d$ in $d$, we have $V\big(s,f(s) + \Delta d \big) -
V\big(s,f(s)\big) \gg RT$, then the quantities under the integral signs, which
are proportional to $e^{-V(s,d)/RT}$, will become negligibly small as soon as
we separate from the constrained subspace $\Sigma$ by any relevant amount.
Moreover, we also know that, close to $\Sigma$, the Taylor expansions in
eqs.~(\ref{eq:VTaylor_2}) and~(\ref{eq:SkTaylor}) can be truncated at low
order. Therefore, for each fixed value of the unconstrained coordinates $s$,
we can substitute the potential energy and the kinetic entropy terms by their
respective low-order Taylor expansions. Also note that this very same argument
can be used if we had chosen to define $\Sigma$ through the minimization of
$F(q)$ instead of $V(w)$.

In principle, there is no reason to truncate at different orders the potential
energy $V(s,d)$ and the kinetic entropy $S^\mathrm{k}(s,d)$.

If we truncate both expansions at \emph{order zero}, we simply have that the
exponential does not depend on the constrained coordinates, and the integral
$\int 1 \mathrm{d}d$ cancels out in the numerator and the denominator of
eq.~(\ref{eq:unconstrPw}), yielding the \emph{(0,0)-stiff} PDF in $\Sigma$:
\begin{equation}
\label{eq:stiff00Ps}
P_\mathrm{s}^{(0,0)}(s) = \frac{e^{-\beta F^{(0,0)}_\mathrm{s}(s)}}
  {\int e^{-\beta F^{(0,0)}_\mathrm{s}(s^\prime)} \mathrm{d}s^\prime} \ ,
\end{equation}
where we have defined
\begin{subequations}
\begin{align}
F^{(0,0)}_\mathrm{s}(s) & := V_\Sigma(s)
 - T S_\Sigma^\mathrm{k}(s) \label{eq:stiff00F} \ , \\
S_\Sigma^\mathrm{k}(s) & :=
  \frac{R}{2} \ln \det G^\prime_\Sigma(s) \label{eq:stiffSk} \ ,
\end{align}
\end{subequations}
being $G^\prime_\Sigma(s)$ the restriction to $\Sigma$ of the matrix
$G^\prime(w)$.

Although this statistical mechanics model is well-defined and may be certainly
considered for some applications (in sec.~\ref{subsec:comparisons} we
mention a number of works in which it has been implicitly used), two issues
lead to try to improve it: On the one hand, the truncation of the Taylor
expansions at order zero means that we are only looking at points
\emph{exactly} on $\Sigma$, and not `close' to it, which was one of the
weaknesses, we argued, of the rigid model. On the other hand, the function we
are substituting $V(s,d)$ with, i.e., $V_\Sigma(s)$, does not have the
important property of becoming very large when the constrained coordinates get
away from $\Sigma$ (because it does not depend on them!). The fact that we
could obtain a finite PDF, $P_\mathrm{s}^{(0,0)}(s)$, is due to the
cancellation of an infinite quantity\footnote{\label{foot:int1dd} At least if
any bond length is included among the constrained coordinates, which is always
the case in practical applications.}, $\int 1 \mathrm{d}d$, in the numerator
and the denominator. However, if we consider the partition function of the
system after performing the substitution, $Z=\int e^{-\beta \left(V_\Sigma(s)
- T S_\Sigma^\mathrm{k}(s)\right)} \mathrm{d}s \mathrm{d}d$, we see that it is
infinite, which is certainly a warning and probably a problem.

If we truncate both expansions at the \emph{first order}, and we assume that
we are dealing with flexible constraints defined by the functions $f(s)$
taking the values of the constrained coordinates that minimize the potential
energy at point $s$, then the first derivatives $\partial_I V_\Sigma(s)$ are
zero, and the leading order in the expansion of $V(s,d)$ is again $V_\Sigma
(s)$. Therefore, for the (1,1)-stiff model, the exponent in the integral is
proportional to $V_\Sigma(s) - T S_\Sigma^\mathrm{k}(s) - T \partial_I
S^\mathrm{k}_\Sigma(s) \big(d^I - f^I(s)\big)$, and, since the sign of the
terms $\partial_I S^\mathrm{k}_\Sigma(s)$ can be anyone, there is no reason
for the integral on the constrained coordinates to be convergent. If we
considered the definition of $\Sigma$ involving the minimization not of $V(w)$
but of $F(q)$, the situation is slightly different but simple, since in this
case $\partial_I V_\Sigma(s) - T \partial_I S^\mathrm{k}_\Sigma(s) = 0$, and
the (1,1)-stiff model would be equivalent to the (0,0) one.

It becomes then clear that the second order terms must be included in the
model if we want to work with finite quantities. Truncating both expansions at
\emph{second order}, and using eq.~(\ref{eq:Gaussian_integral}), we find the
\emph{most general} expression for the integral over the constrained
coordinates that will be considered in this work:
\begin{eqnarray}
\label{eq:stiff22integral}
 & & \int \exp\Big[-\beta \Big( V_\Sigma - T S_\Sigma^\mathrm{k} +
 (\partial_I V_\Sigma - T \partial_I S_\Sigma^\mathrm{k})(d^I-f^I)
  \nonumber \\
 & & \quad \mbox{} + \frac{1}{2} (d^I-f^I)A_{IJ}(d^J-f^J)
 \Big) \Big] \mathrm{d}d \\
 & & = \left(\frac{2\pi}{\beta}\right)^\frac{L}{2}
 \mathrm{det}^{-\frac{1}{2}} A
  \exp\Big[ \frac{\beta}{2}
  (\partial_I V_\Sigma - T \partial_I S_\Sigma^\mathrm{k}) A^{IJ}
  (\partial_J V_\Sigma - T \partial_J S_\Sigma^\mathrm{k}) \Big] \ , \nonumber
\end{eqnarray}
where we have omitted the dependence on $s$, and we have defined
\begin{equation}
\label{eq:A}
A_{IJ} := \mathcal{H}_{IJ}^\Sigma - T \partial^2_{IJ} S_\Sigma^\mathrm{k} \ ,
\end{equation}
in order to lighten the notation. We have also approximated the different
ranges of integration of the constrained coordinates all by the
$(-\infty,\infty)$ range (notice that bond lengths range from 0 to $\infty$,
bond angles from 0 to $\pi$, etc.). The reason why this is expected to be a
good approximation is the same that supports the rest of the calculations,
namely, that the integrated quantity is only non-negligible close to the
constrained subspace $\Sigma$, where the constrained coordinates are typically
far away from the true integration limits.

Despite this formal result, one needs to notice that, for the above integral
to exist, the matrix $A$ has to be positive definite. If we had chosen the
definition of the constrained subspace based on the minimization of $F(q)$,
this property would be satisfied, since $A$ is precisely the Hessian of $F(q)$
at a minimum. In the case treated here, however, the positive definiteness of
$A$ is not guaranteed. If we are in a minimum of the potential energy, the
constrained Hessian $\mathcal{H}_{IJ}^\Sigma$ in eq.~(\ref{eq:HessianV}) is
indeed positive definite, but the presence of the term $- T \partial^2_{IJ}
S_\Sigma^\mathrm{k}$ spoils this property, in general, for the matrix $A$. If
such a thing happens, or even if it is `close' to happening (for example, if
the smallest eigenvalue of $A$ becomes too close to zero), we must consider
the possibility of moving to the model based on the minimization of $F(q)$.

Another issue that may cause the eigenvalues of the matrix $A$ to be small and
thus render the stiff approximation questionable is that the coordinates we
have selected as constrained may actually be a bad choice. Indeed, if the
statistical weights of the conformations do not become significantly small as
we move away from $\Sigma$ at second order in the unconstrained free energy
$F(w):=V(w)-TS^\mathrm{k}(w)$, then we have little reasons to believe that the
situation might change if we include higher orders. Hence, although the
(2,2)-stiff model might not be a very good approximation at every point $s$,
we should see the positive definiteness of the matrix $A$ in eq.~(\ref{eq:A})
and the size of its eigenvalues as a \emph{consistency check}, and regard
those points at which these conditions fail as points in which the (2,2)-stiff
model is a bad approximation of the unconstrained equilibrium, and the
constrained coordinates are probably not `stiff' enough. Although we have
not found conformations in which $A$ (or $\mathcal{H}$; see below) contains
negative eigenvalues in the practical example in sec.~\ref{sec:numerical},
this is probably so due to the small size of the molecule studied. For larger
systems, we have already seen some preliminary indications that these
matrices might in fact become not positive definite depending on the chosen
constrained coordinates.

Now, using eq.~(\ref{eq:stiff22integral}) and recalling that the first
derivatives of the potential energy restricted to $\Sigma$ are zero in the
chosen model, we have that the \emph{(2,2)-stiff} PDF in $\Sigma$ is given by
\begin{equation}
\label{eq:stiff22Ps}
P_\mathrm{s}^{(2,2)}(s) = \frac{e^{-\beta F^{(2,2)}_\mathrm{s}(s)}}
  {\int e^{-\beta F^{(2,2)}_\mathrm{s}(s^\prime)} \mathrm{d}s^\prime} \ ,
\end{equation}
with
\begin{equation}
\label{eq:stiff22F}
F^{(2,2)}_\mathrm{s}(s) := V_\Sigma(s)
 - T S_\Sigma^\mathrm{k}(s) - T S_A^\mathrm{c}(s)
  - \frac{T^2}{2} \partial_I S_\Sigma^\mathrm{k}(s)
                     A^{IJ}(s) \partial_J S_\Sigma^\mathrm{k}(s) \ ,
\end{equation}
being $A^{IJ}$ the entries of $A^{-1}$ and
\begin{equation}
\label{eq:stiffScA}
S_A^\mathrm{c}(s) := - \frac{R}{2} \ln \det A(s) \ ,
\end{equation}
where the letter `c' in the new entropy stands for \emph{conformational}, in
reference to the fact that it appears as the result of eliminating some
positions, and not momenta, being reminiscent of the \emph{conformational}
or \emph{configurational entropies} appearing in quasi-harmonic analysis
\cite{And2001JCP,Kar1981MM,vGu1982MM}. The only difference that we would
find if we chose the definition based on the minimization of $F(q)$ is that
the last term in eq.~(\ref{eq:stiff22F}) would not appear. As we mentioned,
the (2,2)-stiff model is the most general stiff model in this work and it
has never been used before in the literature as far as we know.

The calculation of $\mathcal{H}_\Sigma(s)$, which appears in $A(s)$, can be
performed in many different ways depending on the potential energy we are
dealing with (the computation of the other part of $A$ is described later).
The calculation of $\det G^\prime_\Sigma(s)$, on the other hand, is entirely
`geometric', depending only on the scheme of internal coordinates chosen to
describe the system (assuming that the functions $f(s)$ have already been
calculated). The way to compute $\det G^\prime(w)$ \emph{analytically} has
been described before, using the analogue of eq.~(\ref{eq:gprime}) but without
constraints. However, if a specific set of internal coordinates is chosen, one
can arrive to a much simpler and more explicit expression. In our work in
ref.~\cite{Ech2006JCC3}, we showed that, if the SASMIC internal coordinates
\cite{Ech2006JCC1} for general branched molecules are used (a similar result
will hold for any well-designed scheme of internal coordinates), we have that
\begin{equation}
\label{eq:detGprime_SASMIC}
\det G^\prime(w) = \left( \prod_{\alpha=1}^{n} m_{\alpha}^{3} \right)
         \left( \prod_{\alpha=2}^{n} l_{\alpha}^{4} \right)
          \left( \prod_{\alpha=3}^{n} {\sin}^{2}\theta_{\alpha} \right) \ ,
\end{equation}
where $l_\alpha$ are the bond lengths, and $\theta_\alpha$ the bond angles
associated to atom $\alpha$. The masses can be eliminated, since they come out
of the logarithm as an additive term, which represents a constant change of
reference in the free energy, and has no effect whatsoever in the equilibrium
PDF. G\=o and Scheraga also proved this result long ago using different
mathematics \cite{Go1976MM}.

This expression, together with the additivity of the logarithm, not only
helps to easily calculate $S^\mathrm{k}(w)$ [and therefore its restriction
$S^\mathrm{k}_\Sigma(s)$], but also its derivatives with respect to any
internal coordinate depending on its type:
\begin{subequations}
\label{eq:der_Sk}
\begin{align}
\frac{\partial S^\mathrm{k}}{\partial l_\alpha}(w) & = \frac{2R}{l_\alpha}  
   \ , \label{eq:der_Sk_a} \\
\frac{\partial S^\mathrm{k}}{\partial \theta_\alpha}(w) &
   = R \, \mathrm{cotg} \, \theta_\alpha \ , \label{eq:der_Sk_b} \\
\frac{\partial S^\mathrm{k}}{\partial \phi_\alpha}(w) & = 0 \ ,
   \label{eq:der_Sk_c}
\end{align}
\end{subequations}
where $\phi_\alpha$ denotes the dihedral angle associated to atom $\alpha$.

Using these results, we can rewrite eq.~(\ref{eq:stiff22F}) as
\begin{equation}
\label{eq:stiff22F_2}
F^{(2,2)}_\mathrm{s}(s) := V_\Sigma(s)
 - T S_\Sigma^\mathrm{k}(s) - T S_A^\mathrm{c}(s)
 + U^{(2)}_\mathrm{s}(s) \ ,
\end{equation}
with
\begin{equation}
\label{eq:stiffU2}
U^{(2)}_\mathrm{s}(s) := - \frac{1}{2\beta^2} D\big(f^I(s)\big) A^{IJ}(s)
                          D\big(f^J(s)\big) \ ,
\end{equation}
being $D(d^I)$ the function defined as
\begin{equation}
\label{eq:Dd}
D(d^I) :=
\begin{cases}
2 / d^I &
  \mathrm{if}\ d^I \ \mathrm{is}\ \mathrm{a}\ 
  \mathrm{bond}\ \mathrm{length} \\
\mathrm{cotg}\,d^I &
  \mathrm{if}\ d^I \ \mathrm{is}\ \mathrm{a}\ 
  \mathrm{bond}\ \mathrm{angle} \\
0 &
  \mathrm{if}\ d^I \ \mathrm{is}\ \mathrm{a}\ 
  \mathrm{dihedral}\ \mathrm{angle} \\
\end{cases} \ .
\end{equation}

The entries, $\partial^2_{IJ} S^\mathrm{k}(w)$, of the matrix whose
restriction to $\Sigma$ appears in the definition of $A$ in eq.~(\ref{eq:A})
can also be easily computed, yielding
\begin{equation}
\label{eq:partial2SkIJ}
T \partial^2_{IJ} S^\mathrm{k}(w) = - \frac{1}{\beta} \delta_{IJ} D_2(d^I) \ ,
\end{equation}
where $\delta_{IJ}$ is the Kronecker's delta, and $D_2(d^I)$ is the function
defined as
\begin{equation}
\label{eq:D2d}
D_2(d^I) :=
\begin{cases}
2 / (d^I)^2 &
  \mathrm{if}\ d^I \ \mathrm{is}\ \mathrm{a}\ 
  \mathrm{bond}\ \mathrm{length} \\
1 / \sin^2 d^I &
  \mathrm{if}\ d^I \ \mathrm{is}\ \mathrm{a}\ 
  \mathrm{bond}\ \mathrm{angle} \\
0 &
  \mathrm{if}\ d^I \ \mathrm{is}\ \mathrm{a}\ 
  \mathrm{dihedral}\ \mathrm{angle} \\
\end{cases} \ ,
\end{equation}
allowing to rewrite eq.~(\ref{eq:A}) as
\begin{equation}
\label{eq:A_2}
A_{IJ}(s) := \mathcal{H}_{IJ}^\Sigma(s)
  + \frac{1}{\beta} \delta_{IJ} D_2\big(f^I(s)\big) \ .
\end{equation}

Note that both $D(d^I)$ and $D_2(d^I)$ are singular only in points that have
little physical sense and that will be never reached in a practical
simulation: $l_\alpha = 0$ or $\theta_\alpha = 0,\pi$. Not only the energy
will be very large (even infinite) at these points, but also the change of
coordinates in eq.~(\ref{eq:change_r_to_q}) from the Euclidean to the
curvilinear coordinates becomes singular.

Apart from the (2,2)-stiff model, we can also define two other models which
yield convergent integrals for any point $s$. However, they are derived
truncating the potential energy and the kinetic entropy at different orders;
something which is, in principle, not justified.

Particularizing eq.~(\ref{eq:stiff22integral}), we have that the
\emph{(2,0)-stiff} effective free energy is given by
\begin{equation}
\label{eq:stiff20F}
F^{(2,0)}_\mathrm{s}(s) := V_\Sigma(s)
 - T S_\Sigma^\mathrm{k}(s) - T S_\mathcal{H}^\mathrm{c}(s) \ ,
\end{equation}
with
\begin{equation}
\label{eq:stiffScH}
S_\mathcal{H}^\mathrm{c}(s) :=
 - \frac{R}{2} \ln \det \mathcal{H}^\Sigma(s) \ .
\end{equation}

As we discuss in sec.~\ref{subsec:comparisons}, this model has been
implicitly used in some works, including one by some us \cite{Ech2006JCC2}.

The \emph{(2,1)-stiff} free energy, in turn, includes an additional term:
\begin{equation}
\label{eq:stiff21F}
F^{(2,1)}_\mathrm{s}(s) := V_\Sigma(s)
 - T S_\Sigma^\mathrm{k}(s) - T S_\mathcal{H}^\mathrm{c}(s)
 + U^{(1)}_\mathrm{s}(s) \ ,
\end{equation}
defined as
\begin{equation}
\label{eq:stiffU1}
U^{(1)}_\mathrm{s}(s) := - \frac{1}{2\beta^2} D\big(d^I(s)\big)
                     \mathcal{H}_\Sigma^{IJ}(s)
                     D\big(d^J(s)\big) \ ,
\end{equation}
where we have used again that the first derivatives of the potential energy
are zero in $\Sigma$, and $\mathcal{H}_\Sigma^{IJ}(s)$ are the entries of the
inverse matrix of the constrained Hessian, as usual. If the definition of the
constrained subspace based in the minimization of $F(q)$ is used, then the
(2,0)- and (2,1)-stiff models become identical.

Now, if the modified coordinates $\tilde{q}$ defined in
sec.~\ref{subsubsec:rigid_flex} are used in the calculations in this section,
several points need to be re-examined. First of all, we have the following
relationship between the whole-space MMT in the coordinates $q$ and the one
in the coordinates $\tilde{q}$ (indeed, we have it for any 2-times
covariant tensor \cite{Dub1992Book}):
\begin{equation}
\label{eq:G_q_to_qtilde}
\tilde{G}_{\mu\nu}(\tilde{q}) =
  \frac{\partial Q^\rho(\tilde{q})}{\partial \tilde{q}^\mu}
  G_{\rho\sigma}\big(Q(\tilde{q})\big)
  \frac{\partial Q^\sigma(\tilde{q})}{\partial \tilde{q}^\nu} \ .
\end{equation}

Hence, if we use the fact that the determinant of a product of square
matrices is the product of the determinants, and also the property mentioned
in sec.~\ref{subsubsec:rigid_flex} about the Jacobian determinant of the
change of coordinates being 1, we have that
\begin{equation}
\label{eq:detGtilde}
\det \tilde{G}(\tilde{q}) = \det G\big(Q(\tilde{q})\big) \ .
\end{equation}

Note that this not only guarantees the factorization of the external
coordinates in eq.~(\ref{eq:detG}), but it also allows us to use the explicit
expression~(\ref{eq:detGprime_SASMIC}) in terms of the molecular coordinates
$q$ to compute $\det \tilde{G}(\tilde{q})$ as long as we insert the
transformation functions $Q(\tilde{q})$ into the appropriate functional places
corresponding to bond lengths and bond angles.

Also using the structure of the Jacobian in
eq.~(\ref{eq:Jacobian_q_to_qtilde}), it is easy to prove that the transformed
potential energy,
\begin{equation}
\label{eq:V_qtilde}
\tilde{V}(\tilde{q}) := V\big(Q(\tilde{q})\big) \ ,
\end{equation}
and indeed any scalar function (such as $\det G$ above, and hence
$S^\mathrm{k}$), satisfies
\begin{equation}
\label{eq:derV_qtilde}
\frac{\partial \tilde{V}}{\partial \tilde{d}^I}(\tilde{q}) =
\frac{\partial V}{\partial d^I}\big(Q(\tilde{q})\big) \ ,
\end{equation}
as well as
\begin{equation}
\label{eq:der2V_qtilde}
\frac{\partial^2 \tilde{V}}{\partial \tilde{d}^I
                            \partial \tilde{d}^J}(\tilde{q}) =
\frac{\partial^2 V}{\partial d^I \partial d^J}\big(Q(\tilde{q})\big) \ .
\end{equation}

These transformation rules guarantee that all the final expressions in this
section are identical in the case that we wished to use the modified
coordinates $\tilde{q}$, as long as we `place tildes' over all relevant
objects, such as $\tilde{\mathcal{H}}(\tilde{q})$, $\tilde{A}(\tilde{q})$ or
$\tilde{S^\mathrm{k}}(\tilde{q})$. But not only that, they also allow us to
compute the objects in the original coordinates $q$ and then simply perform
the substitution $Q(\tilde{q})$ to find the transformed quantities. This is
very convenient, since the coordinates $\tilde{q}$ are defined in terms of the
functions $f(s)$, which are typically impossible to express analytically,
while the coordinates $q$ are simple functions of the Euclidean coordinates,
thus admitting compact algorithms for computing quantities such as the Hessian
of the potential energy, or the functions $D(d)$ and $D_2(d)$ in
eqs.~(\ref{eq:Dd}) and~(\ref{eq:D2d}), respectively.

It is also worth remarking that, in the derivation of the stiff models in the
coordinates $\tilde{q}$, the Taylor expansion of the potential energy in
eq.~(\ref{eq:VTaylor}) would be performed at $\tilde{d}=0$ and take the
simpler form
\begin{equation}
\label{eq:VTaylor_qtilde}
\tilde{V}(\tilde{s},\tilde{d}) = \tilde{V}(\tilde{s},0) +
 \left[\frac{\partial \tilde{V}}{\partial \tilde{d}^I}(\tilde{s},0)\right]
 \tilde{d}^I +
 \left[\frac{\partial^2 \tilde{V}}
            {\partial \tilde{d}^I \partial \tilde{d}^J}(\tilde{s},0)\right]
 \tilde{d}^I\tilde{d}^J + O\big(\tilde{d}^3\big) \ ,
\end{equation}
which allows to make contact with some of the potential energy functions in
previous works
\cite{Cot2004BITNM,dOt1998JCP,dOt2000MP,Mor2004ACP,Ral1979JFM,Rei1996PRE},
realizing that the intention of the authors may not be to restrict themselves
to a particular case of the completely general potential considered here, but
just that they are using the modified coordinates $\tilde{q}$, instead of the
molecular coordinates $q$ used in this work. See however the discussion in
sec.~\ref{subsec:comparisons}, about the neglection of the dependence of the
Hessian on $\tilde{s}$, which is not a general feature of
eq.~(\ref{eq:VTaylor_qtilde}), irrespective of the coordinates used.

Finally, to close this section, let us raise an important and subtle point
related to the stiff model and the dynamics of the system: As we have
emphasized in previous sections, the analysis presented in this work is
completely at the equilibrium statistical mechanics level; all approximations
are performed in equilibrium PDFs and all constrained models are defined in
these terms too. In the previous rigid sections~\ref{subsubsec:rigid_flex}
and~\ref{subsubsec:rigid_non-flex}, we briefly commented that the rigid
equilibrium can be sampled using available dynamical algorithms such as SHAKE
\cite{Ryc1977JCoP,Cic1986CPR,Elb2011EPJSTunp} or any of its more modern
descendants (see ref.~\cite{Dub2010JCP} and references therein) for the hard
case; and the algorithms in refs.~\cite{Zho2000JCP,Hes2002JCP,Chr2005JCP} if
the constraints are flexible. That this sampling can be achieved by such
algorithms is easily understandable if we recall that the rigid model is
obtained from a Hamiltonian [the one in eq.~(\ref{eq:rigidH})] and thus the
corresponding discretized Hamilton equations (properly thermostated
\cite{Baj2011EPJSTunp}) will do the trick. Moreover, the fact that the fastest
vibrations are absent from this dynamics makes the integration algorithms
potentially more efficient than the unconstrained dynamics, and the whole
scheme is thus useful.

In the case of the stiff model introduced in this section, the situation is
entirely different. We may of course ask the theoretical question of which
would be a dynamics that directly samples the stiff statistical mechanics
equilibrium, and we may even find some dynamics that do the job. However, any
naive version of such a dynamics will in principle contain the fast vibrations
that we sought to eliminate with the use of constraints, and therefore any
algorithm that implements it would probably be as numerically costly as the
whole unconstrained dynamics (at least). Since the only rationale behind the
development of the stiff model was to provide a good approximation to the
unconstrained system in order to be able to assess the accuracy of the rigid
equilibrium, using a new `stiff' dynamics instead of the unconstrained one
itself seems a rather unprofitable (and convoluted) enterprise. In principle,
the stiff model should be regarded as an approximation to the unconstrained
equilibrium at the level of statistical mechanics, not at the level of the
dynamics. However, as we will discuss in sec.~\ref{subsec:comparisons},
there is a way of sampling the stiff equilibrium using a rigid dynamics with
an additional correcting term (see fig.~\ref{fig:summary} for a schematic
summary of these ideas). Whether or not the calculation of this correcting
term can be performed fast enough so that we end with a dynamics which has
the accuracy of the stiff model while retaining the efficiency of the rigid
one is, however, a topic that we will not analyze in depth in this work.

\subsubsection{Stiff model with hard constraints}
\label{subsubsec:stiff_non-flex}

In principle, we can also consider the stiff model defined using the
hard constraints in eq.~(\ref{eq:constraints_non-flex}). The main
difference with the flexible case appears in the Taylor expansions in
eq.~(\ref{eq:VTaylor_2}) and~(\ref{eq:SkTaylor}), which become
\begin{eqnarray}
\label{eq:VTaylor_non-flex}
V(s,d) & = & \bar{V}_\Sigma(s) + \partial_I \bar{V}_\Sigma (s)
   (d^I - d^I_0) \\
 & + & \frac{1}{2} \bar{\mathcal{H}}_{IJ}^\Sigma(s)
    (d^I - d^I_0)(d^J - d^J_0)
  + O\big((d-d_0)^3\big) \ , \nonumber
\end{eqnarray}
and
\begin{eqnarray}
\label{eq:SkTaylor_non-flex}
S^\mathrm{k}(s,d) & = & \bar{S}^\mathrm{k}_\Sigma(s) + 
   \partial_I \bar{S}^\mathrm{k}_\Sigma(s)
   (d^I - d^I_0) \\
 & + & \frac{1}{2} \partial^2_{IJ} \bar{S}^\mathrm{k}_\Sigma(s)
    (d^I - d^I_0)(d^J - d^J_0) 
  + O\big((d-d_0)^3\big) \ , \nonumber
\end{eqnarray}
where we have used again an over-bar to distinguish objects associated to the
hard constraints from their flexible counterparts (meaning they are
evaluated at $d=d_0$ instead of $d=f(s)$).

Also, since the values $d=d_0$ do not need to minimize the potential energy
at the point $s$ [nor the free energy $F(q)$], the derivatives $\partial_I
\bar{V}_\Sigma(s)$, $\partial_I \bar{S}^\mathrm{k}_\Sigma(s)$ and $\partial_I
\bar{V}_\Sigma(s) - T \partial_I \bar{S}^\mathrm{k}_\Sigma(s)$ are all
non-zero in general, and neither the Hessian $\bar{\mathcal{H}}^\Sigma(s)$ nor
the associated matrix $\bar{A}(s)$ are guaranteed to be positive definite in
the hard case.

All calculations for the hard stiff models are formally identical to
the ones in the previous section as long as we include the terms $\partial_I
\bar{V}_\Sigma(s)$ and $\partial_I \bar{S}^\mathrm{k}_\Sigma(s)$, and
substitute the functions $f(s)$ by the constant values $d_0$.

In particular, the \emph{hard (2,2)-stiff} model has free energy
\begin{equation}
\label{eq:stiff22F_non-flex}
\bar{F}^{(2,2)}_\mathrm{s}(s) := \bar{V}_\Sigma(s)
 - T \bar{S}_\Sigma^\mathrm{k}(s) - T \bar{S}_A^\mathrm{c}(s)
 + \bar{U}^{(2)}_\mathrm{s}(s) \ ,
\end{equation}
with
\begin{subequations}
\label{eq:entropies_non-flex}
\begin{align}
\bar{S}_\Sigma^\mathrm{k}(s) & := \frac{R}{2}
  \ln \det \bar{G}^\prime_\Sigma(s) \ , \label{eq:stiffSk_non-flex} \\
\bar{S}_\mathrm{s}^\mathrm{c}(s) & := - \frac{R}{2}
  \ln \det \bar{A}(s) \ , \label{eq:stiffScA_non-flex}
\end{align}
\end{subequations}
and
\begin{equation}
\label{eq:stiffU2_non-flex}
\bar{U}^{(2)}_\mathrm{s}(s) :=
 - \frac{1}{2}
 \left(\partial_I \bar{V}_\Sigma(s) - \frac{D(d^I_0)}{\beta} \right)
   \bar{A}^{IJ}(s)
 \left(\partial_J \bar{V}_\Sigma(s) - \frac{D(d^J_0)}{\beta} \right) \ ,
 \nonumber
\end{equation}
being
\begin{equation}
\label{eq:A_non-flex}
\bar{A}_{IJ}(s) := \bar{\mathcal{H}}_{IJ}^\Sigma(s)
  + \frac{1}{\beta} \delta_{IJ} D_2(d^I_0) \ .
\end{equation}

As in the flexible case, the substitution of the second order expansion of the
free energy into the integral in eq.~(\ref{eq:unconstrPsfromPw}) is only
justified if $\bar{A}$ is positive definite and has large enough eigenvalues.
In the case of the hard (2,2)-stiff model, the failure of such a
property can be caused not only by the fact that the chosen constrained
coordinates are actually not very `stiff' or because we have decided to
minimize $V(w)$ instead of $F(q)$ (as it happened for the flexible case in the
previous section), but also because of having expanded around a value, $d_0$,
which is not a minimum of the potential energy [or of $F(q)$], thereby
possibly distorting the observed curvature encoded in
$\bar{\mathcal{H}}^\Sigma$ and $\bar{A}$.

The only hard stiff model that presents an equilibrium PDF that is
convergent for all points $s$ (but a partition function that is infinite) is
the (0,0)-stiff model, whose free energy is given by
\begin{equation}
\label{eq:stiff00F_non-flex}
\bar{F}^{(0,0)}_\mathrm{s}(s) := \bar{V}_\Sigma(s)
 - T \bar{S}_\Sigma^\mathrm{k}(s) \ .
\end{equation}

As in the flexible case, the first order truncation produces models that are,
in general, non-convergent, while we can still define two models that
\emph{may} be well-behaved depending on the properties of the hard
constrained Hessian $\bar{\mathcal{H}}^\Sigma$: the (2,0)- and (2,1)-stiff
models. Notice, on the one hand, that these two models were convergent at all
points $s$ in the flexible case, because the flexible constrained Hessian is,
by construction, positive definite. On the other hand, the same caution
applies in the hard case regarding the a priori justification of
truncating the potential energy $V(s,d)$ and the kinetic entropy
$S^\mathrm{k}(s,d)$ at different orders.

The \emph{hard (2,0)-stiff} model has free energy
\begin{equation}
\label{eq:stiff20F_non-flex}
\bar{F}^{(2,0)}_\mathrm{s}(s) := \bar{V}_\Sigma(s)
 - T \bar{S}_\Sigma^\mathrm{k}(s) - T \bar{S}_\mathcal{H}^\mathrm{c}(s)
 + \bar{U}^{(0)}_\mathrm{s}(s) \ ,
\end{equation}
with
\begin{equation}
\label{eq:stiffU0_non-flex}
\bar{U}^{(0)}_\mathrm{s}(s) := - \frac{1}{2}
  \partial_I \bar{V}_\Sigma(s)
   \bar{\mathcal{H}}_\Sigma^{IJ}(s)
  \partial_J \bar{V}_\Sigma(s) \ ,
\end{equation}
and
\begin{equation}
\label{eq:stiffScH_non-flex}
\bar{S}_\mathcal{H}^\mathrm{c}(s) :=
 - \frac{R}{2} \ln \det \bar{\mathcal{H}}^\Sigma(s) \ ,
\end{equation}
whereas the free energy of the \emph{hard (2,1)-stiff} model is
given by
\begin{equation}
\label{eq:stiff21F_non-flex}
\bar{F}^{(2,0)}_\mathrm{s}(s) := \bar{V}_\Sigma(s)
 - T \bar{S}_\Sigma^\mathrm{k}(s) - T \bar{S}_\mathcal{H}^\mathrm{c}(s)
 + \bar{U}^{(1)}_\mathrm{s}(s) \ ,
\end{equation}
being
\begin{equation}
\label{eq:stiffU1_non-flex}
\bar{U}^{(1)}_\mathrm{s}(s) :=
 - \frac{1}{2}
 \left(\partial_I \bar{V}_\Sigma(s) - \frac{D(d^I_0)}{\beta} \right)
   \bar{\mathcal{H}}_\Sigma^{IJ}(s)
 \left(\partial_J \bar{V}_\Sigma(s) - \frac{D(d^J_0)}{\beta} \right) \ .
\end{equation}

It is also worth noting here that, in the hard case treated in this
section, there is no difference between minimizing $V(w)$ or $F(q)$, because
the constrained subspace is not defined through \emph{any} minimization, but
it is postulated to simply be $d=d_0$.

Some of these models have been implicitly used in the literature before, as we
mention in the next section. In fact, being the hard constraints more
popular, their use have been more common than that of their flexible
counterparts in sec.~\ref{subsubsec:stiff_flex}.

\subsection{Summary, comparisons and the Fixman's potential}
\label{subsec:comparisons}

\begin{figure}
\begin{center}
\includegraphics[scale=0.25]{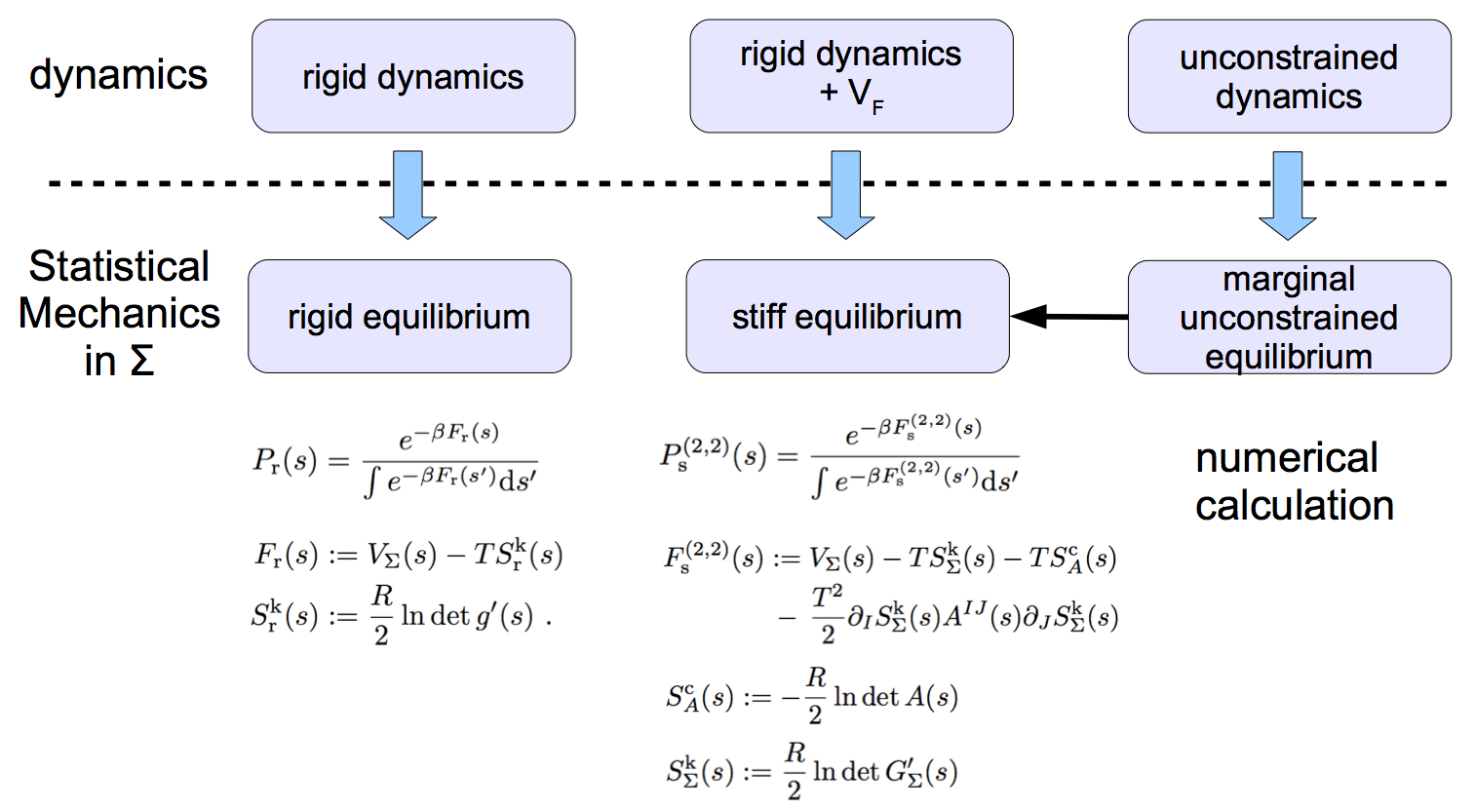}
\caption{\label{fig:summary} Summary and relationships among the
models discussed in this work. For a precise definition of the different
mathematical objects and the meaning of the models, see the main body of
the article.}
\end{center}
\end{figure}

In fig.~\ref{fig:summary}, a schematic depiction of the models discussed in
the previous sections can be found, together with the relationships among
them, as well as the basic mathematical expressions used to describe their
respective statistical mechanics in the constrained subspace. In the figure,
only the flexible versions of the constrained models have been detailed, and
only the more accurate (2,2)-type of stiff model is shown. An equivalent
scheme will be obtained if we define the constrained subspace $\Sigma$ using
the hard relations $d=d_0$ or we truncate the Taylor expansions
defining the stiff model at a different order than 2.

In words, we have the unconstrained dynamics, which, we have argued, is the
reference to which we will judge the accuracy of any constrained model. On the
other end of the scheme, we have the rigid dynamics, which is potentially
faster than the unconstrained one because the fastest vibrations have been
eliminated from the system by constraining the associated degrees of freedom.
Despite this better numerical profile, the rigid dynamics, and its
corresponding statistical mechanics equilibrium in the constrained subspace
$\Sigma$ are based on some questionable assumptions, such as D'Alembert's
principle, or the imposition that the dynamics takes place \emph{exactly} on
$\Sigma$, thus forbidding the system to present velocities that are orthogonal
to this space. These strong approximations require that we assess the accuracy
of the rigid model against the reference unconstrained situation. In this
work, we choose to do so at the level of equilibrium statistical mechanics,
where the key object is the PDF and the free energy that appears in its
exponent. Given the typical complexity of molecular potential energy
functions, the integral needed to marginalize the unconstrained PDF to the
constrained subspace $\Sigma$ and thus be able to compare it with the rigid
one is too complicated to be performed analytically. This is why we have
introduced an approximation to the unconstrained equilibrium, which has
$\Sigma$ as its natural probability space, which is based in weaker
assumptions than the rigid model, and which is termed \emph{stiff model}. Now,
by comparing the rigid equilibrium to the stiff one, the accuracy of the
former can be assessed. That they are indeed appreciably different has long be
known in the literature and many works have discussed it
\cite{Alm1990MP,Ber1983Book,Cha1979JCP,dOt1998JCP,dOt2000MP,Fix1974PNAS,Fre2002Book,Go1969JCP,Go1976MM,Has1974JCP,Hel1979JCP,Hin1994JFM,Lei1995Book,Mor2004ACP,Pas2002JCP,Pat2004JCP,Pea1979JCP,Rei2000PD,Cic1986CPR,Per1985MM,Pet2000Thesis,Ral1979JFM,vGu1980MP,vKa1984AJP}.

The most natural way to perform this assessment is to compare the two
effective free energies that appear in the exponents of their respective PDFs.
For example, if we consider the more general case of flexible constraints in
secs.~\ref{subsubsec:rigid_flex} and~\ref{subsubsec:stiff_flex}, and we use
the most accurate (2,2)-stiff model, we can define the following function of
the unconstrained coordinates $s$:
\begin{equation}
\label{eq:VF}
V_\mathrm{F}(s) := F_\mathrm{s}^{(2,2)}(s) - F_\mathrm{r}(s)
 = T S_\mathrm{r}^\mathrm{k}(s) - T S_\Sigma^\mathrm{k}(s) 
 - T S_A^\mathrm{c}(s) + U^{(2)}_\mathrm{s}(s) \ ,
\end{equation}
which quantifies the difference between the two free energies, and contains
many of the terms related to the MMT and Hessian determinants that we have
introduced in the previous sections (but notice that the constrained potential
energy $V_\Sigma(s)$ has been canceled out).

This function was first introduced by Fixman \cite{Fix1974PNAS} (although in a
much more simplified version) and it is thus normally called \emph{Fixman's
potential}. Apart from quantifying the discrepancies between the equilibrium
statistical mechanics of the rigid model and the, in principle, more accurate
stiff one, it can be easily seen that the addition of $V_\mathrm{F}(s)$ to the
original potential energy $V_\Sigma(s)$ in the rigid dynamics produces an
evolution of the system whose equilibrium is now given by the stiff PDF, at
the same time that the fastest vibrations remain absent from the dynamics
(assuming that $V_\mathrm{F}(s)$ has not reintroduced them). Indeed, if we
modify the rigid Hamiltonian in eq.~(\ref{eq:rigidH}) and use the following
one instead:
\begin{equation}
\label{eq:rigidHplusVF}
H_\mathrm{r}^\mathrm{F}(u,\eta) := \frac{1}{2} \eta_r g^{rs}(u) \eta_s
 + V_\Sigma(s) + V_\mathrm{F}(s) \ ,
\end{equation}
it is immediate to see that the canonical equilibrium PDF of the corresponding
dynamics (after integrating out the momenta and the external coordinates as
we described in the previous sections) will be given by
\begin{equation}
\label{eq:rigidPsplusVF}
P_\mathrm{r}^\mathrm{F}(s) = \frac{e^{-\beta F_\mathrm{r}^\mathrm{F}(s)}}
  {\int e^{-\beta F_\mathrm{r}^\mathrm{F}(s^\prime)} \mathrm{d}s^\prime} \ ,
\end{equation}
where the obtained free energy $F_\mathrm{r}^\mathrm{F}(s)$ can be checked to
be exactly the stiff one after some simple cancelations:
\begin{eqnarray}
\label{eq:rigidFsplusVF}
F_\mathrm{r}^\mathrm{F}(s) & = & V_\Sigma(s)
 - T S_\mathrm{r}^\mathrm{k}(s) + V_\mathrm{F}(s) \nonumber \\
 & = & V_\Sigma(s) - T S_\mathrm{r}^\mathrm{k}(s)
 + T S_\mathrm{r}^\mathrm{k}(s) - T S_\Sigma^\mathrm{k}(s) 
 - T S_A^\mathrm{c}(s) + U^{(2)}_\mathrm{s}(s) \nonumber \\
 & = & V_\Sigma(s) - T S_\Sigma^\mathrm{k}(s) 
 - T S_A^\mathrm{c}(s) + U^{(2)}_\mathrm{s}(s) \nonumber \\
 & =: & F_\mathrm{s}^{(2,2)}(s) \ .
\end{eqnarray}

This practical use of the Fixman's potential to sample the stiff equilibrium
using a modified rigid dynamics has often been discussed in the literature
\cite{Sch2003MP,Pas2002JCP,Rei2000PD,dOt2000MP,dOt1998JCP,Per1985MM,Pea1979JCP,Cha1979JCP,Hel1979JCP,Fix1978JCP,Ber1983Book,Cha1979JCP,Hin1994JFM,Lei1995Book,Per1985MM,Rei1996PRE,vGu1980MP},
and it is probable the main motivation behind its first introduction
\cite{Fix1974PNAS}. In this work, however, no dynamical treatment is going to
be performed, and we will use $V_\mathrm{F}(s)$ simply as a way of quantifying
the difference between the stiff and rigid models at the level of equilibrium
statistical mechanics. As we discussed before, this comparison can be
understood as an assessment of the accuracy of the rigid equilibrium and,
indirectly, as an evaluation of how much error the approximations behind the
rigid model (including D'Alembert's principle) produce in the resulting
statistical mechanics of the system.

We also point out that, if the justification of constrained models is not
based in their comparison to the classical unconstrained case, as in this
work, but to quantum mechanical calculations, then it can be argued that the
equilibrium produced by the rigid dynamics is the `correct' one, and therefore
no corrections such as $V_\mathrm{F}(s)$ are needed \cite{Hes2002JCP}. We
discussed this issue in detail in
sec.~\ref{subsubsec:flexible_vs_nonflexible}.

Apart from the introduction of $V_\mathrm{F}(s)$, Fixman proved in his
celebrated paper \cite{Fix1974PNAS} a theorem that, as we will see later,
allows to simplify some MMT determinants in some cases. This useful result has
been used in some works to compute certain PDFs even when no comparisons
between stiff and rigid models are sought
\cite{Cic2005CPC,Cic1986CPR,Car1989CPL}. Additionally, the issue of the
comparison between constrained models in which the orthogonal velocities to
the constrained subspace are activated and those in which they are assumed to
be zero is central to the interpretation of the methods for computing free
energy differences along a reaction coordinate
\cite{Cic2005CPC,Car1989CPL,Har2007PD,Har2011EPJSTunp,Sch2011EPJSTunp,Sch2003JCP};
however, despite the formal similarity, the reasons for imposing constraints
and the physical considerations behind them are very different in the two
cases. We suspect that these facts add even more confusion and complexity to
some of the accounts about the topic.

It is also worth pointing out that the dynamics produced by the modified
Hamiltonian in eq.~(\ref{eq:rigidHplusVF}) needs not to bear any resemblance
to either the original rigid dynamics or the unconstrained one. The only
convenient features of this dynamics, by construction, are the ones we have
mentioned: (i) its equilibrium statistical mechanics is the stiff one, and (ii)
assuming that $V_\mathrm{F}(s)$ does not introduce new fast oscillations, the
dynamics is still rigid-like, in the sense that the fastest degrees of freedom
have been removed and thus it can be integrated with a larger time-step than
the unconstrained dynamics. In \cite{Bor1995TR2}, for example, it is shown,
under reasonable assumptions, that the averaged dynamics in the infinitely
stiff case satisfies some modified equations of motion that evolve the
trajectory in the constrained subspace, and whose modification \emph{is not}
the force corresponding to the Fixman's potential introduced here. This
discrepancy could have several origins:

\begin{itemize}

\item The differences in the dynamics could be averaged out at equilibrium and
the resulting PDF may turn out to be the same for the two cases [as we said,
the dynamics produced by eq.~(\ref{eq:rigidHplusVF}) are only good for
statistical mechanics sampling]. Since the dynamics in eq.~(12) of
\cite{Bor1995TR2} is, in principle, non-Hamiltonian, the techniques in
\cite{Tuc2001JCP} should be used to obtain its equilibrium PDF.

\item As we mentioned in sec.~\ref{sec:introduction} and we will develop
further later in this section, the calculations in this work and the
definition of Fixman's potential in eq.~(\ref{eq:VF}) are the most general
in the literature as far as we are aware. In every other work,
including \cite{Bor1995TR2}, the version of the Fixman's potential that is
used includes approximations. Therefore, a comparison of the modified
dynamics in \cite{Bor1995TR2} to the one produced by the Hamiltonian in
eq.~(\ref{eq:rigidHplusVF}) could yield different conclusions.

\item Although the truncation of the Taylor expansion that took us to our
definition of the stiff model in sec.~\ref{subsubsec:stiff_flex} seems very
much related to the `infinitely stiff' limit of the equations of motion in
\cite{Bor1995TR2}, there may exist subtle differences that could make the two
physical situations sit on a different basis.

\end{itemize}

We do not pretend to solve this issue here, and the only reason why we have
commented on it is to highlight the fact that dynamical analyses of
constrained models are much more involved than the straightforward approach in
this work, based in equilibrium properties. If one disregards the accuracy of
the trajectories, the fact that the modified Hamiltonian in
eq.~(\ref{eq:rigidHplusVF}) produces the stiff equilibrium through a
constrained dynamics is very easy to prove and uncontroversial, as we have
shown. Since the stiff model does not invoke in its derivation the D'Alembert
principle, it is also clear that $V_\mathrm{F}(s)$ is an appropriate way of
quantifying the accuracy of such an approximation; even if it were not the
appropriate choice for doing the same thing at the dynamical level. It is also
worth remarking that, doing similar dynamical analyses as those in
\cite{Bor1995TR2}, Reich finds a correcting term for the unconstrained
coordinates which \emph{does} seem to agree with the Fixman potential
\cite{Rei1995PD,Rei1996PRE}.

Having these issues in mind, as we mentioned, the calculations in this work
are the most general in the literature as far as we are aware. This is so
because of a number of points:

\begin{itemize}

\item Apart from the requirement that they are `stiff', no special property
has been assumed about the constrained coordinates $d$. On the other hand, in
many works
\cite{Cic1986CPR,Rei1996PRE,Cot2004BITNM,dOt1998JCP,dOt2000MP,Fix1974PNAS,Mor2004ACP,Pet2000Thesis,Ral1979JFM,Sch2003MP},
the constraint functions $\sigma^I(w) := d^I - f^I(s)$ in eq.~(\ref{eq:sigma})
have been chosen as the set $d$. In such a case, as we have discussed in the
previous sections, many simplifications occur. For example, the induced MMT
$g$ in the rigid models in secs.~\ref{subsubsec:rigid_flex}
and~\ref{subsubsec:rigid_non-flex} is actually the unconstrained-unconstrained
sub-block of the whole-space MMT restricted to $\mathcal{K}$, $G_\mathcal{K}$.
Therefore, the theorem by Fixman \cite{Fix1974PNAS} applies, and we have the
following relation, which helps to more efficiently compute the difference $T
S_\mathrm{r}^\mathrm{k}(s) - T S_\Sigma^\mathrm{k}(s)$ in $V_\mathrm{F}(s)$,
and which has been used in many previous works
\cite{Alm1990MP,Ber1983Book,Cic1986CPR,dOt1998JCP,dOt2000MP,Fre2002Book,Mor2004ACP,Pas2002JCP}:
\begin{equation}
\label{eq:Fixman_theorem}
\frac{\det g^\prime(s)}{\det G^\prime_\Sigma(s)} \propto
\frac{\det g(u)}{\det G_\mathcal{K}(u)}	= \det h_\mathcal{K}(u) \ ,
\end{equation}
where $h$ denotes the constrained-constrained sub-block of $G^{-1}$:
\begin{equation}
\label{eq:h}
h^{IJ} := \frac{\partial Q^I}{\partial r^\mu} \frac{1}{m_\mu}
          \frac{\partial Q^J}{\partial r^\mu} \ .
\end{equation}
The reason for expressing the above quotient of determinants as a function of
$\det h$ is that the entries of $h$ are in principle very easy to calculate if
$q$ are the typical molecular coordinates consisting of bond lengths, bond
angles, and dihedral angles \cite{Ech2006JCC1}, since each one of the
functions $Q$ depends only on a few Euclidean coordinates $r$. This is however
not true if the modified coordinates $\tilde{q}$ including the $\sigma$
functions are used. Indeed, if we compute the matrix $h$ in the $\tilde{q}$
coordinates and use the transformation rules in eq.~(\ref{eq:q_to_qtilde}), we
find that
\begin{eqnarray}
\label{eq:h_qtilde}
\tilde{h}^{IJ}(\tilde{q}) & := & 
  \frac{\partial \tilde{Q}^I}{\partial r^\mu}\big(R(\tilde{q})\big) 
  \frac{1}{m_\mu}
  \frac{\partial \tilde{Q}^J}{\partial r^\mu}\big(R(\tilde{q})\big) 
  \nonumber \\
 & = & 
  \frac{\partial \tilde{Q}^I}{\partial q^\nu}\big(Q(\tilde{q})\big)
  \frac{\partial Q^\nu}{\partial r^\mu}\big(R(\tilde{q})\big) 
  \frac{1}{m_\mu}
  \frac{\partial Q^\rho}{\partial r^\mu}\big(R(\tilde{q})\big) 
  \frac{\partial \tilde{Q}^J}{\partial q^\rho}\big(Q(\tilde{q})\big) 
  \nonumber \\
 & = & 
  \frac{\partial f^I}{\partial \tilde{u}^r}
  \frac{\partial Q^r}{\partial r^\mu}
  \frac{1}{m_\mu}
  \frac{\partial Q^s}{\partial r^\mu}
  \frac{\partial f^J}{\partial \tilde{u}^s}
- \frac{\partial Q^I}{\partial r^\mu}
  \frac{1}{m_\mu}
  \frac{\partial Q^s}{\partial r^\mu}
  \frac{\partial f^J}{\partial \tilde{u}^s}
  \nonumber \\
 & & \mbox{}
- \frac{\partial f^I}{\partial \tilde{u}^r}
  \frac{\partial Q^r}{\partial r^\mu}
  \frac{1}{m_\mu}
  \frac{\partial Q^J}{\partial r^\mu}
+ \frac{\partial Q^I}{\partial r^\mu}
  \frac{1}{m_\mu}
  \frac{\partial Q^J}{\partial r^\mu}
  \nonumber \\
 & = & 
  \frac{\partial f^I}{\partial \tilde{u}^r}
  h^{rs} \frac{\partial f^J}{\partial \tilde{u}^s}
- h^{Is} \frac{\partial f^J}{\partial \tilde{u}^s}
- \frac{\partial f^I}{\partial \tilde{u}^r} h^{rJ}
+ h^{IJ} \ ,
\end{eqnarray}
where we have extended the definition in eq.~(\ref{eq:h}) to run in the whole
range $\mu,\nu=1,\ldots,N$, and we have dropped the dependencies towards the
end of the calculation to make the expressions lighter. Now, we can conclude
that, even if in the flexible case one can use the modified coordinates
$\tilde{q}$ and the theorem by Fixman in eq.~(\ref{eq:Fixman_theorem}) still
applies, it is not very useful. This is so because, although the terms related
to the untransformed matrix $h$ in the above expression are still easy to
compute, the derivatives $\partial f / \partial \tilde{u}$ are typically
non-trivial, and they must be found numerically (see sec.~\ref{sec:numerical}
and ref.~\cite{Ech2010Sub}). In the hard case, however, we have $f^I(s) =
d^I_0$, and the relation between $\tilde{h}$ and $h$ becomes
\begin{equation}
\label{eq:h_qtilde_hard}
\tilde{h}^{IJ}(\tilde{q}) = h^{IJ}\big(Q(\tilde{q})\big) \ ,
\end{equation}
I.e., in the hard case, eq.~(\ref{eq:Fixman_theorem}) is not only true, but
also useful. In sec.~II of ref.~\cite{Hel1979JCP}, a very simple model
illustrates how the choice of coordinates discussed in this point might have
implications on the whole formalism apart from the Fixman theorem. In sec.~2.4
of ref.~\cite{Cot2004BITNM}, the authors briefly agree with us about the fact
that the use of the coordinates $\tilde{q}$ assumes that they have already
been identified (computed), but this is non-trivial in general.

\item We have worked in the more general flexible setting, which includes the
more restricted (and much more popular
\cite{Aba1989JBMSD,Abe1984CC,Alm1990MP,Edb1986JCP,Dub2010JCP,Maz1989JBMSD,Maz1991JCoP,Maz1997JCC,Nog1983JPSJ,Ber1983Book,Cha1979JCP,Che2005JCC,Cic1986CPR,Fix1974PNAS,Fre2002Book,Go1976MM,Hin1994JFM,Hin1995PRE,Kar1981MM,Li2009PRL,Pas2002JCP,Pat2004JCP,Pea1979JCP,Per1985MM,Ryc1977JCoP,Sch2003MP,vGu1977MP,vGu1980MP,vGu1982MM,vGu1989Book})
hard models as a particular case. Note that, according to the discussion in
sec.~\ref{subsubsec:stiff_flex}, it is clear that this issue is coupled to the
choice of coordinates discussed in the first point of this list. Indeed, if
the modified coordinates $\tilde{q}$ introduced in the previous sections are
used, we have that the Taylor expansion of the potential energy around the
constrained subspace is given by eq.~(\ref{eq:VTaylor_qtilde}). Since the
orders higher than 2 will also be multiplied by the undisplaced constrained
coordinates $\tilde{d}$, we have that, in the coordinates $\tilde{q}$, the
constraints defined through the minimization of the potential energy are
effectively hard, i.e., $\tilde{d}=0$. However, despite this reformulation,
the vast majority of the works in the literature do not refer to this, but
actually to constraints with the form $d=d_0$ being the coordinates $d$
typically bond angles and bond lengths, which is not a reformulation but a
restriction to a more particular case than the general, flexible one discussed
here.

\item The Taylor expansion used to build the stiff model in
sec.~\ref{subsubsec:stiff_flex} has been truncated at the highest possible
order compatible with the possibility of having analytical integrals, thus
producing more accurate corrections, such as the ones including the
derivatives of the determinant of the whole-space MMT $G$. This Taylor
expansion of the term $- T S^\mathrm{k}(q)$ has not be accounted for in any
previous work as far as we are aware (even in those works in which the
treatment is most general
\cite{Ech2006JCC2,Mor2004ACP,dOt1998JCP,dOt2000MP,Pet2000Thesis,Ral1979JFM}),
thus effectively making the most accurate stiff model considered so far the
(2,0) one in secs.~\ref{subsubsec:stiff_flex}
and~\ref{subsubsec:stiff_non-flex}. Also, it is often assumed in the
literature that the zero-order restriction to $\Sigma$, $- T
S_\Sigma^\mathrm{k}(s)$, does not depend on $s$
\cite{Alm1990MP,Go1976MM,Kar1981MM,Li2009PRL,Pat2004JCP,vGu1980MP,vGu1989Book}.
The first approximation is wrong in general both in the flexible and hard
cases [it is sufficient to look at eq.~(\ref{eq:detGprime_SASMIC}), where the
dependence of $\det G^\prime$ on the typically constrained bond lengths and
bond angles is explicit]. However, if hard constraints are chosen, and all
bond lengths and bond angles are constrained, then $\det G^\prime$ is a
constant and it can be canceled out from the expressions
\cite{Alm1990MP,Go1976MM,Li2009PRL,Pat2004JCP,Pea1979JCP,Per1985MM}. On the
other hand, if at least some bond angles are considered unconstrained (as it
is common in the literature
\cite{Ber1983Book,Fre2009BPJ,Ens2007JMB,Eas2010JCTC1,Hes2008JCTC1,vGu1980MP}),
we have that (i) $- T S^\mathrm{k}(q)$ depends on the constrained coordinates,
thus suggesting the necessity of performing the Taylor expansion in
eq.~(\ref{eq:SkTaylor}), and (ii) even the zero-order restriction to $\Sigma$,
$- T S_\Sigma^\mathrm{k}(s)$, \emph{does} depend on $s$, \emph{also} in the
hard case.

\item The Taylor expansion mentioned in the previous point has also been
performed here on the potential energy, thus producing correcting terms which
include the Hessian matrix. In fact, since the best stiff model that appears
in the previous literature is, as we mentioned, the (2,0) one, the matrix $A$
in secs.~\ref{subsubsec:stiff_flex} and~\ref{subsubsec:stiff_non-flex} is
never found, only the Hessian (at most). This correction related to the
Hessian has not been considered in a number of works, thus reducing the stiff
model used to the (0,0)-type
\cite{Alm1990MP,Cha1979JCP,Cic1986CPR,Fix1974PNAS,Fre2002Book,Hin1994JFM,Kar1981MM,Sch2003MP,vGu1980MP,vGu1989Book}
(sometimes legitimately so, because of the restricted form of the potential
energy used \cite{Pea1979JCP,Pas2002JCP,Pat2004JCP,Per1985MM}), and it has
been included (or at least mentioned) by some others
\cite{Ber1983Book,Cot2004BITNM,dOt1998JCP,dOt2000MP,Go1969JCP,Go1976MM,Hes2002JCP,Li2009PRL,Mor2004ACP,Pet2000Thesis,Ral1979JFM,Rei1996PRE}.
However, even in the second family of papers, it is often assumed that the
Hessian does not depend on the unconstrained coordinates $s$
\cite{Ber1983Book,Go1976MM,Li2009PRL,Rei1996PRE}, which leads to effectively
discard it. It is also worth pointing out that this last approximation is
sometimes justified by the fact that the force field-like potential energy
functions [such as the one in eq.~(\ref{eq:Vff})] typically have
conformation-independent spring constants \cite{Ber1983Book,Go1976MM}.
However, even if we accept to particularize the calculations to classical
force fields, the spring constants can only be assimilated to the Hessian in
the infinite stiffness limit, which is not clear to be reached in force
fields, specially for bond angles. Finally, we can think that the assumption
that the zero-point energy of the quantum harmonic oscillators associated to
the constrained coordinates is $s$-independent in quantum models is somehow
related to this point \cite{Go1969JCP,Go1976MM,Hes2002JCP}.

\item The potential energy used to model the system and to define the
constrained subspace in the flexible case has been considered to be completely
general. It is common in the literature to explicitly or implicitly assume
that a force field-like function [such as the one in eq.~(\ref{eq:Vff})] is
used
\cite{Rei1996PRE,Ber1983Book,Cot2004BITNM,dOt1998JCP,dOt2000MP,Fix1974PNAS,Go1969JCP,Go1976MM,Lei1995Book,Mor2004ACP,Ral1979JFM},
or even simpler ones \cite{Pea1979JCP,Pas2002JCP,Pat2004JCP,Per1985MM}, which
leads to more restricted versions of the objects computed here. For example,
in ref.~\cite{Pea1979JCP}, (and similarly in \cite{Per1985MM}) a polymer is
considered whose potential energy is just the harmonic terms in
eq.~(\ref{eq:Vff}), i.e.,
\begin{equation}
\label{eq:V_Pearl}
V(w) := \sum_I \frac{1}{2} K_I (d^I - d^I_0)^2 \ .
\end{equation}
In such a case, the flexible constraints, as defined in this work, become
$d=d_0$, i.e., \emph{exactly} hard; with the associated simplifications
discussed in the previous points. In the celebrated paper by Fixman
\cite{Fix1974PNAS} a similar choice is made. Also, as we mentioned in
sec.~\ref{subsubsec:stiff_flex}, the choice of coordinates discussed in the
first point of this list is related to the issue of the form potential energy.
The Taylor expansion of the potential energy around the constrained subspace
in the modified coordinates $\tilde{q}$ is given by
eq.~(\ref{eq:VTaylor_qtilde}), which is in agreement with the potential used
in some previous works
\cite{Cot2004BITNM,dOt1998JCP,dOt2000MP,Go1969JCP,Lei1995Book,Mor2004ACP,Ral1979JFM},
implying that maybe the authors were not considering a restriction to a force
field-like function, but simply using the coordinates $\tilde{q}$. Remember
however what we discussed about these coordinates in the first point, and also
note that in some of these studies it is typically assumed that the Hessian is
independent from the conformation \cite{Lei1995Book,Rei1996PRE}, which is not
a general feature of eq.~(\ref{eq:VTaylor_qtilde}). In sec. II of
ref.~\cite{Hel1979JCP}, the relationship between the coordinates chosen and
the form of the potential energy is clearly illustrated using a simple
2-dimensional system, and see also
sec.~\ref{subsubsec:flexible_vs_nonflexible} for more implications about this
point in relation to flexible constraints.

\end{itemize}

In summary, in this work, all quantities appearing in the expressions: the MMT
and Hessian determinants, the functions $f$, the derivatives of the potential
energy, etc., have been considered to change with every variable that they can
in principle depend on. No `constancy' assumption has been made in our
development. They have been expanded to the highest possible order compatible
with analytical results, and no special form of the potential energy has been
assumed.

The first issue in the above list affects the form of the final expressions
and therefore complicates the `translation' between our results and those in
other works, but it does not introduce additional approximations. The rest of
the assumptions (that we did not make here) effectively restrict the
generality of the treatment, therefore potentially introducing modeling
inaccuracies. Both types of issues typically produce simpler expressions and
algorithms than the ones discussed here, being probably this practical benefit
the original reason behind them. The most radical example of this is the case
of a freely jointed chain with constrained bond lengths and bond angles, and
no other potential energy than the harmonic oscillators in charge of enforcing
these constraints
\cite{Pas2002JCP,Pat2004JCP,Pea1979JCP,Per1985MM,vGu1980MP,vGu1989Book}. In
such a situation, all terms in $V_\mathrm{F}(s)$ become independent from the
conformation $s$ (the torsional angles in this case), except for
$S^\mathrm{k}_\mathrm{r}(s)$; and therefore all the difference between the
rigid and stiff models can be attached to $\det g$, which, in addition, can be
very rapidly computed in terms of banded matrices
\cite{Pas2002JCP,Pat2004JCP}.

Finally, it is worth remarking that, apart from the works that assume some of
the approximations discussed above, the vast majority of constrained MD
simulations in the literature do not consider any correcting term to
$V_\Sigma$ \emph{at all}; sometimes giving analytical or numerical reasons for
it, sometimes simply not mentioning the issue
\cite{Aba1989JBMSD,Edb1986JCP,Fee1999JCC,Dub2010JCP,Maz1989JBMSD,Maz1991JCoP,Maz1997JCC,Che2005JCC,Hes2002JCP,Hin1995PRE,Tox2009JCP,vGu1977MP,vGu1982MM,Zho2000JCP}.

\section{Numerical examples}
\label{sec:numerical}

In this section, we provide a numerical example of the application of the
general formalism introduced in this work to a real molecular system, using a
classical force field.

It is worth mentioning at this point that many numerical studies about the
equilibrium of constrained systems exist in the literature, each one of them
containing some of the approximations which we carefully enumerated in the
previous section, and which are avoided in this work.

Regarding the articles in which the stiff and rigid equilibrium distributions
are compared, we can mention ref.~\cite{Alm1990MP}, where a simplified model
of $N$-butane is studied and the Fixman potential is found to be negligible;
they use hard constraints, and they neglect both the Hessian and the
$s$-dependence of $\det G$, effectively using the (0,0)-stiff model introduced
in the previous sections. In ref.~\cite{Cha1979JCP}, the same molecule is
studied, under the same approximating assumptions; they find that the
probability of the trans-gauche transition state is changed in a 20-30\% by
the Fixman correction. $N$-butane is also analyzed in ref.~\cite{Per1985MM},
where they do not see an appreciable effect in transition rates or relaxation
times if the Fixman potential is included; they assume again that the Hessian
and the $s$-dependence of $\det G$ are negligible, and they impose hard
constraints. Under the same assumptions, in ref.~\cite{vGu1980MP}, the
equilibrium of $N$-butane is shown to be altered by the inclusion of the
Fixman correction if bond angles are constrained, but not if only bond lengths
are (see below for more works discussing the bond angles issue). In
ref.~\cite{Hel1979JCP}, the equilibrium distribution of certain angles in
simplified models of three and four beads is found to be different between the
stiff and rigid models. The same four-beads model is simulated in
ref.~\cite{Pea1979JCP}, where the same comparison is performed and the
rigid-plus-Fixman simulation is shown to be both equivalent to the
unconstrained one, and different from the rigid simulation without correcting
terms. In both ref.~\cite{Hel1979JCP} and ref.~\cite{Pea1979JCP}, the
simplicity of the model and the potential energy used to describe its
behaviour make the hard and (0,0) assumptions exact. In
ref.~\cite{Pas2002JCP}, a linear freely jointed chain is considered, and the
stiff torsional angles distribution is shown to be recovered from rigid
simulations if the Fixman potential is included, which they find to have a
non-negligible effect; they assume again the approximations that produce the
(0,0)-stiff model, and they use hard constraints. The same system is studied
in ref.~\cite{Pat2004JCP}, where Fixman's potential effects are measured as a
function of the chain length, finding it non-negligible; again, the Hessian
and the dependence of $\det G$ on $s$ are not considered.

A different but related family of works have numerically compared flexible and
hard constraints. Among them, we can mention ref.~\cite{Chr2007CPC}, were
they measure appreciable discrepancies between the two types of constraints in
the calculation of the free energy difference between liquid water and
methanol. In ref.~\cite{Hes2002JCP}, the unconstrained dynamics is taken as
the reference to which the flexible and hard rigid models are compared in the
computation of the velocity autocorrelation function of liquid water; the
authors find that the predictions of the unconstrained and flexible rigid
models are very similar, and different from those of the hard rigid case. In
ref.~\cite{Zho2000JCP} the same conclusion is drawn. In
ref.~\cite{Lei1995Book}, a simple four-beads system is studied and it is shown
that the unconstrained average variation in bond angles is recovered if
flexible constraints are used. A similar study in $N$-butane can be found in
ref.~\cite{Rei1996PRE}, where the unconstrained distribution of the central
torsion angle is shown to be well reproduced by the flexible rigid (plus
Fixman) simulation, but not by the hard one. Finally, in
ref.~\cite{Rei1995PD}, flexible constraints are proved to better reproduce the
low-frequency part of the vibrational spectrum than hard ones in a simple toy
model. The reader should notice that only in ref.~\cite{Rei1996PRE} the
Fixman potential is used to correct the rigid dynamics, which makes the
agreement between the flexible rigid simulations and the unconstrained ones
in the rest of works somehow surprising, probably suggesting that the
Fixman correction is not significantly important in the particular systems
that have been explored.

It is also worth to mention at this point the general agreement in the
literature about bond angle constraints changing too much the equilibrium
distribution with respect to the unconstrained case. Many works have studied
this issue and have recommended not to constrain bond angle coordinates on the
basis of their numerical findings
\cite{Ber1983Book,Che2005JCC,Hin1995PRE,Kar1981MM,vGu1977MP,vGu1982MM}.
However, all the simulations in these works have been performed (i) using hard
constraints, and (ii) without including the Fixman correction. We have showed
in this work that both these two assumptions should be considered
approximations that might potentially compromise the accuracy of the final
results. Therefore, we agree with the author of ref.~\cite{Rei1996PRE} in
suggesting the possibility that, perhaps, constraining bond angles can be made
accurately if flexible constraints are used and the most general form of the
Fixman correction introduced in this work is used to produce stiff averages
(the numerical cost of such an approach is very relevant too, but it is
independent from the accuracy issue). The checking of this hypothesis in
biologically relevant molecules is currently in progress and it shall be
reported elsewhere.

\begin{figure}
\begin{center}
\includegraphics[scale=0.25]{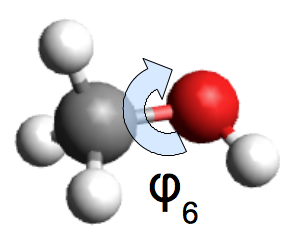}
\caption{\label{fig:methanol} The methanol molecule studied in this section.
The only unconstrained coordinate, the dihedral angle $\varphi_6$, is
indicated with a blue arrow.}
\end{center}
\end{figure}

In the example calculation presented in this section, we have deliberately
chosen a simple system in order to illustrate the theoretical concepts
introduced in the previous sections. We study here the methanol molecule
schematically depicted in fig.~\ref{fig:methanol}, and we compute its
potential energy using the AMBER 96 force field
\cite{Cor1995JACS,Kol1997Book}. The numeration of the atoms and the definition
of the internal coordinates follow the SASMIC scheme, which is specially
adapted to deal with constrained molecular systems \cite{Ech2006JCC1}. All the
internal coordinates have been constrained, except for the principal dihedral
angle, $\varphi_6$, which then becomes the only variable parameterizing the
internal constrained subspace $\Sigma$. In order to produce the conformations
in which all the quantities introduced in this work have been measured, we
have scanned this dihedral angle from $0^\mathrm{o}$ to $180^\mathrm{o}$, in
steps of $10^\mathrm{o}$. At each point, we have generated the hard
conformations simply by setting all the constrained coordinates to the
constant values appearing in the force field, and the flexible conformations
have been produced by minimizing the potential energy with respect to the
constrained coordinates at fixed $\varphi_6$. These minimizations in internal
coordinates, as well as the computation of the potential energy Hessian, have
been performed with Gaussian 03 \cite{Gaussian03E01}. All correcting terms to
the different statistical mechanics models discussed in the previous section
have been computed using home-made programs.

\begin{figure}
\begin{center}
\includegraphics[scale=0.40]{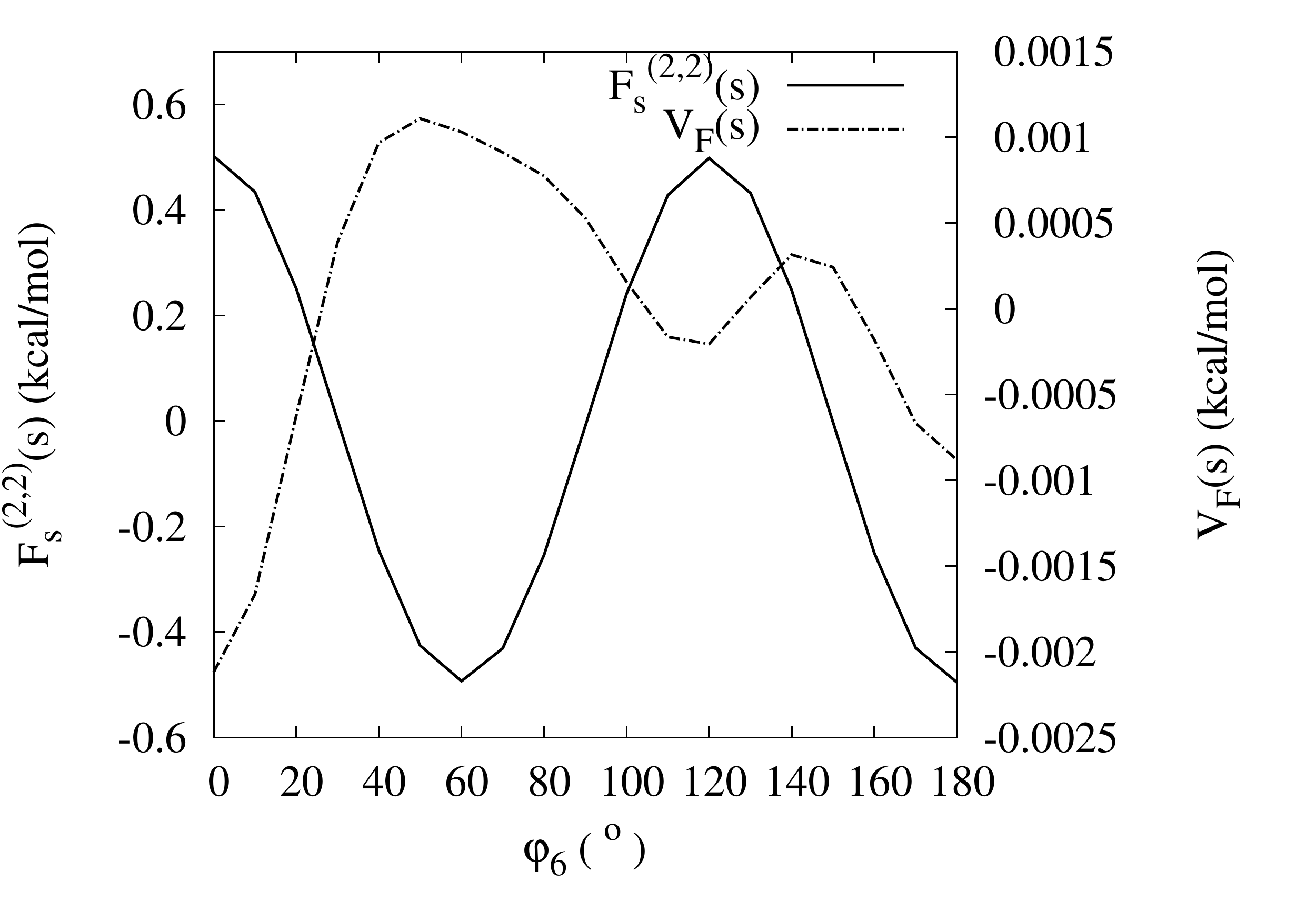}
\caption{\label{fig:fixman} Free energy, $F_\mathrm{s}^{(2,2)}(s)$, of the
flexible (2,2)-stiff model, and Fixman's potential, $V_\mathrm{F}(s)$, as a
function of the dihedral angle $\varphi_6$ in methanol. Both quantities have
been added an irrelevant energy reference in order for them to have zero
average.}
\end{center}
\end{figure}

\begin{table}[!ht]
\begin{center}
\begin{tabular}{c@{\hspace{30pt}}rrrr}
\hline\\[-8pt]
 & \multicolumn{2}{c}{Flexible} & 
   \multicolumn{2}{c}{Hard} \\[3pt]
  & Std.$^{a}$ & Max.$^{b}$ &
    Std.$^{a}$ & Max.$^{b}$ \\[3pt]
\hline\\[-8pt]
$V_\Sigma$                    & 0.37100819 & 1.02289469 &
                                0.37164847 & 1.02456429 \\
$-TS_\mathrm{r}^\mathrm{k}$   & 0.01116055 & 0.03295965 &
                                0.00000028 & 0.00000098 \\
$-TS_\Sigma^\mathrm{k}$       & 0.00008127 & 0.00022596 &
                                0.00000000 & 0.00000000 \\
$-TS_\mathcal{H}^\mathrm{c}$  & 0.01079596 & 0.03198016 &
                                0.01071757 & 0.03179323 \\
$-TS_A^\mathrm{c}$            & 0.01078666 & 0.03146866 &
                                0.01070879 & 0.03176306 \\
$U_\mathrm{s}^{(0)}$          & N/A & N/A &
                                0.00358154 & 0.01322298 \\
$U_\mathrm{s}^{(1)}$          & 0.00000004 & 0.00000013 &
                                0.00358563 & 0.01325454 \\
$U_\mathrm{s}^{(2)}$          & 0.00000004 & 0.00000013 &
                                0.00357814 & 0.01322534 \\[3pt]
\hline
\end{tabular}
\end{center}
\caption{\label{tab:sizes} $^{a}$Standard deviation, and $^{b}$maximum
variation in the conformational space of the methanol molecule of the
different terms introduced in sec.~\ref{subsec:SM_models} which affect the
equilibrium free energy of constrained models. All values are in kcal/mol.}
\end{table}

First of all, we measured the most accurate version of the flexible stiff free
energy in sec.~\ref{subsubsec:stiff_flex}, the quantity
$F_\mathrm{s}^{(2,2)}(s)$, as a function of $\varphi_6$, as well as the Fixman
potential, $V_\mathrm{F}(s)$, in eq.~(\ref{eq:VF}), which quantifies the
difference between the stiff equilibrium and the rigid one. The results,
depicted in fig.~\ref{fig:fixman}, indicate that the variations in
$V_\mathrm{F}(s)$ are at least two orders of magnitude smaller than those in
$F_\mathrm{s}^{(2,2)}(s)$, and indeed much smaller than the thermal noise $RT
\simeq 0.6$ kcal/mol at room temperature, allowing us to conclude that, in
this case, it is safe to neglect the Fixman correction. However, if we take a
look at the variation of the different terms involved in the definition of
$V_\mathrm{F}(s)$ in tab.~\ref{tab:sizes}, we see that
$-TS_\mathrm{r}^\mathrm{k}(s)$ and $-TS_A^\mathrm{c}(s)$ show a variation
which is one order of magnitude larger than that of $V_\mathrm{F}(s)$. What is
happening is that these two quantities are much correlated in methanol and
have opposite signs, thus almost canceling out. Preliminary data in larger
systems indicates that this feature is only present in small molecules such as
the one studied here. Also, the significance of $V_\mathrm{F}(s)$ appears to
grow with molecular size. Therefore, the particular results in methanol should
not be extrapolated in face value, but rather seen as an illustration of the
concepts involved (probably the same can be said about the works discussed at
the beginning of this section which analyze small toy systems and $N$-butane).

\begin{figure}
\begin{center}
\includegraphics[scale=0.26]{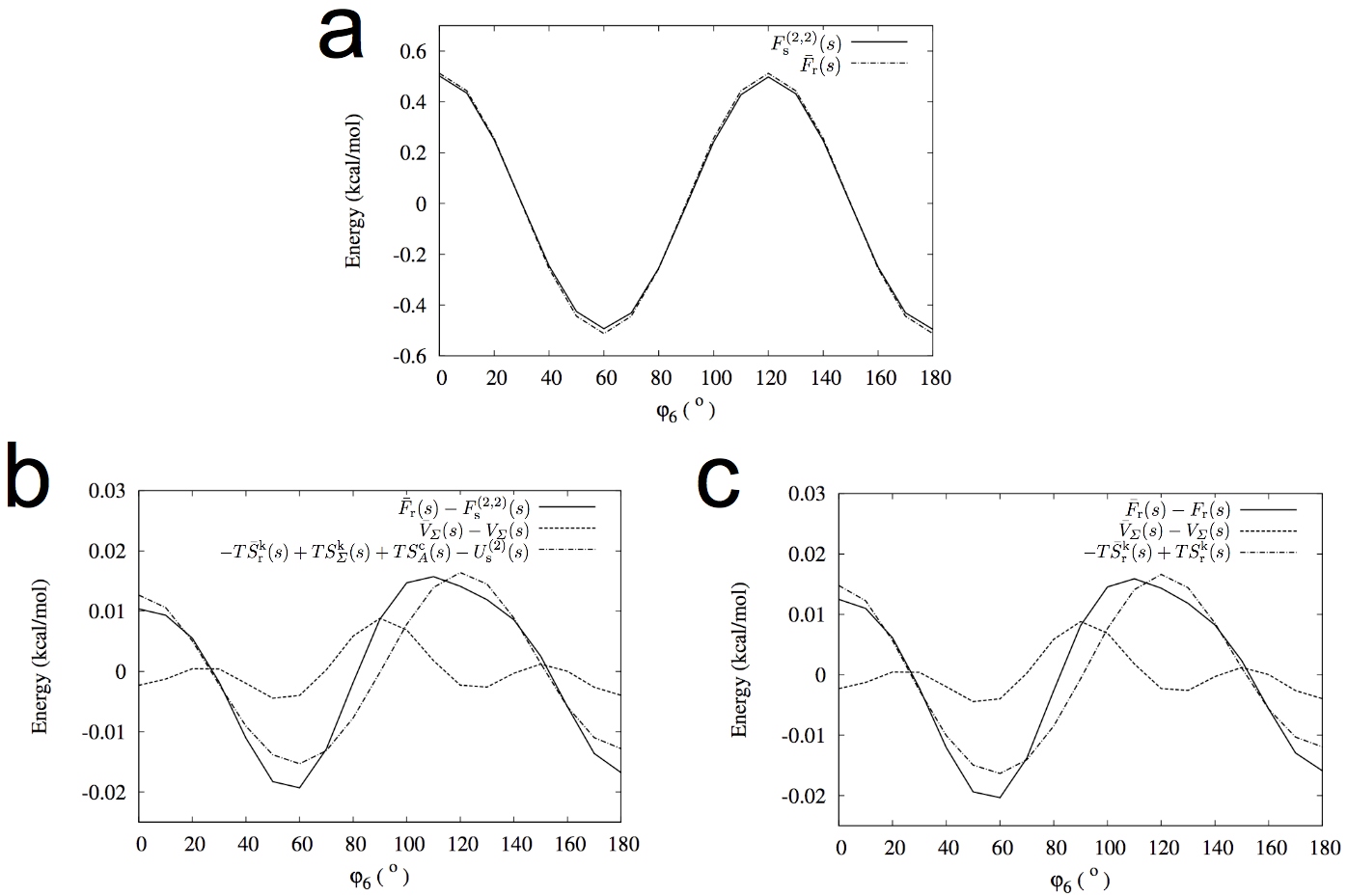}
\caption{\label{fig:flex_vs_hard} Assessment of the hard rigid model for the
methanol molecule. \textbf{(a)} Comparison between the flexible (2,2)-stiff
free energy, $F_\mathrm{s}^{(2,2)}(s)$, and the hard rigid one,
$\bar{F}_\mathrm{r}(s)$ as a function of the dihedral angle $\varphi_6$.
\textbf{(b)} Difference between the two free energies in (a), as well as the
difference between the potential energies, $\bar{V}_\Sigma(s) - V_\Sigma(s)$,
and the correcting terms involved in the two models. \textbf{(b)} Difference
between the flexible rigid free energy, $F_\mathrm{r}(s)$, and the hard rigid
one, $\bar{F}_\mathrm{r}(s)$. Again the difference between the potential
energies, $\bar{V}_\Sigma(s) - V_\Sigma(s)$, and the difference between the
respective correcting terms are also depicted.}
\end{center}
\end{figure}

The difference between flexible and hard constraints in this molecule is
larger than the difference between the stiff and rigid models, as we can
appreciate in fig.~\ref{fig:flex_vs_hard}, where the hard rigid model is
compared both to the flexible (2,2)-stiff one (fig.~\ref{fig:flex_vs_hard}b)
and to the flexible rigid one (fig.~\ref{fig:flex_vs_hard}c). In both cases,
the maximum variation of the difference in the conformational space of
methanol is approximately $0.04$ kcal/mol, an order of magnitude larger than
the variation of $V_\mathrm{F}(s)$ in fig.~\ref{fig:fixman}. We can also see
that both the difference between the flexible and hard potential energies,
$\bar{V}_\Sigma(s) - V_\Sigma(s)$, and the difference between the correcting
terms in the respective models contribute to the total discrepancy between the
flexible and hard cases, being the second contribution somewhat larger than
the first one. The fact that the graphs in figs.~\ref{fig:flex_vs_hard}b
and~\ref{fig:flex_vs_hard}c seem identical to the eye (they are not exactly
the same numerically) is again caused by the already mentioned correlation
between the terms $-TS_\mathrm{r}^\mathrm{k}(s)$ and $-TS_A^\mathrm{c}(s)$ in
methanol. Preliminary data in larger molecules suggest that these two
comparisons are different in general, and also that the difference
$\bar{V}_\Sigma(s) - V_\Sigma(s)$ becomes increasingly more important with
system size if the bond angles remain among the constrained coordinates. This
is probably due to the possibility of more steric clashes in longer, more
flexible chains, and it is in agreement with the works discussing the bond
angles issue that we have summarized at the beginning of this section.

\begin{figure}
\begin{center}
\includegraphics[scale=0.27]{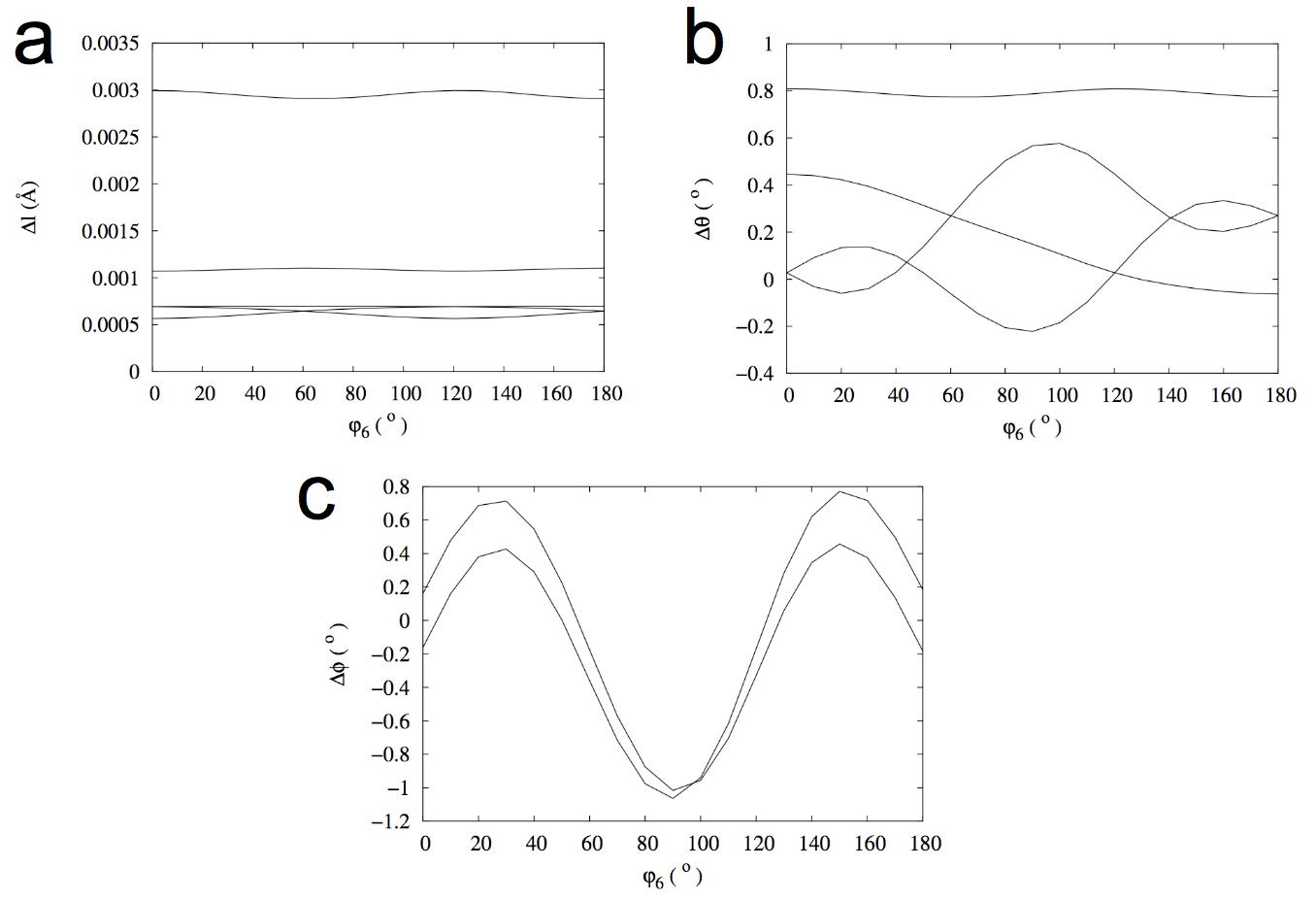}
\caption{\label{fig:constrained_coords_variation} Variation of the constrained
coordinates in the flexible case in methanol, as a function of the
unconstrained dihedral angle $\varphi_6$. \textbf{(a)} Variation of the bond
lengths, \textbf{(b)} bond angles, and \textbf{(c)} phase dihedral angles with
respect to the constant values appearing in the force field and defining the
hard models.}
\end{center}
\end{figure}

Finally, we can see in fig.~\ref{fig:constrained_coords_variation} that the
constrained coordinates depend on the conformation in the more general
flexible case, as expected from the discussion in
sec.~\ref{subsubsec:flexible_vs_nonflexible}.

\section{Conclusions and future lines of research}
\label{sec:conclusions}

In this review, we have attempted to provide a unifying view of the
statistical equilibrium of constrained classical mechanics models of molecular
systems. To this end, we have introduced the most general formalism (i.e.,
including the fewest approximations) compatible with the possibility of having
analytical results. From this advantageous standpoint, we have been able to
rationalize most of the previous works in the literature, clearly identifying
the underlying (often implicit) assumptions behind the different analyses.
Also, we have tried to provide the reader with a coherent vocabulary which we
hope facilitates future comparisons and reviews. Finally, we have shown a
practical example of the different theoretical concepts.

Along the text, we have suggested a number of possible future lines of
research that we believe could be interesting for distinct reasons, and some
of which we have already started to pursue. We conclude this account by
briefly collecting and commenting them:

\begin{itemize}

\item Define the constrained coordinates in an \emph{adaptive} way, instead
of simply picking whole sets of internal molecular coordinates of the same
\emph{type} (e.g., all bond angles involving hydrogens). If properly done,
such an scheme could not only help to make calculations more accurate, but 
also to assess how good are the simpler approaches followed at the moment in
the literature.

\item Develop a rigorous and general quantum mechanical derivation of
constrained molecular models that could help decide which one of the classical
models discussed in this work is more appropriately justified on physical
grounds (if any of them).

\item Compute the equilibrium PDF of the non-Hamiltonian, flexibly constrained
dynamics described in ref.~\cite{Hes2002JCP}, maybe using the technique in
ref.~\cite{Tuc2001JCP}.

\item Compare the option chosen in this work for defining the flexible
constraints based on the minimization of the potential energy $V(w)$ to the
alternative discussed in sec.~\ref{subsubsec:flexible_vs_nonflexible}, which
uses the free energy $F(q)$ that appears in the exponent of the marginal
PDF of the unconstrained dynamics in the space of the positions.

\item Use techniques for the computation of free energy differences in order
to numerically calculate the marginal PDF of the unconstrained dynamics in the
space of the unconstrained coordinates $u$ [see eq.~(\ref{eq:unconstrPs})],
and compare it to the approximation to it provided by the stiff model.

\item Solve the discrepancy between the correction to the rigid dynamics found
from statistical analyses like the one in this work (i.e., the Fixman
potential), and those obtained in some studies based on dynamical
considerations, such as ref.~\cite{Bor1995TR2}. See also
sec.~\ref{subsec:comparisons} for further details.

\item Extend the analysis in sec.~\ref{sec:numerical} to larger, more relevant
biological molecules.

\item Explore the hypothesis, mentioned in sec.~\ref{sec:numerical}, that
flexibly constraining bond angles and including the most general form of the
Fixman potential to correct the rigid equilibrium may result in more accurate
constrained simulations than the ones reported so far in the literature, where
constraining bond angles is typically not recommended.

\end{itemize}

\begin{acknowledgement}
We thank J. L. Alonso, I. Calvo, G. Ciccotti, F. Falceto and B. Hess for
illuminating discussions and advices regarding the mathematical formalism and
the scope of the discussion presented here.

This work has been supported by the research projects E24/3 (DGA, Spain),
FIS2009-13364-C02-01 (MICINN, Spain), 200980I064 (CSIC, Spain) and ARAID and
Ibercaja grant for young researchers (Spain). P. G.-R. is supported by a JAE
PREDOC grant (CSIC, Spain).
\end{acknowledgement}


\begin{thebibliography}{154}

\bibitem{And2009JCTC}
X.~Andrade, A.~Castro, D.~Zueco, J.L. Alonso, P.~Echenique, F.~Falceto,
  A.~Rubio, J. Chem. Theory Comput. \textbf{5}, 728 (2009)

\bibitem{Rap2004Book}
D.C. Rapaport, \emph{The art of molecular dynamics simulation}, 2nd~edn.
  (Cambridge University Press, 2004)

\bibitem{All2005Book}
M.P. Allen, D.J. Tildesley, \emph{Computer simulation of liquids} (Clarendon
  Press, Oxford, 2005)

\bibitem{Hun2011EPJSTunp}
T.~Hundertmark, S.~Reich, Eur.\ Phys.\ J. ST  (2011), this issue.

\bibitem{Fre2002Book}
D.~Frenkel, B.~Smit, \emph{Understanding molecular simulations: From algorithms
  to applications}, 2nd~edn. (Academic Press, Orlando FL, 2002)

\bibitem{Ech2007CoP}
P.~Echenique, Contemp. Phys. \textbf{48}, 81 (2007)

\bibitem{Ens2007JMB}
D.L. Ensign, P.M. Kasson, V.S. Pande, J. Mol. Biol. \textbf{374}, 806 (2007)

\bibitem{Fre2009BPJ}
P.L. Freddolino, K.~Schulten, Biophys. J. \textbf{97}, 2338 (2009)

\bibitem{Sha2009Proc}
D.E. Shaw, R.O. Dror, J.K. Salmon, J.P. Grossmann, K.M. Mackenzie, J.A. Bank,
  C.~Young, M.M. Deneroff, B.~Batson, K.J. Bowers et~al.,
  \emph{Millisecond-scale molecular dynamics simulations on {A}nton}, in
  \emph{Proceedings of the {ACM/IEEE Conference on Supercomputing (SC09)}}
  (Portland, Oregon, 2009), pp. 14--20

\bibitem{Ech2008JCC}
P.~Echenique, J.L. Alonso, J. Comput. Chem. \textbf{29}, 1408 (2008)

\bibitem{Kle2009COSB}
J.L. Klepeis, K.~Lindorff-Larsen, R.O. Dror, D.E. Shaw, Curr. Opin. Struct.
  Biol. \textbf{19}, 1 (2009)

\bibitem{Mar2000TR}
D.~Marx, J.~Hutter, \emph{Ab initio molecular dynamics: {T}heory and
  implementation}, in \emph{Modern Methods and Algorithms of Quantum
  Chemistry}, edited by J.~Grotendorst ({John von Neumann Institute for
  Computing}, J{\"u}lich, 2000), Vol.~3, pp. 329--477

\bibitem{Alo2008PRL}
J.L. Alonso, X.~Andrade, P.~Echenique, F.~Falceto, D.~Prada-Gracia, A.~Rubio,
  Phys. Rev. Lett. \textbf{101}, 096403 (2008)

\bibitem{Nik2006PRL}
A.M.N. Niklasson, C.J. Tymczak, M.~Challacombe, Phys. Rev. Lett. \textbf{97},
  123001 (2006)

\bibitem{Ani2009CPC}
V.M. Anisimov, V.L. Bugaenko, C.N. Cavasotto, ChemPhysChem \textbf{10}, 3194
  (2009)

\bibitem{Bro2009JCC}
B.R. Brooks, C.L. Brooks~III, A.D. MacKerell, L.~Nilsson, R.J. Petrella,
  B.~Roux, Y.~Won, G.~Archontis, C.~Bartels, S.~Boresch et~al., J. Comput.
  Chem. \textbf{30}, 1545 (2009)

\bibitem{Jor1988JACS}
W.L. Jorgensen, J.~Tirado-Rives, J. Am. Chem. Soc. \textbf{110}, 1657 (1988)

\bibitem{Jor1996JACS}
W.L. Jorgensen, D.S. Maxwell, J.~Tirado-Rives, J. Am. Chem. Soc. \textbf{118},
  11225 (1996)

\bibitem{Pon2003APC}
J.W. Ponder, D.A. Case, Adv. Prot. Chem. \textbf{66}, 27 (2003)

\bibitem{AMBER10}
D.A. Case, T.A. Darden, T.E. Cheatham~III, C.L. Simmerling, J.~Wang, R.E. Duke,
  R.~Luo, M.~Crowley, R.C. Walker, W.~Zhang et~al., \emph{Amber 10}, University
  of California, San Francisco (2008)

\bibitem{Pea1995CoPC}
D.A. Pearlman, D.A. Case, J.W. Caldwell, W.R. Ross, T.E. Cheatham~III,
  S.~DeBolt, D.~Ferguson, G.~Seibel, P.~Kollman, Comp. Phys. Commun.
  \textbf{91}, 1 (1995)

\bibitem{Jen1998Book}
F.~Jensen, \emph{Introduction to Computational Chemistry} (John Wiley \& Sons,
  Chichester, 1998)

\bibitem{Ech2007MP}
P.~Echenique, J.L. Alonso, Mol. Phys. \textbf{105}, 3057 (2007)

\bibitem{Liw2010JCTC}
A.~Liwo, S.~Oldziej, C.~Czaplewski, D.S. Kleinerman, P.~Blood, H.A. Scheraga,
  J. Chem. Theory Comput. p. Articles ASAP (2010)

\bibitem{Cza2009JCTC}
C.~Czaplewski, S.~Kalinowski, A.~Liwo, H.A. Scheraga, J. Chem. Theory Comput.
  \textbf{5}, 627 (2009)

\bibitem{Emp2008BPJ}
A.~Emperador, O.~Carrillo, M.~Rueda, M.~Orozco, Biophys. J. \textbf{95}, 2127
  (2008)

\bibitem{Han2008JCTC}
W.~Han, C.K. Wan, Y.D. Wu, J. Chem. Theory Comput. \textbf{4}, 1891 (2008)

\bibitem{dlT2011EPJSTunp}
J.A. de~la Torre, G.~Ciccotti, P.~Espa\~{n}ol, M.~Ferrario, Eur.\ Phys.\ J. ST
  (2011), this issue.

\bibitem{Car1985PRL}
R.~Car, M.~Parrinello, Phys. Rev. Lett. \textbf{55}, 2471 (1985)

\bibitem{Hut2005CPC}
J.~Hutter, A.~Curioni, ChemPhysChem \textbf{6}, 1788 (2005)

\bibitem{Car1989CPL}
E.A. Carter, G.~Ciccotti, J.T. Hynes, R.~Kapral, Chem. Phys. Lett. \textbf{5},
  472 (1989)

\bibitem{Har2011EPJSTunp}
C.~Hartmann, G.~Ciccotti, Eur.\ Phys.\ J. ST  (2011), this issue.

\bibitem{Sch2011EPJSTunp}
J.~Schlitter, Eur.\ Phys.\ J. ST  (2011), this issue.

\bibitem{Dil1999PS}
K.A. Dill, Prot. Sci. \textbf{8}, 1166 (1999)

\bibitem{Dill1997NSB}
K.A. Dill, H.S. Chan, Nat. Struct. Biol. \textbf{4}, 10 (1997)

\bibitem{Lev1969Proc}
C.~Levinthal, \emph{How to fold graciously}, in \emph{Mossbauer Spectroscopy in
  Biological Systems}, edited by J.T.P. DeBrunner, E.~Munck (University of
  Illinois Press, Allerton House, Monticello, Illinois, 1969), pp. 22--24

\bibitem{Dag2003NRMCB}
V.~Daggett, A.~Fersht, Nat. Rev. Mol. Cell Biol. \textbf{4}, 497 (2003)

\bibitem{Kar2002NSB}
M.~Karplus, J.A. McCammon, Nat. Struct. Biol. \textbf{9}, 646 (2002)

\bibitem{Sch1997ARBPBMS}
T.~Schlick, E.~Barth, M.~Mandziuk, Annu. Rev. Biophys. Biomol. Struct.
  \textbf{26}, 181 (1997)

\bibitem{Bor1995TR2}
F.A. Bornemann, C.~Sch{\"u}tte, Tech. rep., Konrad-Zuse-Zentrum f{\"u}r
  Informationstechnik Berlin (1995)

\bibitem{Lei1995Book}
B.J. Leimkuhler, S.~Reich, R.D. Skeel, in \emph{Mathematical approaches to
  biomolecular structure and dynamics}, edited by J.P. Mesirov, K.~Schulten
  (Springer, 1996)

\bibitem{Cha1985SIAMJNA}
M.M. Chawla, SIAM J. Numer. Anal. \textbf{22}, 127 (1985)

\bibitem{Fee1999JCC}
K.A. Feenstra, B.~Hess, H.J.C. Berendsen, J. Comput. Chem. \textbf{20}, 786
  (1999)

\bibitem{Gol2002Book}
H.~Goldstein, C.~Poole, J.~Safko, \emph{Classical Mechanics}, 3rd~edn.
  (Addison-Wesley, 2002)

\bibitem{Ech2006JCC1}
P.~Echenique, J.L. Alonso, J. Comput. Chem. \textbf{27}, 1076 (2006)

\bibitem{Dub1992Book}
B.A. Dubrovin, A.T. Fomenko, S.P. Novikov, \emph{Modern Geometry --- Methods
  and Applications} (Springer, Berlin, 1992)

\bibitem{Wei2009Web1}
E.W. Weisstein, \emph{Ordinary differential equations}, from MathWorld -- A
  Wolfram Web Resource.
  \url{http://mathworld.wolfram.com/OrdinaryDifferentialEquations.html} (last
  accessed on 03/17/09)

\bibitem{Ech2006JCC2}
P.~Echenique, I.~Calvo, J.L. Alonso, J. Comput. Chem. \textbf{27}, 1748 (2006)

\bibitem{Hes2002JCP}
B.~Hess, H.~Saint-Martin, H.J.C. Berendsen, J. Chem. Phys. \textbf{116}, 9602
  (2002)

\bibitem{Zho2000JCP}
J.~Zhou, S.~Reich, B.R. Brooks, J. Chem. Phys. \textbf{112}, 7919 (2000)

\bibitem{Jos1998Book}
J.V. Jos{\'e}, E.J. Saletan, \emph{Classical dynamics: a contemporary approach}
  (Cambridge University Press, 1998)

\bibitem{Bal1975Book}
R.~Balescu, \emph{Equilibrium and nonequilibrium statistical mechanics} (John
  Wiley \& Sons, New York, 1975)

\bibitem{Pet2010Misc}
K.B. Petersen, M.S. Pedersen, \emph{The matrix cookbook}, Downoladed from
  \url{http://matrixcookbook.com} (2010)

\bibitem{Go1969JCP}
N.~G{\={o}}, H.A. Scheraga, J. Chem. Phys. \textbf{51}, 4751 (1969)

\bibitem{Ech2006JCC3}
P.~Echenique, I.~Calvo, J. Comput. Chem. \textbf{27}, 1733 (2006)

\bibitem{Eas2010JCTC1}
P.~Eastman, V.S. Pande, J. Chem. Theory Comput. \textbf{6}, 434 (2010)

\bibitem{Hes2008JCTC1}
B.~Hess, J. Chem. Theory Comput. \textbf{4}, 116 (2008)

\bibitem{Che2005JCC}
J.~Chen, W.~Im, C.L. Brooks~III, J. Comput. Chem. \textbf{26}, 1565 (2005)

\bibitem{Hin1995PRE}
K.~Hinsen, G.R. Kneller, Phys. Rev. E \textbf{52}, 6868 (1995)

\bibitem{Mat1994PSFG}
A.M. Mathiowetz, A.~Jain, N.~Karasawa, W.A. {Goddard III}, PROTEINS: Struct.
  Funct. Gen. \textbf{20}, 227 (1994)

\bibitem{Maz1989JBMSD}
A.K. Mazur, R.A. Abagyan, J. Biomol. Struct. Dyn. \textbf{6}, 815 (1989)

\bibitem{Maz1997JCC}
A.K. Mazur, J. Comput. Chem. \textbf{18}, 1354 (1997)

\bibitem{Cic1986CPR}
G.~Ciccotti, J.P. Ryckaert, Comput. Phys. Rep. \textbf{4}, 345 (1986)

\bibitem{Chr2007CPC}
M.~Christen, C.D. Christ, W.F. {van Gunsteren}, ChemPhysChem \textbf{8}, 1557
  (2007)

\bibitem{Chr2005JCP}
M.~Christen, W.F. {van Gunsteren}, J. Chem. Phys. \textbf{122}, 144106 (2005)

\bibitem{Sai2004JCP}
H.~Saint-Martin, B.~Hess, H.J.C. Berendsen, J. Chem. Phys. \textbf{120}, 11133
  (2004)

\bibitem{Cot2004BITNM}
C.J. Cotter, S.~Reich, BIT Num. Math. \textbf{44}, 439 (2004)

\bibitem{Sto2003MS}
U.~Stocker, D.~Juchli, W.F. {van Gunsteren}, Mol. Simul. \textbf{29}, 123
  (2003)

\bibitem{Rei1996PRE}
S.~Reich, Phys. Rev. E \textbf{53}, 53 (1996)

\bibitem{Rei1998NA}
S.~Reich, Num. Alg. \textbf{19}, 213 (1998)

\bibitem{Pea1979JCP}
M.R. Pear, J.H. Weiner, J. Chem. Phys. \textbf{71}, 212 (1979)

\bibitem{Go1976MM}
N.~G{\={o}}, H.A. Scheraga, Macromolecules \textbf{9}, 535 (1976)

\bibitem{Hel1979JCP}
E.~Helfand, J. Chem. Phys. \textbf{71}, 5000 (1979)

\bibitem{Pec1980JCP}
P.~Pechukas, J. Chem. Phys. \textbf{72}, 6320 (1980)

\bibitem{Alm1990MP}
N.G. Almarza, E.~Enciso, J.~Alonso, F.J. Bermejo, M.~{\'A}lvarez, Mol. Phys.
  \textbf{70}, 485 (1990)

\bibitem{Ber1983Book}
H.J.C. Berendsen, W.F. Van~Gunsteren, in \emph{The Physics of Superionic
  Conductors and Electrode Materials}, edited by J.W. Perram (Plenum Press,
  1983), Vol. {NATO ASI Series B92}, pp. 221--240

\bibitem{Per1985MM}
D.~Perchak, J.~Skolnick, R.~Yaris, Macromolecules \textbf{18}, 519 (1985)

\bibitem{Dub2010JCP}
D.~Dubbeldam, G.A.E. Oxford, R.~Krishna, L.J. Broadbelt, R.Q. Snurr, J. Chem.
  Phys. \textbf{133}, 034114 (2010)

\bibitem{Mor2004ACP}
D.C. Morse, Adv. Chem. Phys. \textbf{128}, 65 (2004)

\bibitem{Hin1994JFM}
E.J. Hinch, J. Fluid Mech. \textbf{271}, 219 (1994)

\bibitem{vKa1984AJP}
N.G. Van~Kampen, J.J. Lodder, Am. J. Phys. \textbf{52}, 419 (1984)

\bibitem{Pet2000Thesis}
E.A.J.F. Peters, Ph.D. thesis, Technische Universiteit Delft (2000)

\bibitem{Ral1979JFM}
J.M. Rallison, J. Fluid Mech. \textbf{93}, 251 (1979)

\bibitem{Gal2007Book}
G.~Gallavotti, \emph{The Elements of Mechanics} (Ipparco Editore, 2007),
  available at \url{http://ipparco.roma1.infn.it/pagine/libri.html}

\bibitem{Cha1979JCP}
D.~Chandler, B.J. Berne, J. Chem. Phys. \textbf{71}, 5386 (1979)

\bibitem{Pas2002JCP}
M.~Pasquali, D.C. Morse, J. Chem. Phys. \textbf{116}, 1834 (2002)

\bibitem{Pat2004JCP}
A.~Patriciu, G.S. Chirikjian, R.V. Pappu, J. Chem. Phys. \textbf{121}, 12708
  (2004)

\bibitem{dOt1998JCP}
W.K. Den~Otter, W.J. Briels, J. Chem. Phys. \textbf{109}, 4139 (1998)

\bibitem{dOt2000MP}
W.K. Den~Otter, W.J. Briels, Mol. Phys. \textbf{98}, 773 (2000)

\bibitem{Kar1981MM}
M.~Karplus, J.N. Kushick, Macromolecules \textbf{14}, 325 (1981)

\bibitem{vGu1982MM}
W.F. Van~Gunsteren, M.~Karplus, Macromolecules \textbf{15}, 1528 (1982)

\bibitem{Maz2001CTPS}
A.K. Mazur, B.G. Sumpter, D.W. Noid, Comput. Theor. Polym. Sci. \textbf{11}, 35
  (2001)

\bibitem{Sai1963SSP}
N.~Sait{\^o}, K.~Okano, S.~Iwayanagi, T.~Hideshima, Solid State Phys.
  \textbf{14}, 343 (1963)

\bibitem{Tir1996CPL}
I.G. Tironi, R.M. Brunne, W.F. van Gunsteren, Chem. Phys. Lett. \textbf{250},
  19 (1995)

\bibitem{Maz2001Book}
A.K. Mazur, in \emph{Computational Biochemistry and Biophysics}, edited by O.M.
  Becker, A.D. {MacKerell Jr.}, B.~Roux, M.~Watanabe (Marcel Dekker Inc., 2001)

\bibitem{Fro2001CMP}
R.~Froese, I.~Herbst, Commun. Math. Phys. \textbf{220}, 489 (2001)

\bibitem{dCo1982PRA}
R.C.T. da~Costa, Phys. Rev. A \textbf{25}, 2893 (1982)

\bibitem{Alv1998MMTS}
R.F. {\'A}lvarez-Estrada, Macromol. Theory Simul. \textbf{7}, 457 (1998)

\bibitem{Alv2000MMTS}
R.F. {\'A}lvarez-Estrada, Macromol. Theory Simul. \textbf{9}, 83 (2000)

\bibitem{Alv2002MP}
R.F. {\'A}lvarez-Estrada, G.F. Calvo, Mol. Phys. \textbf{100}, 2957 (2002)

\bibitem{Alv2004JPCM}
R.F. {\'A}lvarez-Estrada, G.F. Calvo, J. Phys.: Condens. Matter \textbf{16},
  S2037 (2004)

\bibitem{Cal2005JPCM}
G.F. Calvo, R.F. {\'A}lvarez-Estrada, J. Phys.: Condens. Matter \textbf{17},
  7755 (2005)

\bibitem{Cal2008JPCM}
G.F. Calvo, R.F. {\'A}lvarez-Estrada, J. Phys.: Condens. Matter \textbf{20},
  035101 (2008)

\bibitem{Alv2011EPJSTunp}
R.F. {\'A}lvarez-Estrada, G.F. Calvo, Eur.\ Phys.\ J. ST  (2011), this issue.

\bibitem{Fix1978JCP}
M.~Fixman, J. Chem. Phys. \textbf{69}, 1527 (1978)

\bibitem{Bor1995TR3}
F.A. Bornemann, C.~Sch{\"u}tte, Tech. rep., Konrad-Zuse-Zentrum f{\"u}r
  Informationstechnik Berlin (1995)

\bibitem{vKa1985PR}
N.G. van Kampen, Phys. Rev. \textbf{124}, 69 (1985)

\bibitem{Ske2011EPJSTunp}
R.D. Skeel, R.~S., Eur.\ Phys.\ J. ST  (2011), this issue.

\bibitem{Baj2011EPJSTunp}
J.~Bajars, J.~Frank, B.~Leimkuhler, Eur.\ Phys.\ J. ST  (2011), this issue.

\bibitem{Mad2011EPJSTunp}
J.~Maddocks, C.~Hartmann, Eur.\ Phys.\ J. ST  (2011), this issue.

\bibitem{Fla2005AJP}
M.R. Flannery, Am. J. Phys. \textbf{73}, 265 (2005)

\bibitem{Koo1997ReMP}
W.S. Koon, J.E. Marsden, Rep. Math. Phys. \textbf{40}, 21 (1997)

\bibitem{vdS1994ReMP}
A.J. van~der Schaft, B.M. Maschke, Rep. Math. Phys. \textbf{34}, 225 (1994)

\bibitem{Mol2010Prep}
M.~Molina-Becerra, E.~Freire, J.G. Vioque, \emph{Equations of motion of
  nonholonomic {H}amiltonian systems}, Preprint obtained from
  \url{http://www.matematicaaplicada2.es/data/pdf/1276179170_1811485430.pdf}
  (2010)

\bibitem{Tuc2001JCP}
M.E. Tuckerman, Y.~Liu, G.~Ciccotti, G.J. Martyna, J. Chem. Phys. \textbf{115},
  1678 (2001)

\bibitem{Ech2010Sub}
P.~Echenique, C.N. Cavasotto, M.~De~Marco, P.~Garc{\'{\i}}a-Risue{\~n}o, J.L.
  Alonso, \emph{An exact expression to calculate the derivatives of
  position-dependent observables in molecular simulations with flexible
  constraints}, Submitted (2010)

\bibitem{Wei2009Ebook}
T.~Weise, \emph{Global optimization algorithms --theory and application--},
  Retrieved from \url{http://www.it-weise.de} (2009)

\bibitem{Kir1983Sci}
S.~Kirkpatrick, C.D. Gelatt, M.P. Vecchi, Science \textbf{220}, 671 (1983)

\bibitem{Yu2001JMS}
C.H. Yu, M.A. Norman, L.~Sch{\"{a}}fer, M.~Ramek, A.~Peeters, C.~van Alsenoy,
  J. Mol. Struct. \textbf{567--568}, 361 (2001)

\bibitem{Sch1995BP}
L.~Sch{\"{a}}fer, M.~Cao, M.J. Meadows, Biopolymers \textbf{35}, 603 (1995)

\bibitem{Sch1997JMS}
L.~Sch{\"{a}}fer, M.~Cao, M.~Ramek, B.J. Teppen, S.Q. Newton, K.~Siam, J. Mol.
  Struct. \textbf{413--414}, 175 (1997)

\bibitem{Hag1979JACS}
A.T. Hagler, P.S. Stern, R.~Sharon, J.M. Becker, F.~Naider, J. Am. Chem. Soc.
  \textbf{101}, 6842 (1978)

\bibitem{Aba1989JBMSD}
R.A. Abagyan, A.K. Mazur, J. Biomol. Struct. Dyn. \textbf{6}, 833 (1989)

\bibitem{Abe1984CC}
H.~Abe, W.~Braun, N.~G{\={o}}, Comp. Chem. \textbf{8}, 239 (1984)

\bibitem{Edb1986JCP}
R.~Edberg, D.J. Evans, G.P. Morris, J. Chem. Phys. \textbf{84}, 6933 (1986)

\bibitem{Maz1991JCoP}
A.K. Mazur, V.E. Dorofeev, R.A. Abagyan, J. Comput. Phys. \textbf{92}, 261
  (1991)

\bibitem{Nog1983JPSJ}
T.~Noguti, N.~G{\={o}}, J. Phys. Soc. Japan \textbf{52}, 3685 (1983)

\bibitem{Fix1974PNAS}
M.~Fixman, Proc. Natl. Acad. Sci. USA \textbf{71}, 3050 (1974)

\bibitem{Li2009PRL}
D.W. Li, R.~Br{\"u}schweiler, Phys. Rev. Lett. \textbf{102}, 118108 (2009)

\bibitem{Ryc1977JCoP}
J.P. Ryckaert, G.~Ciccotti, H.J.C. Berendsen, J. Comput. Phys. \textbf{23}, 327
  (1977)

\bibitem{Sch2003MP}
J.~Schlitter, M.~Kl{\"{a}}n, Mol. Phys. \textbf{101}, 3439 (2003)

\bibitem{vGu1977MP}
W.F. {van Gunsteren}, H.J.C. Berendsen, Mol. Phys. \textbf{34}, 1311 (1977)

\bibitem{vGu1980MP}
W.F. {van Gunsteren}, Mol. Phys. \textbf{40}, 1015 (1980)

\bibitem{vGu1989Book}
W.F. Van~Gunsteren, in \emph{Computer Simulations of Biomolecular Systems},
  edited by W.F. Van~Gunsteren, P.K. Weiner (Escom science publishers,
  Netherlands, 1989), pp. 27--59

\bibitem{Hes2004JCP}
B.~Hess, H.~Saint-Martin, H.J.C. Berendsen, J. Chem. Phys. \textbf{120}, 11133
  (2004)

\bibitem{Sza1996Book}
A.~Szabo, N.S. Ostlund, \emph{Modern Quantum Chemistry: Introduced to Advanced
  Electronic Structure Theory} (Dover Publications, New York, 1996)

\bibitem{Fer1985MP}
M.~Ferrario, J.P. Ryckaert, Mol. Phys. \textbf{54}, 587 (1985)

\bibitem{Hes2011Priv}
B.~Hess, Private communication (2011)

\bibitem{Yu2011EPJSTunp}
T.Q. Yu, M.~Tuckerman, Eur.\ Phys.\ J. ST  (2011), this issue.

\bibitem{Elb2011EPJSTunp}
R.~Elber, B.~Hess, Eur.\ Phys.\ J. ST  (2011), this issue.

\bibitem{Rei1995PD}
S.~Reich, Physica D \textbf{89}, 28 (1995)

\bibitem{Arn1989Book}
V.I. Arnold, \emph{Mathematical Methods of Classical Mechanics}, Graduate Texts
  in Mathematics, 2nd~edn. (Springer, New York, 1989)

\bibitem{Rub1957CPAM}
H.~Rubin, P.~Ungar, Commun. Pure Appl. Math. \textbf{X}, 65 (1957)

\bibitem{Cic2005CPC}
G.~Ciccotti, R.~Kapral, E.~Vanden-Eijnden, ChemPhysChem \textbf{6}, 1809 (2005)

\bibitem{Tor1977JCoP}
G.M. Torrie, J.P. Valleau, J. Comput. Phys. \textbf{23}, 187 (1977)

\bibitem{Sch2003JCP}
J.~Schlitter, M.~Kl{\"{a}}n, J. Chem. Phys. \textbf{118}, 2057 (2003)

\bibitem{And2001JCP}
I.~Andricioaei, M.~Karplus, J. Chem. Phys. \textbf{115}, 6289 (2001)

\bibitem{Has1974JCP}
O.~Hassager, J. Chem. Phys. \textbf{60}, 2111 (1974)

\bibitem{Rei2000PD}
S.~Reich, Physica D \textbf{118}, 210 (2000)

\bibitem{Har2007PD}
C.~Hartmann, C.~Sch{\"u}tte, Physica D \textbf{228}, 59 (2007)

\bibitem{Tox2009JCP}
S.~Toxvaerd, O.J. Heilmann, T.~Ingebrigtsen, T.B. Schr{\o}der, J.C. Dyre, J.
  Chem. Phys. \textbf{131}, 064102 (2009)

\bibitem{Cor1995JACS}
W.D. Cornell, P.~Cieplak, C.I. Bayly, I.R. Gould, J.~Merz, K.~M., D.M.
  Ferguson, D.C. Spellmeyer, T.~Fox, J.W. Caldwell, P.A. Kollman, J. Am. Chem.
  Soc. \textbf{117}, 5179 (1995)

\bibitem{Kol1997Book}
P.~Kollman, R.~Dixon, W.~Cornell, T.~Fox, C.~Chipot, A.~Pohorill, in
  \emph{Computer Simulations of Biomolecular Systems}, edited by W.F.
  Van~Gunsteren, P.K. Weiner, A.J. Wilkinson (Kluwer Academic Publishing,
  Dordrecht, 1997), Vol.~3, pp. 83--96

\bibitem{Gaussian03E01}
M.J. Frisch, G.W. Trucks, H.B. Schlegel, G.E. Scuseria, M.A. Robb, J.R.
  Cheeseman, J.A. Montgomery, Jr., T.~Vreven, K.N. Kudin, J.C. Burant et~al.,
  \emph{Gaussian 03, {R}evision {E}.01} (2007), {G}aussian, Inc., Wallingford,
  CT.

\end{thebibliography}

\end{document}